\documentclass[%
 reprint,
 superscriptaddress, oneside,
 amsmath,amssymb,
 aps,
 pra,
 floatfix,
]{revtex4-2}
\usepackage[table]{xcolor}
\usepackage{rotating}
\usepackage{graphicx}
\usepackage{colortbl}
\definecolor{zebragray}{gray}{0.94}
\usepackage{dcolumn}
 \usepackage{mathrsfs}
\usepackage{bm}

\usepackage{dsfont}
\usepackage{placeins}
\usepackage{xcolor}
\usepackage{amsthm}
\usepackage{longtable}
\usepackage{multirow}
\usepackage{booktabs}
\usepackage{hyperref}

\usepackage[linesnumbered, ruled, vlined]{algorithm2e}
\SetKwInput{KwInput}{Input}
\SetKwInput{KwOutput}{Output}
\usepackage{graphicx} 
\usepackage{makecell}
\newtheorem{definition}{Definition}

\newtheorem{lemma}{Lemma}
\usepackage{array}
\usepackage{amsmath, amssymb}
\usepackage{mathtools}
\usepackage{amsmath}
\usepackage{amsfonts}
\usepackage{amssymb}
\usepackage{amsthm}
\usepackage{makecell}
\usepackage{physics}
\newcommand{\F}{\mathbb{F}}
\newcommand{\SL}{\mathrm{SL}}

\usepackage[textsize=tiny]{todonotes}

\newtheorem*{remark}{Remark}
\theoremstyle{definition}
\newtheorem{example}{Example}
\theoremstyle{definition}
\usepackage{amsmath, amssymb, amsthm, verbatim, graphicx, xcolor, bbm, braket, hyperref}
\usepackage{tikz}
\usetikzlibrary{shapes, arrows, positioning, decorations.pathreplacing}
\definecolor{nodefill}{RGB}{221, 238, 243}
\definecolor{border}{RGB}{100, 149, 175}
\definecolor{arrowcolor}{RGB}{135, 206, 235}
\definecolor{annotationcolor}{RGB}{100, 100, 100}
\usepackage{amsmath, amssymb}
\newcommand\editcolor[1]{#1}
\usepackage[textsize=tiny]{todonotes}

\usepackage{listings}
\usepackage{color}
\definecolor{mygreen}{rgb}{0,0.6,0}
\definecolor{myred}{rgb}{0.9,0.2,0.3}
\definecolor{mymauve}{rgb}{0.28,0,0.92}
\definecolor{mygray}{rgb}{0.5,0.5,0.5}

\lstdefinelanguage{JuliaREPL}{
  keywords={julia},
  keywordstyle=\color{mygreen}\bfseries,
  keywords=[2]{Stabilizer, GeneralizedStabilizer, naive_encoding_circuit, PauliOperator, MixedDestabilizer, project, CliffordOperator, apply, stabilizerview, destabilizerview, logicalxview, logicalzview, MixedStabilizer, traceout, random_stabilizer, canonicalize, naive_syndrome_circuit, parity_checks, Steane7, pftrajectories, pfmeasurements, PauliError, code_n},
  keywordstyle=[2]\color{mymauve}\bfseries,
  identifierstyle=\color{black},
  sensitive=true,
  mathescape=true,
  comment=[l]{\#},
  morecomment=[s]{/*}{*/},
  commentstyle=\color{mygray}\ttfamily,
  stringstyle=\color{myred}\ttfamily,
  morestring=[b]',
  morestring=[b]"
}

\DeclareMathOperator{\wt}{wt}

\begin{document}

\title{Quantum Tanner Codes at Moderate Blocklength}

\author{Feroz Ahmed Mian}
\email{fmian@umass.edu}
\affiliation{Manning College of Information and Computer Sciences, University of Massachusetts Amherst, 140 Governors Drive Amherst, Massachusetts 01003, USA}
\author{Vaishnavi L. Addala}
\affiliation{Research Laboratory of Electronics, Massachusetts Institute of Technology, Cambridge, MA 02139, USA}
\affiliation{Department of Electrical Engineering and Computer Science,
Massachusetts Institute of Technology, Cambridge, MA 02139, USA}
\author{Arman Meraj}
\affiliation{Manning College of Information and Computer Sciences, University of Massachusetts Amherst, 140 Governors Drive Amherst, Massachusetts 01003, USA}
\author{Adhiraj Chadha}
\affiliation{Manning College of Information and Computer Sciences, University of Massachusetts Amherst, 140 Governors Drive
Amherst, Massachusetts 01003, USA}
\author{Stefan Krastanov}
\email{skrastanov@umass.edu}
\affiliation{Manning College of Information and Computer Sciences, University of Massachusetts Amherst, 140 Governors Drive Amherst, Massachusetts 01003, USA}
\affiliation{Department of Physics, University of Massachusetts Amherst, 710 North Pleasant Street Amherst, Massachusetts 01003, USA}
\date{\today} 
\begin{abstract}
We present explicit constructions of quantum Tanner (QT) codes with good rate and distance, obtained through two complementary approaches: the left-right Cayley complex (LRCC) description and the ``lifting'' perspective, in which a seed Calderbank--Shor--Steane (CSS) code is lifted by commuting left-right regular actions of a finite group~$\mathcal{G}$. Through an extensive search over non-abelian groups from GAP's \texttt{SmallGrp} library, we investigate the moderate blocklength regime ($n \in [500, 1000]$) and identify several new code instances with distance upper bounds exceeding $20$. These include $[[480, 8,\, (\le 21,\, \le 21)]]$, $[[504, 4,\, (\le 36,\, \le 27)]]$, $[[672, 4,\, (\le 48,\, \le 28)]]$, $[[720, 6,\, (\le 30,\, \le 30)]]$, and $[[864, 8,\, (\le 39,\, \le 31)]]$, with these bounds obtained using up to $350$ million trials of \texttt{sQetch}, a randomized distance estimator. The code instances presented have check weights ranging from $9$ to $20$. Using the Tesseract decoder, we estimate pseudo-thresholds of $3.6$--$4.6\%$ under phenomenological noise and $0.14$--$0.27\%$ under circuit-level noise, comparable to prior results at shorter code lengths. We also provide \texttt{QuantumExpanders.jl}, an open-source \texttt{Julia} library for constructing QT codes and explicit constructions of Ramanujan graphs.
\end{abstract}
\maketitle

\section{Introduction}

\editcolor{Quantum low-density parity-check (qLDPC) codes have been central to fault-tolerant quantum computing (FT-QC) since Kitaev's surface code~\cite{kitaev2003fault}. A quantum code with parameters $[[n,k,d]]$ encodes $k$ logical qubits into $n$ physical qubits with code distance $d$; for CSS codes, one further distinguishes the distances $d_x$ and $d_z$ against phase-flip and bit-flip errors, respectively, with $d = \min\{d_x, d_z\}$. The surface code achieves $[[n,1,(\sqrt{n}, \sqrt{n})]]$ using geometrically local, two-dimensional nearest-neighbor checks. For codes restricted to such 2D-local interactions, Bravyi, Poulin, and Terhal proved the bound $kd^2 \le cn$ for some constant $c$ (the BPT bound)~\cite{PhysRevLett.104.050503}; this bound is tight up to a constant factor, achieved using $k$ independent copies of the surface code. Escaping this trade-off requires relaxing geometric locality while retaining the sparsity (bounded-weight checks and bounded-degree qubits) that defines qLDPC codes more broadly. Tillich and Z\'emor's hypergraph product (HP) construction~\cite{6671468, tillichquantum} was the first to exploit this relaxation, achieving constant rate with $[[n,\Theta(n),\Theta(\sqrt{n})]]$ codes. Gottesman showed that constant-rate qLDPC codes admitting a suitable decoder can achieve FT-QC with constant overhead~\cite{gottesman2013fault}. Fawzi, Grospellier, and Leverrier applied this framework to quantum expander codes using the HP technique, proving an efficient, single-shot decoder that remains robust even under noisy syndrome measurements~\cite{fawzi2020constant}; even $\Theta(\sqrt{n})$-distance qLDPC codes therefore already met this practical benchmark. The $\sqrt{n}$ distance barrier itself nonetheless persisted for over a decade until Hastings, Haah, and O'Donnell's fiber bundle codes~\cite{hastings2021fiber} pushed the distance to $\Omega(n^{3/5}/\mathrm{polylog}(n))$, and was finally broken by Panteleev and Kalachev~\cite{Panteleev_2022,panteleev2022asymptotically}, whose asymptotically good qLDPC codes, based on the lifted product construction, achieve $[[n,\Theta(n),\Theta(n)]]$.}

\editcolor{Following the breakthrough of asymptotically good constant-rate and linear-distance qLDPC codes~\cite{panteleev2022asymptotically}, additional constructions of good codes have been proven~\cite{leverrier2022quantum, dinur2023good}, several extending the HP framework~\cite{tillichquantum} via increasingly polished chain-complex product constructions, and fast decoders have been developed~\cite{gu2022efficient, leverrier2023efficient, dinur2023good}, demonstrating that practical error correction is achievable at scale. Among these, the QT codes of Leverrier \textit{et al.}~\cite{leverrier2022quantum} offer a conceptually simpler route to asymptotically good quantum codes: rather than building on chain-complex products, the code is obtained by placing classical Tanner codes directly on the two graphs underlying a square complex proposed by Dinur \textit{et al.} \cite{dinur2022locally}. For near-term devices, small explicit instances of QT codes have recently been identified~\cite{radebold2025explicit, wang2026check, leverrier2025small}, with competitive parameters such as the weight-9 $[[144,12,(11,11)]]$ code achieving performance comparable to Gross's weight-6 $[[144,12,(12,12)]]$ code~\cite{leverrier2025small}, demonstrating the practical relevance of this code family. A central question raised by~\cite{leverrier2025small} nevertheless remains open: \textit{can codes with competitive parameters be found in the moderate blocklength regime ($n \in [500, 1000]$)?} While Wang~\textit{et al.}~\cite{wang2026check} identify explicit QT codes in this regime, their search was restricted to a limited set of groups. We present a comprehensive search over many new groups, with particular emphasis on non-abelian groups, to address this gap.}

\editcolor{Recent construction of QT codes adopt the ``lifting" point of view from \cite{leverrier2025small}: a QT code is obtained from small local codes and two lifts associated with multisets $\mathcal{A}, \mathcal{B}$ of a finite group $\mathcal{G}$. While the original construction provides bounds on the rate and code distances, the lifting perspective has proven more useful to QT code exploration. The key challenge is that computing the rate and distance parameters of a QT code directly from its local structure and choice of lift appears intractable in general. As discussed in Ref.~\cite{leverrier2025small}, only special cases admit closed-form solutions, for instance when one local code is the $[2,1,2]$ repetition code \cite[see Theorem 1]{leverrier2025small}. This barrier fundamentally motivates an extensive search approach: without direct formulas for dimension and code distance, one must systematically explore many choices of local codes and underlying groups.}

\editcolor{In this work, we greatly expand this search to new groups from GAP's \texttt{SmallGrp} \cite{SmallGrp} library,  using \texttt{sQetch}~\cite{bhardwaj2026high} and \texttt{QDistRnd}~\cite{pryadko2023qdistrnd} for cross-validation of distance estimates. We conduct an extensive search over these groups to construct QT codes via the left-right Cayley complex (LRCC), a two-dimensional combinatorial structure labeled by two generating multisets $\mathcal{A},\mathcal{B}$ of a finite group $\mathcal{G}$, and via lifted QT codes constructed from commuting left-right group actions. By combining both code construction approaches with our extensive search over non-abelian groups, we address the open question of whether competitive moderate-blocklength codes exist in the $n \in [500, 1000]$ regime.}

\editcolor{The paper is organized as follows. Section~\ref{sec:preliminaries} covers the background on graph theory, group theory, and stabilizer codes needed for the QT construction; readers familiar with this material can go directly to Section~\ref{sec:quantumtannercodes}, which describes QT codes from both the LRCC and lifting perspectives. Section~\ref{sec:numerical-results-main-text} presents our search heuristics over groups from GAP's \texttt{SmallGrp}~\cite{SmallGrp} library and benchmarks four representative codes under phenomenological and circuit-level noise. We conclude with a discussion in Section~\ref{sec:discussion}.}
\section{Preliminaries}
\label{sec:preliminaries}
This section develops the prerequisites for the QT code construction in Section~\ref{sec:quantumtannercodes}. We begin with expander graphs and their expansion properties, leading to Ramanujan graphs and then to Tanner and expander codes. Since the explicit Ramanujan families used here are built from Cayley graphs, and the LRCC is the combinatorial backbone of the original QT code construction, we also recall the necessary background on generating sets and Cayley graphs. We conclude by recalling stabilizer and Calderbank--Shor--Steane (CSS) codes. Readers familiar with this material can skip ahead to Section~\ref{sec:quantumtannercodes}.
\subsection{Expander Graphs}
\begin{figure*}[ht!]
    \includegraphics[width=1.0\linewidth]{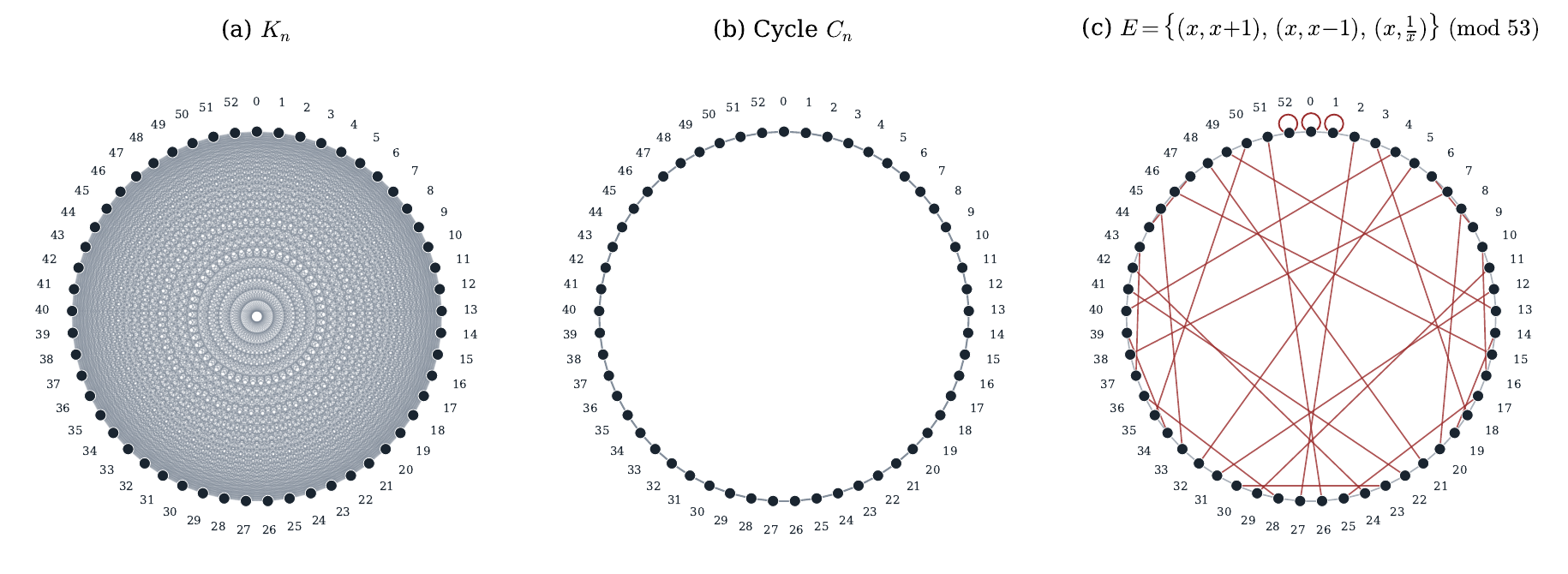}
     \caption{Expander graphs must be simultaneously sparse and highly connected; the three graphs on $n = 53$ vertices shown here separate these two requirements to illustrate why neither alone suffices. (a) The complete graph $K_n$ attains optimal edge and vertex expansion but requires $\Theta(n^2)$ edges. (b) The cycle $C_n$ has only $n$ edges but is a poor expander: removing a contiguous arc $S$ of $n/2$ vertices severs just two edges, so $h(S) = O(1/n)$. (c) Adjoining the edges $\{(x, 1/x) : x \in \mathbb{Z}_n\}$ (with $1/0 := 0$) to the cycle, with arithmetic taken modulo the prime $n = 53$, keeps the degree at most $3$ while restoring $\varepsilon$-edge and vertex expansion; the three self-loops mark the fixed points $0, 1, n-1 \in \mathbb{Z}_n$ of $x \mapsto 1/x$, which therefore have only two distinct neighbors. This construction is due to Margulis \cite{Mar73}.}
\label{fig:expander-example}
\end{figure*}
Expander graphs are sparse graphs that are nonetheless highly connected: every subset of vertices has many neighbors outside it. 

Formally, let $\Gamma = (V,E)$ be a finite, undirected, $r$-regular graph with $|V|=n$. In the bipartite setting, $\Gamma$ has vertex bipartition $V = I \cup O$ with $|I| = |O| = n$ (so $|V| = 2n$), and every edge connects a vertex in $I$ to a vertex in $O$.

For $S, T \subseteq V$, write
\begin{equation}
    E(S,T) = \bigl\{\, (s,t) : s \in S,\ t \in T,\ \{s,t\} \in E \,\bigr\}
\end{equation}
for the edges between $S$ and $T$, and $\bar{S} = V \setminus S$ for the complement of $S$.

The \emph{edge boundary} of $S$ is $\partial S = E(S, \bar S)$, the edges leaving $S$. The \emph{vertex boundary} of $S$ is
\begin{equation}
    N(S) = \bigl\{\, v \in \bar S \;\big|\; \exists\, u \in S : \{u,v\} \in E \,\bigr\},
\end{equation}
the vertices outside $S$ adjacent to some vertex of $S$; in the bipartite case ($S \subseteq I$), $N(S) \subseteq O$.

There are two standard combinatorial measures of expansion, distinguished by which boundary is tracked: the number of edges leaving $S$, or the number of vertices outside $S$ reachable from it.

\begin{definition}[Vertex expansion]
$\Gamma$ is an $\varepsilon$-\emph{vertex expander} if
\begin{equation}
    \frac{|N(S)|}{|S|} \;\ge\; \varepsilon
\end{equation}
for every $S \subseteq V$ (or $S \subseteq I$ in the bipartite case) with
$|S| \le n/2$.
\end{definition}

\begin{definition}[Edge expansion]
The edge expansion of $S$ is
\begin{equation}
    h(S) = \frac{|\partial S|}{|E(S,V)|},
\end{equation}
the fraction of edges incident to $S$ that leave it. $\Gamma$ is an $\varepsilon$-\emph{edge expander} if $h(S) \ge \varepsilon$ for every $S \subseteq V$ with $|S| \le n/2$.
\end{definition}

The restriction to $|S| \le n/2$ compares the boundary against the smaller side of the cut $(S, \bar S)$; for $|S| > n/2$ no nontrivial bound is possible, since $S$ could itself constitute the larger component of a disconnected graph. The two notions are related but not equivalent: both formalize the requirement that small sets cannot be separated from the rest of the graph while the number of edges incident to them remains bounded.

\begin{example}
The complete graph $K_n$ attains the optimal value of both notions, a $\tfrac{1}{2}$-edge expander and a $1$-vertex expander, since every vertex outside $S$ is already adjacent to $S$; however, $K_n$ has $\Theta(n^2)$ edges, motivating the question of whether sparse graphs ($r = O(1)$) can also expand well. The cycle graph in Fig. \ref{fig:expander-example} illustrates that sparsity alone does not suffice: removing a set $S$ of size $n/2$ severs only two edges regardless of $n$, so $h(S) = O(1/n)$. Adding a small number of additional edges per vertex remedies this. On $V = \mathbb{Z}_p$ for prime $p$, the graph with edge set $E = \{(x, x{+}1), (x, x{-}1), (x, x^{-1})\}$ (with $0^{-1} := 0$) satisfies $\varepsilon$-edge and $\varepsilon'$-vertex expansion, for some constants $\varepsilon, \varepsilon' > 0$. This construction is due to Margulis \cite{Mar73}, and establishing its expansion requires number-theoretic results (Selberg's $3/16$ theorem \cite{selberg1965estimation}, Kazhdan's property~(T) \cite{kavzdan1967connection}) rather than combinatorial arguments.
\end{example}

A more analytic measure of expansion comes from the spectrum of the graph. Let $A \in \mathbb{R}^{V \times V}$ be the normalized adjacency operator of $\Gamma$, with $A_{uv} = 1/r$ if $(u,v) \in E$ and $A_{uv} = 0$ otherwise. Since $A$ is symmetric and nonnegative, its eigenvalues $1 = \lambda_1 \ge \lambda_2 \ge \cdots \ge \lambda_n \ge -1$ are real, with $\lambda_1 = 1$ attained by the all-ones vector.

\begin{definition}[Spectral expander]
$\Gamma$ is a $\lambda$-spectral expander if
\begin{equation}
    \lambda(\Gamma) \;:=\; \max\bigl(|\lambda_2|, |\lambda_n|\bigr) \;\le\; \lambda.
\end{equation}
\begin{figure}[ht]
    \centering
    \includegraphics[width=1.0\linewidth]{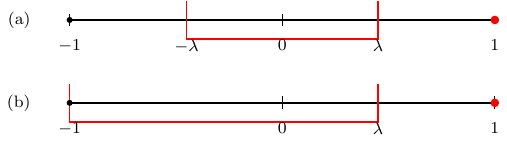}
\caption{Location of the non-trivial eigenvalues of $\Gamma$ relative to the trivial eigenvalue $\lambda_1=1$: (a) two-sided spectral expansion, $|\lambda_2|,|\lambda_n|\le\lambda$; (b) one-sided spectral expansion, where only $\lambda_2\le\lambda$ is required, with no control on $\lambda_n$.}
\label{fig:spectral}
\end{figure}
(see Fig.~\ref{fig:spectral}(a)). $\Gamma$ is a \emph{one-sided} $\lambda$-spectral expander if only the bound $\lambda_2 \le \lambda$ is required, with no control on $\lambda_n$ (see
Fig.~\ref{fig:spectral}(b)).
\end{definition}
\begin{remark}
The one-sided notion is strictly weaker: the complete bipartite graph is a one-sided expander for $\lambda=0$ but not two-sided, since its most negative eigenvalue is $\lambda_n=-1$. The one-sided notion is the one frequently used for higher-dimensional expanders.
\end{remark}

Smaller $\lambda$ corresponds to a larger spectral gap $1 - \lambda$ and hence to stronger expansion. The remarkable connection between the spectral and combinatorial pictures is the expander mixing lemma: a small spectral gap forces every pair of large sets to see approximately the number of edges between them that a random $r$-regular graph would.

\begin{lemma}[Expander mixing lemma~\cite{alon1986eigenvalues}]
\label{lem:mixing}
Let $\Gamma$ be an $r$-regular $\lambda$-spectral expander on $n$ vertices, and let $S, T \subseteq V$. Writing $E(S,T)$ for the number of edges with one endpoint in $S$ and one in $T$,
\begin{equation}
    \left| \, E(S,T) - \frac{r\,|S|\,|T|}{n} \, \right| \;\le\; \lambda r \sqrt{|S|\,|T|}.
\end{equation}
\end{lemma}

The term $r|S||T|/n$ is exactly the expected edge count between $S$ and $T$ in a random $r$-regular graph on $n$ vertices, so Lemma~\ref{lem:mixing} states that a spectral expander behaves like a pseudorandom graph at the level of edge counts.

\subsection{Ramanujan Graphs}

Ramanujan graphs are $r$-regular graphs whose eigenvalue spectrum is as clustered near zero as possible, making them spectrally optimal expanders. Formally, a connected $r$-regular graph $\Gamma$ is a \emph{Ramanujan graph} if its second-largest adjacency eigenvalue $\lambda(\Gamma)$ satisfies
\begin{equation}
\label{eq:expander_inequality}
    \lambda(\Gamma) \leq 2\sqrt{r-1}.
\end{equation}
By the Alon--Boppana theorem \cite{alon1986eigenvalues}, $2\sqrt{r-1}$ is the asymptotically optimal lower bound on $\lambda(\Gamma)$ for any infinite family of $r$-regular graphs, so Ramanujan graphs achieve the best possible spectral gap. This spectral optimality translates directly into combinatorial expansion~\cite{morgenstern1994existence}: any Ramanujan graph is an $(n, r, d)$-expander with $d = \Theta(r)$, satisfying Eq.~\eqref{eq:expander_inequality} for every admissible $S$.

\begin{example}[LPS construction \cite{lubotzky1986explicit}: $X^{5,29}$]
Take $p=5$ and $q=29$, with $p,q\equiv 1\pmod{4}$. Since
$\left(\frac{5}{29}\right)=1$, the LPS construction gives a Cayley graph of
$\mathrm{PSL}_2(\mathbb{F}_{29})$. Thus,
\begin{align}
\left|\mathrm{PSL}_2(\mathbb{F}_{29})\right|
&=\frac{29(29^2-1)}{2}=12180.
\end{align}
The $p+1=6$ generators are obtained by filtering the $8(p+1)=48$ integer representations of $p$ as a sum of four squares,
\begin{align}
5&=a_0^2+a_1^2+a_2^2+a_3^2,
\end{align}
to those satisfying $a_0>0$ odd and $a_1,a_2,a_3$ even, as prescribed by the LPS construction. For each such solution, let $\alpha=a_0+a_1\mathbf{i}+a_2\mathbf{j}+a_3\mathbf{k}$. Choose $u\in\mathbb{F}_{29}$ satisfying $u^2=-1$. The corresponding matrix is
\begin{align}
M_\alpha&=
\begin{bmatrix}
a_0+u a_1 & a_2+u a_3\\
-a_2+u a_3 & a_0-u a_1
\end{bmatrix},
\end{align}
which satisfies $\det(M_\alpha)=a_0^2+a_1^2+a_2^2+a_3^2=5$. Since $\left(\frac{5}{29}\right)=1$, there exists $s\in\mathbb{F}_{29}$ with $s^2=5$. Hence $\det(s^{-1}M_\alpha)=s^{-2}\det(M_\alpha)=1$, so $s^{-1}M_\alpha\in\mathrm{SL}_2(\mathbb{F}_{29})$. Its image in $\mathrm{PSL}_2(\mathbb{F}_{29})$ gives a generator of the Cayley graph.

The resulting graph is $6$-regular and therefore has $6\cdot12180/2=36540$ edges. Its largest nontrivial adjacency eigenvalue in absolute value is approximately $4.4420$, which is below the Ramanujan bound $2\sqrt{5}\approx4.4721$. Thus $X^{5,29}$ is a $6$-regular Ramanujan graph.
\end{example}
\subsection{Tanner Codes}
Tanner codes build a large code out of a graph and a small code: bits live on the edges of the graph, and a string is a codeword exactly when, at every vertex, the bits on its incident edges form a codeword of a fixed small code.

Formally, let $G = (L, R, E)$ be an $r$-regular bipartite graph with $|L| = |R| = n$, so $|E| = nr = m$. Bits are assigned to edges, making codewords binary strings of length $m$. For each vertex $v \in L \cup R$,
let $\mathcal{E}(v)$ denote its $r$ incident edges with a fixed ordering, and write $c|_{\mathcal{E}(v)} \in \mathbb{F}_2^r$ for the restriction of $c$ to those edges. Let $\mathcal{C}_0 \subseteq \{0,1\}^r$ be a linear base code of length $r$, rate $R_0$, and relative minimum distance $\Delta_0$.

\begin{definition}[Tanner code~\cite{tanner1981recursive}]
\begin{align}
    \mathcal{T}(G, \mathcal{C}_0) = \bigl\{\, c \in \{0,1\}^{m} \;\big|\;
    \forall\, v \in L \cup R,\; c\big|_{\mathcal{E}(v)} \in \mathcal{C}_0
    \,\bigr\}.
\end{align}
\end{definition}

Since $\mathcal{C}_0$ is linear, $\mathcal{T}(G, \mathcal{C}_0)$ is linear: $(c + c')|_{\mathcal{E}(v)} = c|_{\mathcal{E}(v)} + c'|_{\mathcal{E}(v)} \in \mathcal{C}_0$ at every vertex. Each vertex contributes $r(1-R_0)$ independent parity checks across $2n$ vertices, giving
\begin{align}
    \dim\bigl(\mathcal{T}(G, \mathcal{C}_0)\bigr) \;\geq\;
    m - 2n\cdot r(1-R_0) \;=\; m(2R_0-1),
\end{align}
so a positive rate requires $R_0 > \tfrac{1}{2}$. When $G$ is
$\varepsilon$-pseudorandom with $\varepsilon < \Delta_0$, the minimum
distance satisfies
\begin{align}
    \Delta\bigl(\mathcal{T}(G, \mathcal{C}_0)\bigr)
    \;\geq\; \Delta_0(\Delta_0 - \varepsilon)\,m.
\end{align}

\begin{figure*}
    \centering
    \includegraphics[width=1.0\linewidth]{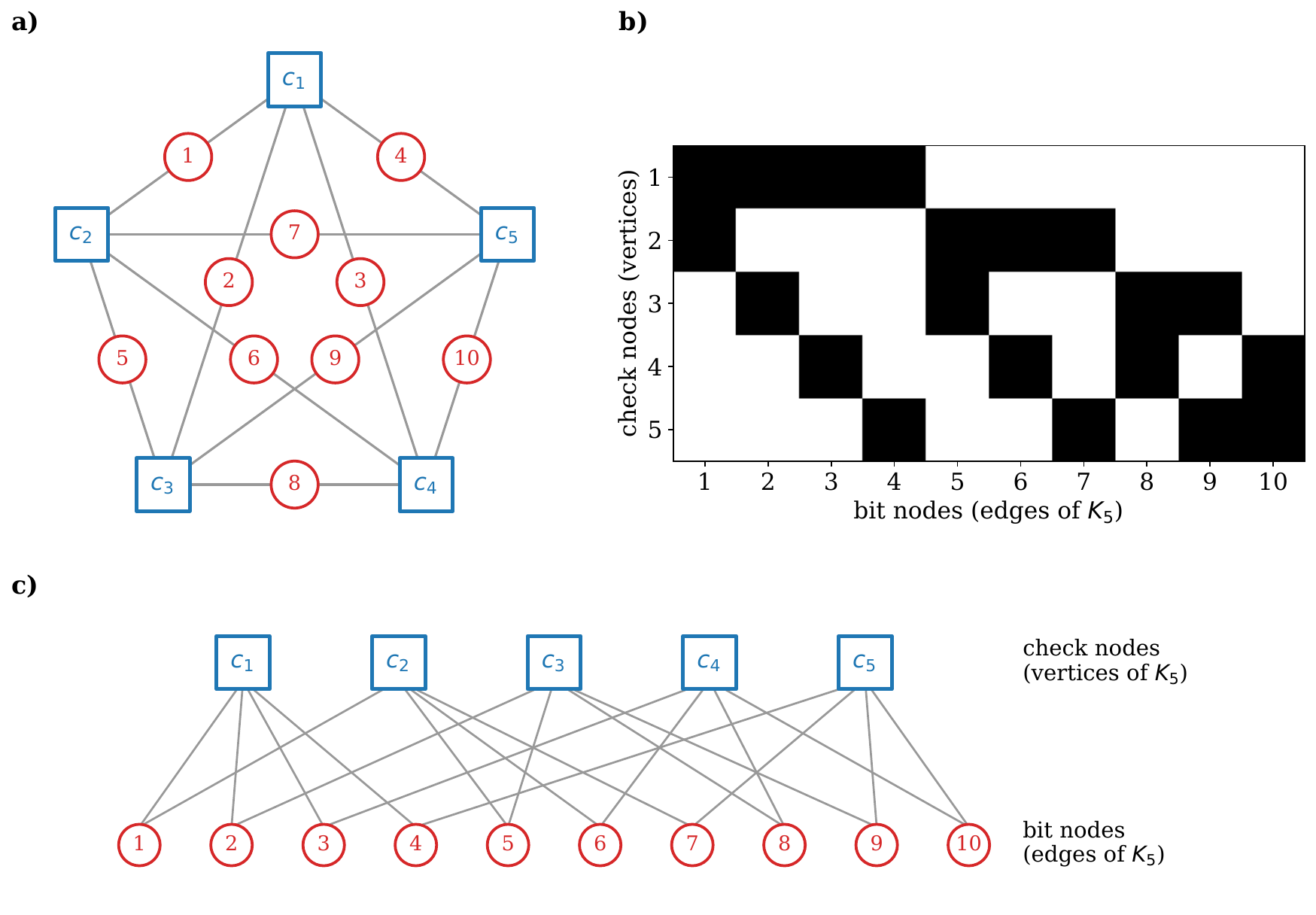}
    \caption{Tanner code construction from $K_5$~\cite{tanner1981recursive}. (a)~Bipartite graph: square nodes are check nodes (vertices of $K_5$, degree four) and circular nodes are bit nodes (edges of $K_5$, degree two), each incident to exactly two check nodes. (b)~Parity-check matrix $H\in\mathbb{F}_2^{5\times 10}$; entry $H_{ij}=1$ if and only if bit $j$ is incident to check $i$. The kernel of $H$ is the $[10,6,3]$ code. (c) Bipartite Tanner graph of the code: each bit node (bottom) is connected to the two check nodes (top) corresponding to the endpoints of its edge in $K_5$; the edges of this bipartite graph are in one-to-one correspondence with the nonzero entries of $H$. Since $K_5$ is $4$-regular, the local view of each vertex is a length-four bit string, one bit per incident edge, which is constrained to lie in the local code $C_0 \subseteq \mathbb{F}_2^4$.}
    \label{fig:tanner_code_example}
\end{figure*}

\begin{example}
Let $\mathcal{G}$ be the Tanner graph of $K_5$: five check nodes of degree four and $\binom{5}{2}=10$ bit nodes of degree two (Fig.~\ref{fig:tanner_code_example}a). At each vertex $v$, the four incident bits $c|_{\mathcal{E}(v)}$ must satisfy a simple parity check, i.e.\
$\sum_{e \in \mathcal{E}(v)} c_e = 0 \pmod{2}$. The code $\mathcal{T}(G,\mathcal{C}_0)$ is the set of binary strings of length~10 satisfying all five vertex checks simultaneously, yielding the $[10,6,3]$ code~\cite{tanner1981recursive}, whose parity-check matrix
$H\in\mathbb{F}_2^{5\times 10}$ is shown in Fig.~\ref{fig:tanner_code_example}b.
Applying the same construction to $K_{11}$ with $\mathcal{C}_0 = [10,6,3]$ yields a $[55,11,18]$ code whose tree bound predicts only $d\geq 6$, yet the true minimum distance is $18$.
\end{example}

\subsection{Expander Codes}

An expander code encodes information on the left vertices of a bipartite expander and uses the right vertices as parity checks, with the expansion property guaranteeing that every small error set has a unique neighbor, meaning a violated check touching exactly one corrupt bit, which forces linear distance and enables a local iterative decoder.

\begin{figure}[ht]
\centering
\begin{tikzpicture}[
    vertex/.style={circle, fill=black, inner sep=0pt, minimum size=2.5pt},
    ellipseset/.style={draw=black, thick, fill=blue!8, ellipse, minimum width=2.4cm, minimum height=3.8cm}
]
    \node[ellipseset] (L) at (0,0) {};
    \node[ellipseset] (R) at (5.5,0) {};
    \node at (0,-2.3) {$L$};
    \node at (5.5,-2.3) {$R$};
    \node at (0,-1.6) {\small $n$};
    \node at (5.5,-1.6) {\small $m$};
    \node[vertex] (lv) at (0.3,1.1) {};
    \node[vertex] (r1) at (2.0, 1.1)  {};
    \node[vertex] (r2) at (2.0, 0.7)  {};
    \node[vertex] (r3) at (2.0, 0.3)  {};
    \node[vertex] (r4) at (2.0,-0.1)  {};
    \draw[thin] (lv) -- (r1);
    \draw[thin] (lv) -- (r2);
    \draw[thin] (lv) -- (r3);
    \draw[thin] (lv) -- (r4);
    \node at (1.9,-0.55) {\small $\rho$-regular};
    \node[vertex] (rv) at (5.2,-1.1) {};
    \node[vertex] (l1) at (3.5, 0.1)  {};
    \node[vertex] (l2) at (3.5,-0.3)  {};
    \node[vertex] (l3) at (3.5,-0.7)  {};
    \node[vertex] (l4) at (3.5,-1.1)  {};
    \draw[thin] (l1) -- (rv);
    \draw[thin] (l2) -- (rv);
    \draw[thin] (l3) -- (rv);
    \draw[thin] (l4) -- (rv);
    \node at (3.6,0.55) {\small $\sigma$-regular};
\end{tikzpicture}
\caption{A bipartite graph $\mathcal{G} = (L, R, E)$, $\rho$-regular on the left and $\sigma$-regular on the right, with $|L| = n$ and $|R| = m$. Every left vertex has $\rho$ incident edges (illustrated for one vertex); every right vertex has $\sigma$ incident edges.}
\label{fig:expander-bipartite}
\end{figure}
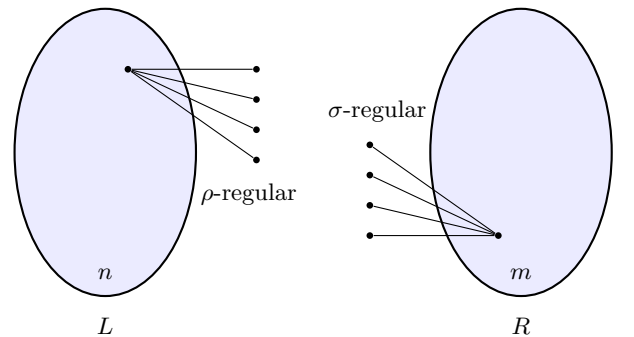

Formally, let $\mathcal{G} = (L, R, E)$ be a bipartite graph that is $\rho$-regular on the left and $\sigma$-regular on the right (denoting the left- and right-degree respectively), with $|L| = n$ and $|R| = \rho n/\sigma$ (Figure~\ref{fig:expander-bipartite}). For $v \in R$ (resp.\ $S \subseteq L$), let $\mathcal{N}(v)$ (resp.\ $\mathcal{N}(S)$) denote its neighborhood in $\mathcal{G}$. Suppose $\mathcal{G}$ is an $(\alpha, \rho(1-\varepsilon))$-expander, i.e., $|\mathcal{N}(S)| \geq \rho(1-\varepsilon)|S|$ for every $S \subseteq L$ with $|S| \leq \alpha n$.

\begin{definition}[Expander code~\cite{sipser1996expander}]
\begin{equation}
    \mathcal{C}(\mathcal{G}) = \Bigl\{\, x \in \{0,1\}^{n} \;\Big|\;
    \forall\, v \in R,\;
    \bigoplus_{i \in \mathcal{N}(v)} x_i \;\equiv\; 0 \pmod{2}
    \,\Bigr\}.
\end{equation}
\end{definition}
Each right vertex imposes one parity check, giving
\begin{equation}
    \mathrm{rate}\bigl(\mathcal{C}(\mathcal{G})\bigr) \;\geq\;
    1 - \frac{|R|}{n} \;=\; 1 - \frac{\rho}{\sigma}.
\end{equation}

For $\varepsilon < \tfrac{1}{2}$, the minimum distance satisfies $\delta\bigl(\mathcal{C}(\mathcal{G})\bigr) \geq 2\alpha(1-\varepsilon)$. Moreover, for the stricter range $\varepsilon < \tfrac{1}{4}$, $\mathcal{C}(\mathcal{G})$ admits a linear-time iterative decoder: repeatedly flip any left vertex incident to more than $\rho/2$ violated constraints. The expansion property guarantees such a vertex exists whenever $\Delta(x, \mathcal{C}(\mathcal{G})) < \alpha(1-\varepsilon)n$, so each flip strictly decreases the total number of violated constraints, and the algorithm terminates in $O(n)$ steps, correcting any error pattern of weight less than $\alpha(1-\varepsilon)n$~\cite{sipser1996expander}.
\begin{figure*}[ht]
    \includegraphics[width=1.0\linewidth]{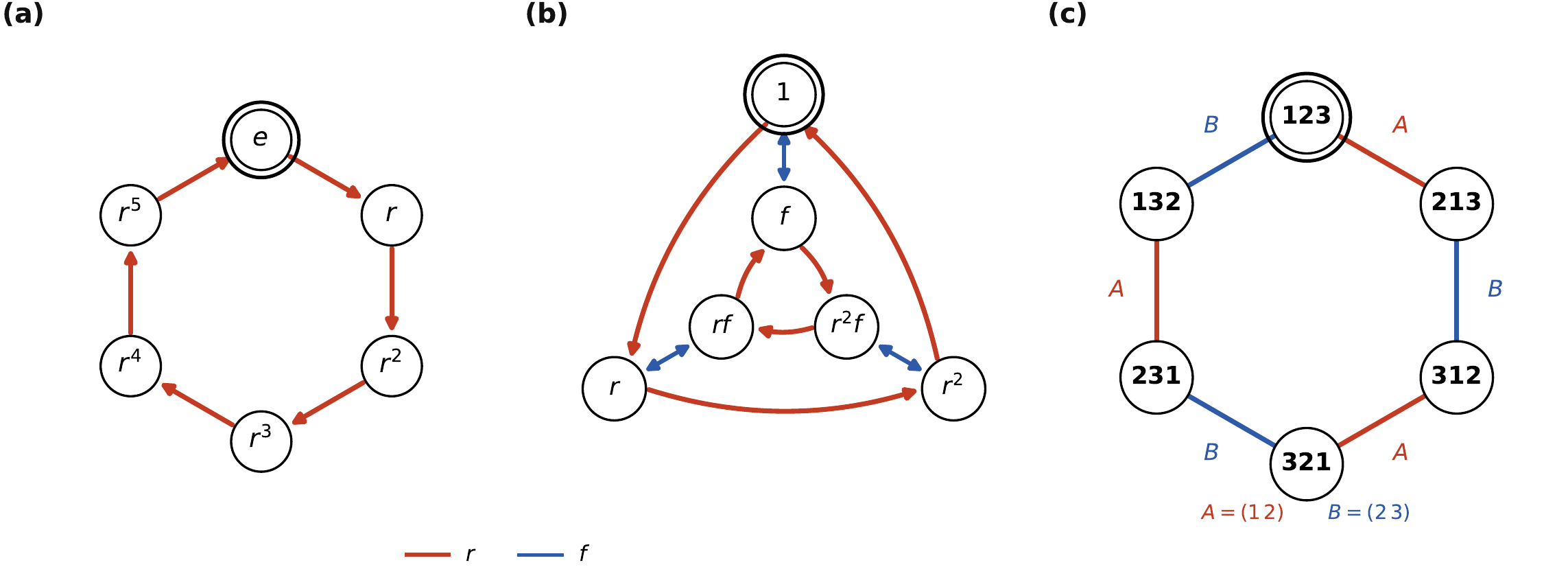}
\caption{Three Cayley graphs, with generators shown by edge color. (a) $\mathbb{Z}_6 = \langle r \rangle$: starting from the identity $e$, six applications of $r$ trace out a directed hexagon and land back on $e$, since $r^6=e$. (b) $S_3 = \mathrm{Tri} = \langle r, f \rangle$, generated by a rotation $r$ and a reflection $f$. The rotation subgroup $\{1,r,r^2\}$ sits on the outer triangle, its coset $\{f,rf,r^2f\}$ on the inner one; red arrows are right multiplication by $r$, blue double arrows by $f$. Notice the inner triangle rotates the opposite way from the outer one. That's because $r$ and $f$ don't commute ($rf=fr^{-1}$), so there's no single direction that works for both. (c) The same group $S_3$, but generated instead by two transpositions, $A=(1\,2)$ and $B=(2\,3)$, acting on permutations of $\{1,2,3\}$. Both generators are their own inverse, so every edge is undirected, and the graph collapses into a plain hexagon rather than the nested triangles of (b). This highlights how the same group can produce a completely different-looking graph if paired with different generators.}
\label{fig:cayley-graphs}
\end{figure*}

\subsection{Generating Sets}
The main intuition behind a generating set is that a group can be specified compactly, without enumerating every element of $\mathcal{G}$: one specifies a subset $\mathcal{S} \subseteq \mathcal{G}$ from which every element of $\mathcal{G}$ can be obtained as a product of elements of $\mathcal{S}$. 

The original construction of QT codes~\cite{leverrier2022quantum} is based on an LRCC, whose combinatorial structure is entirely determined by a group $\mathcal{G}$ and two symmetric generating sets $\mathcal{A}$ and $\mathcal{B}$. We therefore recall the relevant definitions.

Let $\mathcal{G}$ be a group and $\mathcal{S} \subseteq \mathcal{G}$ a subset. Since the intersection of any family of subgroups of $\mathcal{G}$ is again a subgroup, and since $\mathcal{G}$ itself contains $\mathcal{S}$, there exists a smallest subgroup of $\mathcal{G}$ containing $\mathcal{S}$. Here, smallest is with respect to inclusion in the sense that it is contained in every subgroup of $\mathcal{G}$ that contains $\mathcal{S}$, called the \emph{subgroup generated by $\mathcal{S}$} and denoted
\begin{equation}
    \langle \mathcal{S} \rangle \;=\; \bigcap_{\substack{\mathcal{H} \leq \mathcal{G} \\ \mathcal{S} \subseteq \mathcal{H}}} \mathcal{H}.
\end{equation}
Equivalently, $\langle \mathcal{S} \rangle$ is the set of all finite products
\begin{equation}
    g_{\alpha_1}^{n_1} g_{\alpha_2}^{n_2} \cdots g_{\alpha_k}^{n_k}, \qquad g_{\alpha_i} \in \mathcal{S},\; n_i \in \mathbb{Z},
\end{equation}
since this set of products is itself a subgroup containing $\mathcal{S}$, and is therefore contained in every subgroup that contains $\mathcal{S}$; the two characterizations thus coincide. When $\mathcal{S} = \{g_1, \dots, g_n\}$ is finite we write $\langle g_1, \dots, g_n \rangle$. If $\langle \mathcal{S} \rangle = \mathcal{G}$, the elements of $\mathcal{S}$ are called \emph{generators} of $\mathcal{G}$, and $\mathcal{G}$ is said to be \emph{finitely generated} if it admits a finite generating set.

\smallskip
\begin{example}
Let $\mathcal{G} = \mathbb{Z}_6 = \mathbb{Z}/6\mathbb{Z}$. The subgroup generated by $2$ is $\langle 2 \rangle = \{0, 2, 4\}$: it must contain $0$, $2$, and $2+2=4$, and these three elements already form a subgroup, so $\langle 2 \rangle \neq \mathcal{G}$. In contrast, $\langle 1 \rangle = \mathbb{Z}_6$, since every element of $\mathbb{Z}_6$ is a sum of copies of $1$. This illustrates that not every subset generates the whole group, and that the generating power of $\mathcal{S}$ depends on which elements it contains, not merely on $|\mathcal{S}|$.
\end{example}
\subsection{Cayley Graphs}
\editcolor{The main intuition behind Cayley graphs is that they provide an algebraic construction of graphs from groups. Given a group $\mathcal{G}$ and a subset $\mathcal{S} \subseteq \mathcal{G}$, the vertex set of the associated graph is taken to be $\mathcal{G}$ itself, while the edge set encodes left multiplication by elements of $\mathcal{S}$: a vertex $g \in \mathcal{G}$ is connected to $sg$ for each $s \in \mathcal{S}$. This yields an edge set}
\editcolor{\begin{align}
    E_S = \bigl\{\, \{g,\, sg\} : g \in \mathcal{G},\; s \in \mathcal{S} \,\bigr\}.
\end{align}}
\editcolor{The resulting graph is undirected when $\mathcal{S}$ is symmetric, i.e.\ $\mathcal{S}^{-1} = \mathcal{S}$, since in this case the transition $g \mapsto sg$ admits an inverse transition $sg \mapsto g$ via multiplication by $s^{-1} \in S$. Depending on whether $\mathcal{S}$ is taken to act on $\mathcal{G}$ by left or right multiplication, one obtains the left or right Cayley graph. Fig.~\ref{fig:cayley-graphs} illustrates both regimes on two small groups: the cyclic group $\mathbb{Z}_6$ with the non-symmetric generating set $\{r\}$ gives a purely directed Cayley graph, while $S_3$ with the mixed generating set $\{r, f\}$ combines directed edges (from $r$) with undirected edges (from the self-inverse reflection $f$). Generating $S_3$ instead by two involutions $\{A,B\}$ gives a purely undirected Cayley graph, since both generators are self-inverse.}

In the original QT code construction~\cite{leverrier2022quantum}, two such graphs $\mathrm{Cay}(\mathcal{G},\mathcal{A})$ and $\mathrm{Cay}(\mathcal{G},\mathcal{B})$ serve as the left and right \emph{skeletons} of the complex: their spectral properties directly control the minimum distance of the resulting quantum code, and the Ramanujan condition
\begin{align}
    \lambda\bigl(\mathrm{Cay}(\mathcal{G},\mathcal{A})\bigr) \leq 2\sqrt{|\mathcal{A}|-1}, \\
    \lambda\bigl(\mathrm{Cay}(\mathcal{G},\mathcal{B})\bigr) \leq 2\sqrt{|\mathcal{B}|-1}
\end{align}
is what is needed to guarantee a linear minimum distance.
Formally, we additionally require $S$ to be finite, to generate $G$, and to satisfy $1 \notin S$, so that no vertex has a self-loop. Under these conditions $\mathrm{Cay}(G,S) = (G, E_S)$, with $E_S$ as defined above, is a well-defined undirected graph. Moreover, since left multiplication by each $s \in S$ is injective, the $|S|$ neighbors $\{sg : s \in S\}$ of any vertex $g \in G$ are pairwise distinct, so $\mathrm{Cay}(G,S)$ is $|S|$-regular.
\subsection{Quantum Codes}
\label{subsec:quantumcodes}
 
We now recall stabilizer codes and, in particular, CSS codes, which reduce a quantum code to a pair of classical parity-check matrices subject to a compatibility condition.

A stabilizer code protects quantum information by listing operators that fix the encoded states. Its code space is the simultaneous $+1$ eigenspace of a set of commuting Pauli operators, and a Pauli error is detected through the operators it anticommutes with, whose measurement outcomes flip to $-1$. Formally, the $n$-qubit Pauli group $\mathcal{P}_n$ consists of all operators $\alpha E_1 \otimes \cdots \otimes E_n$ with $\alpha \in \{\pm 1, \pm i\}$ and $E_i \in \{I, X, Y, Z\}$; the weight $|E|$ of such an operator is its number of non-identity factors. Any two elements of $\mathcal{P}_n$ either commute or anticommute, and modulo phases each is represented by a pair $(u \mid v) \in \mathbb{F}_2^{2n}$ noting the supports of its $X$- and $Z$-components.
\begin{definition}[Stabilizer code]
\label{def:stabiliser}
Given an abelian subgroup $\mathcal{S} \leq \mathcal{P}_n$ with $-I \notin \mathcal{S}$, the associated stabilizer code is
\begin{equation}
\mathcal{C} \;=\; \{\, \ket{\varphi} \in (\mathbb{C}^2)^{\otimes n} : E\ket{\varphi} = \ket{\varphi} \ \text{for all } E \in \mathcal{S} \,\}.
\end{equation}
\end{definition}

If $\mathcal{S}$ has $r$ independent generators, then $\mathcal{C}$ encodes $k = n - r$ logical qubits. Errors in $\mathcal{S}$ act trivially on $\mathcal{C}$, while errors outside the centralizer $C(\mathcal{S})$ anticommute with some generator and are detected by measuring the generators. The harmful errors are thus the logical operators in $C(\mathcal{S}) \setminus \mathcal{S}$, and the minimum distance of the code is
\begin{align}
\label{eq:stab-distance}
d \;=\; \min\{\, |E| : E \in C(\mathcal{S}) \setminus \mathcal{S} \,\},
\end{align}
giving parameters $[[n, k, d]]$.

A CSS code is a stabilizer code admitting generators of pure $X$-type or pure $Z$-type, specified by binary matrices $H_X \in \mathbb{F}_2^{r_X \times n}$ and $H_Z \in \mathbb{F}_2^{r_Z \times n}$ whose rows record the generator supports and which satisfy the commutativity condition
\begin{align}
\label{eq:css-orthogonality}
H_X H_Z^{\intercal} = 0.
\end{align}
Its dimension is $k = n - \operatorname{rank} H_X - \operatorname{rank} H_Z$. The matrices $H_X$ and $H_Z$ may be viewed as parity-check matrices of two classical linear codes $\mathcal{C}_X = \ker H_X$ and $\mathcal{C}_Z = \ker H_Z$, whose dual codes $\mathcal{C}_X^{\perp}$ and $\mathcal{C}_Z^{\perp}$ are the row spaces of $H_X$ and $H_Z$ respectively. In this language, condition Eq.~\eqref{eq:css-orthogonality} becomes $\mathcal{C}_X^{\perp} \subseteq \mathcal{C}_Z$, or equivalently $\mathcal{C}_Z^{\perp} \subseteq \mathcal{C}_X$. An error $(u \mid v)$ has syndromes $H_X v^{\intercal}$ and $H_Z u^{\intercal}$; it has zero syndrome, and thus lies in $C(\mathcal{S})$, if and only if $v \in \mathcal{C}_X$ and $u \in \mathcal{C}_Z$, and it lies in $\mathcal{S}$ if and only if moreover $v$ and $u$ lie in the row spaces of $H_Z$ and $H_X$, i.e.\ $v \in \mathcal{C}_Z^{\perp}$ and $u \in \mathcal{C}_X^{\perp}$. The code distance Eq.~\eqref{eq:stab-distance} therefore specializes to
\begin{equation}
\label{eq:css-distance}
d \;=\; \min\{\, \wt(x) : x \in (\mathcal{C}_X \setminus \mathcal{C}_Z^{\perp}) \cup (\mathcal{C}_Z \setminus \mathcal{C}_X^{\perp}) \,\}.
\end{equation}

\begin{figure*}
    \centering
    \includegraphics[width=1.0\linewidth]{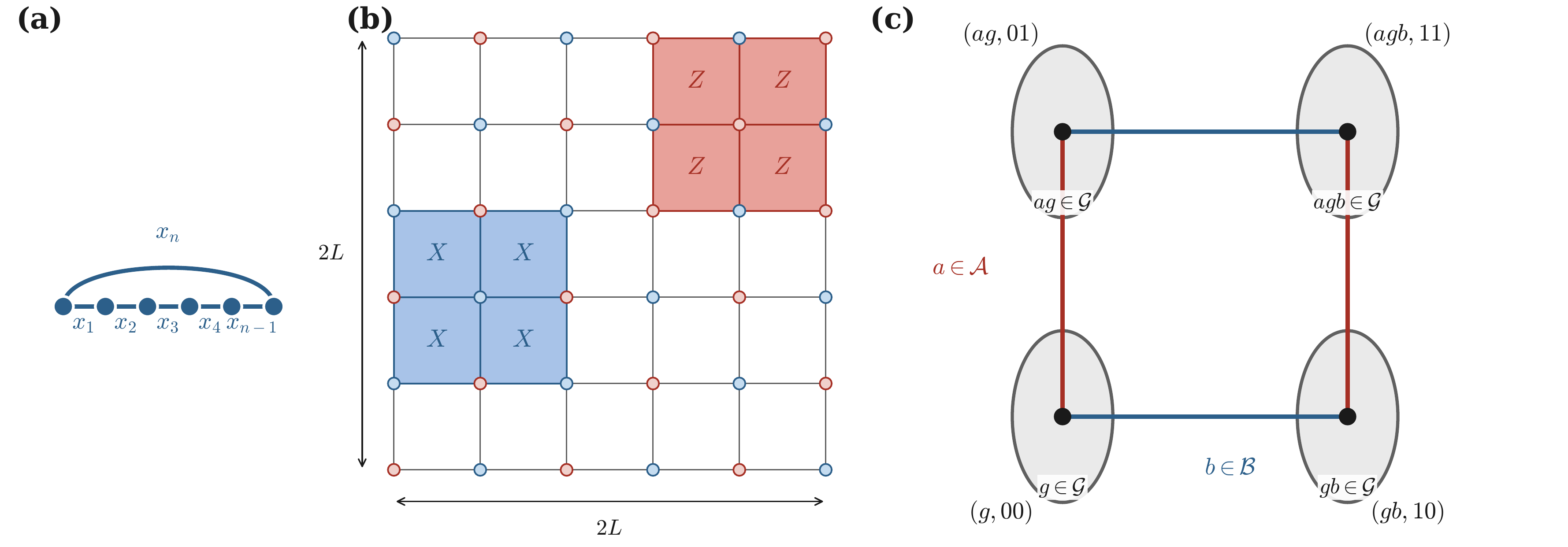}
\caption{\textbf{(a)} The repetition code on a 1D cycle: bits $x_1,\ldots,x_n$ live on the edges and each vertex enforces the constraint that the two incident bits are equal, i.e., $(x_i,x_{i+1})\in C_0=\{00,11\}$. \textbf{(b)} Taking the product of two such 1D graphs yields a 2D grid --- an object with vertices, edges, and squares (plaquettes). Qubits are placed on the squares and constraints on the vertices. Two types of vertices are required since we define a CSS code: blue ($X$-type) and red ($Z$-type). At each vertex the local code is the product code $C_0\otimes C_0=\bigl\{\bigl[\begin{smallmatrix}0&0\\0&0\end{smallmatrix}\bigr],\,\bigl[\begin{smallmatrix}1&1\\1&1\end{smallmatrix}\bigr]\bigr\}$; the non-trivial codeword $\bigl[\begin{smallmatrix}1&1\\1&1\end{smallmatrix}\bigr]$ defines the generator support (blue faces for $X$-support, red for $Z$-support, purple for the shared overlap). Adjacent $X$- and $Z$-generators always share an \emph{even} number of qubits, so they commute. On a $2L\times 2L$ grid with periodic boundary conditions, this yields the rotated toric code with parameters $[[4L^2,2,2L]]$. The code is local in 2D: at each vertex we have the product of two length-2 repetition codes (one vertical, one horizontal). QT codes apply the same recipe but with local codes $C_\mathcal{A}=[\Delta,\rho\Delta,\delta_A\Delta]$ and $C_\mathcal{B}=[\Delta,(1-\rho)\Delta,\delta_B\Delta]$ of length $\Delta>2$, which requires a 2D complex with larger vertex degree. \textbf{(c)} The LRCC replaces the 2D grid. Given a group $\mathcal{G}$ and two multisets $\mathcal{A},\mathcal{B}\subseteq\mathcal{G}$ with $|\mathcal{A}|=|\mathcal{B}|=\Delta$, the vertex set consists of four copies of $\mathcal{G}$, indexed $(g,ij)$ for $i,j\in\{0,1\}$ and $g\in\mathcal{G}$. $\mathcal{A}$-edges link $(g,00)\sim(ag,01)$ and $(gb,10)\sim(agb,11)$ for $a\in\mathcal{A}$; $\mathcal{B}$-edges link $(g,00)\sim(gb,10)$ and $(ag,01)\sim(agb,11)$ for $b\in\mathcal{B}$. Each square is the four-tuple $\{(g,00),(ag,01),(agb,11),(gb,10)\}$; qubits are placed on the squares and constraints on the vertices. The neighborhood $Q(v)$ of any vertex $v=(g,ij)$ is in bijection with $\mathcal{A}\times\mathcal{B}$, forming a $\Delta\times\Delta$ grid of squares. A defining property of the quadripartite LRCC is that two horizontally adjacent vertices share a \emph{column} of $\mathcal{A}\times\mathcal{B}$, while two vertically adjacent vertices share a \emph{row}. $X$-generators at vertices $(g,00)$ and $(agb,11)$ have support equal to codewords of $C_\mathcal{A}\otimes C_\mathcal{B}$; $Z$-generators at $(ag,01)$ and $(gb,10)$ have support equal to codewords of $C_\mathcal{A}^\perp\otimes C_\mathcal{B}^\perp$. Since horizontally adjacent $X$- and $Z$-generators share a column where one sees a codeword of $C_\mathcal{A}$ and the other a codeword of $C_\mathcal{A}^\perp$, orthogonality ensures commutativity; the same argument applied to rows and $C_\mathcal{B}$, $C_\mathcal{B}^\perp$ covers vertically adjacent pairs.}
    \label{fig:lrcc-overview}
\end{figure*}

\begin{figure*}
    \centering
    \includegraphics[width=0.8\linewidth]{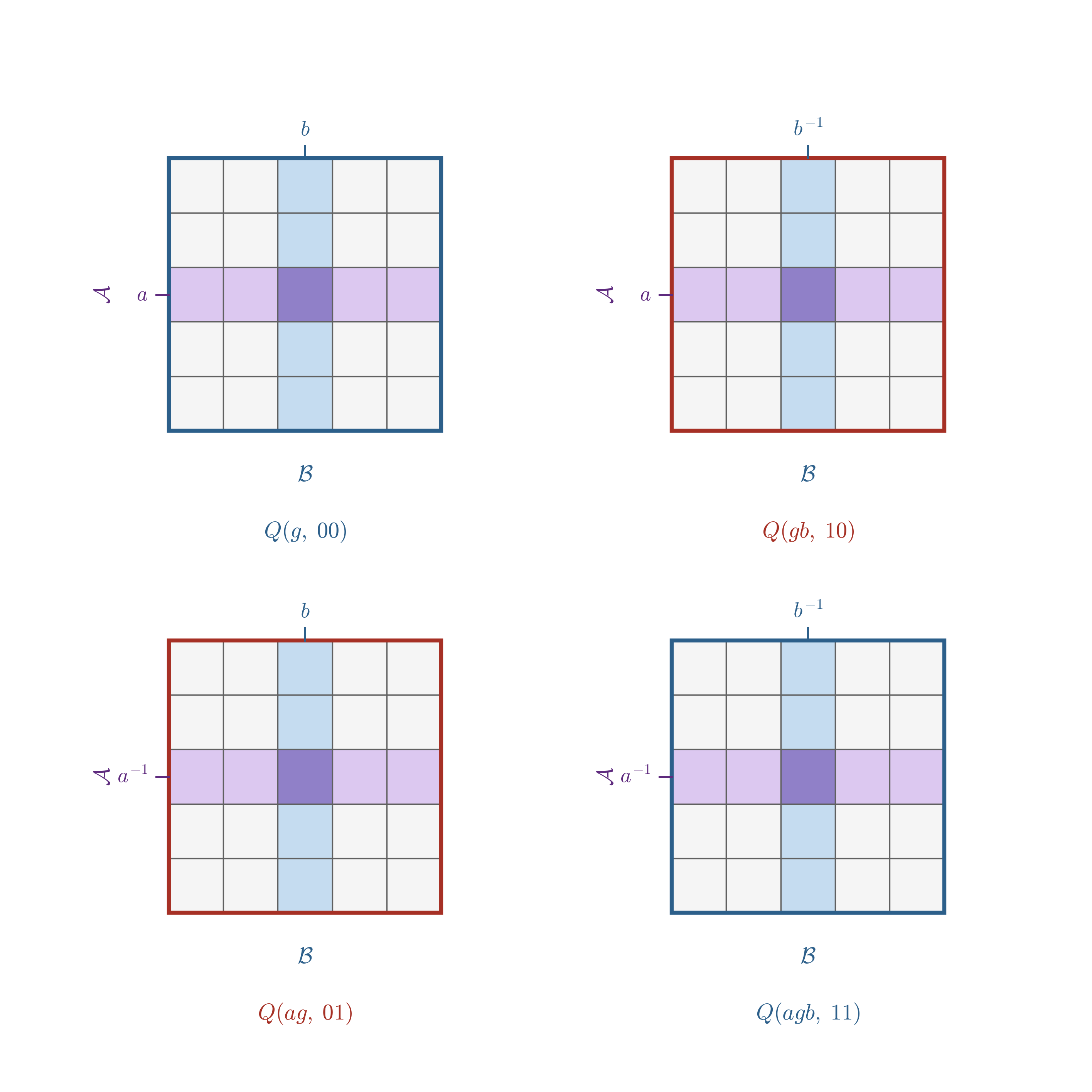}
    \caption{The neighborhood $Q(v)$ of each vertex $v=(g,ij)$ in the LRCC is in bijection with $\mathcal{A}\times\mathcal{B}$ via the map $\varphi_v\colon (a,b)\mapsto\{v,\,av,\,vb,\,avb\}$, so the local view at $v$ can be represented as a $|\mathcal{A}|\times|\mathcal{B}|$ array of squares (qubits), with rows indexed by $\mathcal{A}$ and columns by $\mathcal{B}$. A defining property of the LRCC is that adjacent vertices share an entire column or row of their local view: for a $\mathcal{B}$-edge $(v,vb)$, the shared squares occupy column $b$ in $Q(v)$ and column $b^{-1}$ in $Q(vb)$ (blue); for an $\mathcal{A}$-edge $(v,av)$, they occupy row $a$ in $Q(v)$ and row $a^{-1}$ in $Q(av)$ (purple). $X$-generators are supported on the $|\mathcal{A}|\times|\mathcal{B}|$ local view of vertices $(g,00)$ and $(agb,11)$ (blue borders), and are chosen to be codewords of $C_{\mathcal{A}}\otimes C_{\mathcal{B}}$; $Z$-generators are supported on the local view of $(gb,10)$ and $(ag,01)$ (red borders), and are codewords of $C_{\mathcal{A}}^{\perp}\otimes C_{\mathcal{B}}^{\perp}$. Any pair of adjacent $X$- and $Z$-generators overlaps on a shared column, where one sees a codeword of $C_{\mathcal{A}}$ and the other a codeword of $C_{\mathcal{A}}^{\perp}$, so they are orthogonal; an analogous argument using $C_{\mathcal{B}}\perp C_{\mathcal{B}}^{\perp}$ applies to vertically adjacent pairs, and all generators therefore commute by construction. }
    \label{fig:lrcc-local}
\end{figure*}

\section{Quantum Tanner Codes}
\label{sec:quantumtannercodes}
\editcolor{Quantum Tanner codes admit two complementary descriptions that we draw on throughout this work. The original \emph{square-complex description} places the code in a geometric setting via the LRCC, and remains the most convenient framework for establishing theoretical bounds on the distance~\cite{leverrier2022quantum}. The \emph{lifting description}, by contrast, constructs the same code algebraically by expanding a small base CSS code through the action of a finite group~\cite{leverrier2025small}. 

Beyond its practical value for systematic code search, the lifting description offers something the square-complex picture cannot easily provide: a concrete framework for tracing the structure of logical operators. Understanding logicals is essential not only for characterizing code parameters, but also for implementing fault-tolerant gates. In the base code, this structure is remarkably clean: a symplectic basis of logicals always exists in whichever $X$- and $Z$-type operator is supported on a single row or column of the qubit array~\cite[see Lemma 1]{leverrier2025small}. The lift inherits part of this structure, as codewords of the classical Tanner codes defined along individual rows and columns of the lifted code automatically give rise to logical operators of the quantum code~\cite{leverrier2025small}. Yet this account is incomplete, since these locally supported operators do not span the full logical space, and identifying the remaining logical operators remains an open problem. In what follows, we describe both perspectives with this question in mind, before turning to an explicit search for well-performing QT code instances.}

\subsection{Original Description}
\label{subsec:lrccconstruction}

The LRCC, introduced in~\cite{dinur2022locally}, provides the algebraic framework for QT codes. The goal of using the LRCC arises from applying the classical expander-code recipe~\cite{sipser1996expander} to the quantum code setting.

In a classical expander code, one places bits on the edges of a $\Delta$-regular graph and imposes on each vertex $v$ the constraint that the $\Delta$-bit string indexed by the incident edges is a codeword of some local code $C_0$ of length $\Delta$, as illustrated for the repetition code on a 1D cycle in Fig.~\ref{fig:lrcc-overview}(a). These classical constraints can be imposed simultaneously without any compatibility condition. In a CSS code, however, every $X$-type generator must commute with every $Z$-type generator. If qubits were placed on the edges of an ordinary cycle graph, any two generators of opposite type at adjacent vertices would share exactly one qubit and therefore anticommute; no classical graph suffices.

The resolution is to replace the graph with a \emph{square complex} \cite{wise2007complete}, a combinatorial object carrying vertices, edges, and \emph{squares} (faces), and to place qubits on squares rather than edges. Fig.~\ref{fig:lrcc-overview}(b) illustrates this for the special case $\Delta = 2$: the product of two 1D cycle graphs yields a 2D periodic grid whose squares carry the qubits of the rotated toric code, with the length-$2$ repetition code as the local code. At $\Delta = 2$ the grid structure is sufficient, but for $\Delta > 2$ one needs a square complex in which the link $Q(v)$ of every vertex remains identified with $\mathcal{A}\times\mathcal{B}$, and a 2D grid no longer provides this. The LRCC, shown in Fig.~\ref{fig:lrcc-overview}(c), achieves this by constructing the complex algebraically from a group $\mathcal{G}$ and two multisets $\mathcal{A}, \mathcal{B} \subseteq \mathcal{G}$, guaranteeing the grid structure at every vertex regardless of $\Delta$ (see also Fig.~\ref{fig:lrcc-local}). This is why \emph{product codes} are the appropriate local codes: a product code $C_\mathcal{A}\otimes C_\mathcal{B}$ acts on an $\mathcal{A}\times\mathcal{B}$ grid by requiring every column to be a codeword of $C_\mathcal{A}$ and every row to be a codeword of $C_\mathcal{B}$.

\paragraph{Bipartite construction.}
Given a group $\mathcal{G}$ and two sets of symmetric generating $\mathcal{A} = \mathcal{A}^{-1}$ and $\mathcal{B} = \mathcal{B}^{-1}$ with $|\mathcal{A}| = |\mathcal{B}| = \Delta$, the bipartite LRCC has a vertex set $V = V_0 \cup V_1$ with $V_i = G \times \{i\}$. The $\mathcal{A}$-edges connect $(g,0)$ to $(ag,1)$ for each $g \in G, a \in \mathcal{A}$, giving the bipartite graph $\mathcal{G}_\mathcal{A} = (V, E_\mathcal{A})$. Similarly, $\mathcal{B}$-edges connect $(g,0)$ to $(gb,1)$, giving $\mathcal{G}_\mathcal{B} = (V, E_\mathcal{B})$. Each of these is a bipartite $\Delta$-regular graph and serves as the double cover of the respective Cayley graph $\mathrm{Cay}(G,\mathcal{A})$ and $\mathrm{Cay}(G,\mathcal{B})$.

The \emph{squares} of the complex are the $4$-subsets of vertices of the form
\begin{align}
    q(g,a,b) = \bigl\{(g,0),\,(ag,1),\,(gb,1),\,(agb,0)\bigr\}
\end{align}
for $g \in \mathcal{G}$, $a \in \mathcal{A}$, $b \in \mathcal{B}$.
For a vertex $v \in V$, the \emph{link} $Q(v)$ is the set of all squares incident to $v$. The \emph{Total No-Conjugacy} (TNC) condition,
\begin{equation}
    \forall\, a \in \mathcal{A},\; b \in \mathcal{B},\; g \in \mathcal{G}, \quad ag \neq gb,
\end{equation}
ensures that every square contains exactly four distinct vertices and that the link of every vertex satisfies $|Q(v)| = \Delta^2$. Under TNC, the map $(a,b) \mapsto q(g,a,b)$ is a bijection from $\mathcal{A} \times \mathcal{B}$ to $Q(v)$ for every $v = (g,i)$~\cite{dinur2022locally}, so the link of any vertex is canonically identified with the grid $A \times B$.

Two further graphs are defined on the complex. The union graph $\mathcal{G}^\cup = (V, E_A \cup E_B)$ has degree $2\Delta$. The square graph $\mathcal{G}^\square = (V, E^\square)$ places an edge between $(g,i)$ and $(agb,i)$ for all $g \in \mathcal{G}, a \in \mathcal{A}, b \in \mathcal{B}, i \in \{0,1\}$; it is $\Delta^2$-regular and decomposes into two connected components
$\mathcal{G}_0^\square$ on $V_0$ and $\mathcal{G}_1^\square$ on $V_1$. If both $\mathcal{G}_A$ and $\mathcal{G}_B$ are Ramanujan, their spectral expansion is inherited by the derived graphs~\cite[see Lemma 4]{leverrier2022quantum}:
\begin{align}
    \lambda(\mathcal{G}^\cup) \leq 4\sqrt{\Delta}, \qquad
    \lambda(\mathcal{G}_0^\square) \leq 4\Delta, \qquad
    \lambda(\mathcal{G}_1^\square) \leq 4\Delta.
\end{align}

\paragraph{Quadripartite construction.}
The TNC condition can be a significant restriction. The quadripartite variant of the LRCC~\cite{panteleev2022asymptotically} avoids it entirely by partitioning the vertex set into four copies of $\mathcal{G}$:
\begin{align}
    V = V_{00} \cup V_{01} \cup V_{10} \cup V_{11},
 V_{ij} = G \times \{i,j\},\quad i,j \in \{0,1\}.
\end{align}
The edge sets are defined by
\begin{align}
    \mathcal{A}\text{-edges:} &\quad \{(g,00),(ag,01)\} \;\text{and}\;
    \{(g,10),(ag,11)\}, \\
    \mathcal{B}\text{-edges:} &\quad \{(g,00),(gb,10)\} \;\text{and}\;
    \{(g,01),(gb,11)\},
\end{align}
for $g \in \mathcal{G}$, $a \in \mathcal{A}$, $b \in \mathcal{B}$. Squares are the $4$-subsets
\begin{align}
    q(g,a,b) = \bigl\{(g,00),\,(ag,01),\,(gb,10),\,(agb,11)\bigr\}.
\end{align}
Since the four vertices of every square lie in distinct parts $V_{00}, V_{01}, V_{10}, V_{11}$, no degeneracy can arise regardless of the group relations or whether $\mathcal{A} = \mathcal{B}$. Consequently, the link $Q(v)$ of any vertex $v$ is isomorphic to $\mathcal{A} \times \mathcal{B}$ without any TNC assumption.
Fig.~\ref{fig:lrcc-local} depicts the local view at each of the four vertex types $(g,00)$, $(ag,01)$, $(gb,10)$, $(agb,11)$: each is an $|\mathcal{A}|\times|\mathcal{B}|$ grid of squares, and adjacent vertices of opposite type share an entire row or column of this grid, which is what ensures that $X$- and $Z$-generators commute.
The spectral bounds above continue to hold in the quadripartite case~\cite{leverrier2022quantum}, and the QT code construction extends to this setting without change.

With the LRCC in place, place one qubit on each square of the complex, so that $n = |Q| = \tfrac{1}{2}\Delta^2|\mathcal{G}|$, and choose two classical binary codes $C_\mathcal{A} \subset \mathbb{F}_2^\mathcal{A}$ and $C_\mathcal{B} \subset \mathbb{F}_2^\mathcal{B}$ of length $\Delta$, with $\dim C_\mathcal{A} = \rho\Delta$ and $\dim C_\mathcal{B} = (1-\rho)\Delta$. These induce two \emph{local} tensor codes on the $\mathcal{A}\times\mathcal{B}$ grid, $C_0 = C_\mathcal{A}\otimes C_\mathcal{B}$ and $C_1 = C_\mathcal{A}^\perp \otimes C_\mathcal{B}^\perp$, which by construction are mutually orthogonal on any shared row or column. Using $\phi_v$ to identify $Q(v)$ with $\mathcal{A}\times\mathcal{B}$, each basis codeword of $C_0$ at a vertex $v \in V_0$ defines a $Z$-stabilizer supported on $Q(v)$, and each basis codeword of $C_1$ at $v \in V_1$ defines an $X$-stabilizer supported on $Q(v)$; opposite-type generators share at most one row or column of the grid and therefore commute. The CSS code $\mathcal{Q} = (\mathcal{C}_X, \mathcal{C}_Z)$ is then a pair of classical Tanner codes on the two square graphs, with local codes given by the duals of the tensor codes above:
\begin{equation}
    \label{eq:cx-cz-tanner}
    \mathcal{C}_Z \;=\; T\bigl(\mathcal{G}_0^\square,\, C_0^\perp\bigr),
    \qquad
    \mathcal{C}_X \;=\; T\bigl(\mathcal{G}_1^\square,\, C_1^\perp\bigr),
\end{equation}
where $T(\mathcal{G}, C) = \{x \in \mathbb{F}_2^Q : x_v \in C \text{ for all } v \in \mathcal{G}\}$. Counting generators gives $k/n \geq (2\rho-1)^2$; each stabilizer has weight at most $\Delta^2$ and each qubit lies in at most $4\rho(1-\rho)\Delta^2$ stabilizers, so with $\Delta$ fixed and $|\mathcal{G}|\to\infty$ the family is LDPC, and asymptotically good whenever the Cayley graphs are sufficiently expanding and $C_0^\perp, C_1^\perp$ are sufficiently robust~\cite{leverrier2022quantum,panteleev2022asymptotically}.

\begin{figure*}
    \centering
    \includegraphics[width=1.0\linewidth]{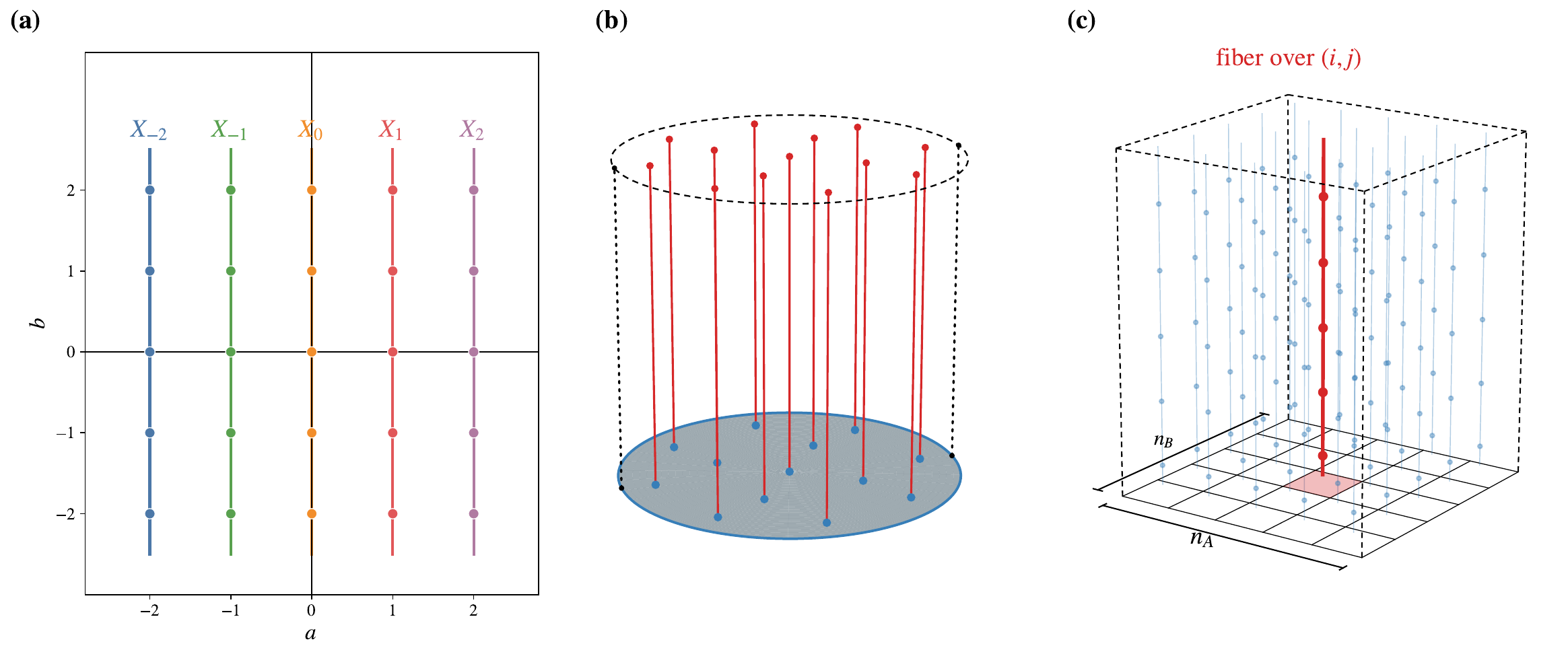}
\caption{The notion of a fiber in three settings. In each case a total space projects onto a base space, and the fiber over a base point is its preimage under the projection. (a)~For the projection $f:\mathbb{Z}\times\mathbb{Z}\to\mathbb{Z}$, $f(a,b)=a$, the fiber over $a$ is $X_a=f^{-1}(a)=\{(a,b): b\in\mathbb{Z}\}$, the vertical line of lattice points with first coordinate $a$ (colored lines). (b)~A cylinder, viewed as a disk with a fiber over each of its points. (c)~The lifted quantum Tanner code of Eq.~\eqref{eq:lifted}, with base space the $n_A\times n_B$ qubit grid of the base code of Lemma 1 ~\cite{leverrier2025small}. The lift replaces each base qubit $(i,j)$ by a fiber of $|\mathcal{G}|$ qubits indexed by $(i,j,g)$ with $g\in\mathcal{G}$; one fiber is highlighted in red.}
\label{fig:fibers}
\end{figure*}
\subsection{Quantum Tanner Code via Lifts}
\label{subsec:liftedtannercodes}
\editcolor{While the LRCC description provides a perspective useful for proving lower bounds on distance, the lifting view point offers a more direct approach for constructing explicit instances. In this section, we describe the lifting view point recently developed in~\cite{leverrier2025small}, which interprets QT codes as CSS codes obtained by lifting a base code through commuting group actions.}

\editcolor{The lifted code construction begins with a base CSS code on $n_A n_B$ qubits. Let $C_0, C_1 \subseteq \F_2^{n_A}$ and $C_0', C_1' \subseteq \F_2^{n_B}$ denote binary linear codes with parity-check matrices $H_0, H_1$ and $H_0', H_1'$. We associate generator matrices $G_0, G_1$ and $G_0', G_1'$ such that $\ker G_i = C_i^\perp$ and $\ker G_i = {C_i'}^\perp$.} \editcolor{The base CSS code is constructed by arranging qubits as an $n_A \times n_B$ grid and defining:}
\editcolor{\begin{align}\label{eq:base}
H_X^{\mathrm{base}}=
\begin{pmatrix}
H_0\otimes G_0'\\
H_1\otimes G_1'
\end{pmatrix},
\quad
H_Z^{\mathrm{base}}=
\begin{pmatrix}
G_0\otimes H_1'\\
G_1\otimes H_0'
\end{pmatrix}.
\end{align}}
\editcolor{This local template (a.k.a the base code) encodes the structure that will be lifted to larger quantum codes through group actions. Closed-form expressions for the $[[n,k,d]]$ parameters of this base code are given in \cite[Lemma~1]{leverrier2025small}.}

\editcolor{To construct a QT code, we enlarge the base code through a finite group $\mathcal{G}$ and two multisets $\mathcal{A} = (a_1, \ldots, a_{n_A})$ and $\mathcal{B} = (b_1, \ldots, b_{n_B})$ of group elements. The lifting procedure expands each base qubit at position $(i,j) \in [n_A] \times [n_B]$ into a fiber of $|\mathcal{G}|$ qubits, one for each $g \in \mathcal{G}$. The total number of qubits is $n = n_A n_B |\mathcal{G}|$.}
The lifted code is obtained by acting on the base code with commuting left and right group actions. The resulting parity-check matrices are given by~\cite{leverrier2025small}:
\begin{equation}\label{eq:lifted}
\begin{aligned}
H_X^{\mathrm{lifted}} &=
\begin{pmatrix}
H_0\otimes G_0'\otimes I_{|\mathcal{G}|}\\[2pt]
\bigl(H_1\otimes G_1'\otimes I_{|\mathcal{G}|}\bigr)\,L_A R_B
\end{pmatrix},\\[6pt]
H_Z^{\mathrm{lifted}} &=
\begin{pmatrix}
\bigl(G_0\otimes H_1'\otimes I_{|\mathcal{G}|}\bigr)\,R_B\\[2pt]
\bigl(G_1\otimes H_0'\otimes I_{|\mathcal{G}|}\bigr)\,L_A
\end{pmatrix}.
\end{aligned}
\end{equation}
where $L_A$ and $R_B$ are block-diagonal matrices defined by the multisets $\mathcal{A}$ and $\mathcal{B}$. We refer the reader to Ref. ~\cite{leverrier2025small} for detail about these permutations of left and right group actions.

Eq.\eqref{eq:lifted} admits a natural geometric picture. Label each qubit by a triple $(i,j,g)\in[n_A]\times[n_B]\times\mathcal{G}$, so that the full qubit array forms a three-dimensional grid of shape $n_A\times n_B\times|\mathcal{G}|$. Each generator of $H_X^{\mathrm{lifted}}$ and $H_Z^{\mathrm{lifted}}$ is then supported on a $\mathcal{G}$-translate of a horizontal slice: a copy of the base $n_A\times n_B$ grid whose $\mathcal{G}$-coordinate depends on $(i,j)$ through $L_A$, $R_B$, both, or neither, as dictated by the corresponding block. The lift is illustrated in Fig.~\ref{fig:fibers}(c), alongside two more elementary instances of the same notion of a fiber.
Within such a slice, the support of each generator is itself a codeword of a product code: the $X$-type generators correspond to product codewords in $C_i^{\perp}\otimes C_i'$ for $i=0,1$, while the $Z$-type generators correspond to product codewords in $C_0\otimes C_1'^{\perp}$ and $C_1\otimes C_0'^{\perp}$~\cite{leverrier2025small}. This slice structure has a clean combinatorial description: as noted in~\cite{leverrier2025small}, it is ``naturally encoded by the LRCC associated with $(\mathcal{A},\mathcal{B},\mathcal{G})$.'' In particular, an $X$-generator and a $Z$-generator can overlap on at most a single row or column of such a slice, where orthogonality of the local codes guarantees their commutativity.
\label{sec:numerical}
\definecolor{mycolor1}{RGB}{255, 140, 0}
\definecolor{mycolor2}{RGB}{180, 70, 255}
\definecolor{mycolor3}{RGB}{50, 205, 50}
\definecolor{mycolor4}{RGB}{30, 90, 200}
\definecolor{mycolor5}{RGB}{174, 217, 69}
\definecolor{mycolor6}{RGB}{255, 20, 147}
\definecolor{mycolor7}{RGB}{0, 220, 220}
\definecolor{mycolor8}{RGB}{255, 230, 0}
\definecolor{mycolor9}{RGB}{80, 255, 80}
\definecolor{mycolor10}{RGB}{255, 80, 80}
\definecolor{mycolor11}{RGB}{0, 170, 255}
\definecolor{mycolor12}{RGB}{255, 0, 120}
\definecolor{mycolor13}{RGB}{0, 255, 127}
\definecolor{mycolor14}{RGB}{255, 99, 71}
\definecolor{mycolor15}{RGB}{135, 206, 250}
\definecolor{mycolor16}{RGB}{255, 105, 180}
\definecolor{mycolor17}{RGB}{124, 252, 0}
\definecolor{mycolor18}{RGB}{255, 165, 0}
\definecolor{mycolor19}{RGB}{138, 43, 226}
\definecolor{mycolor20}{RGB}{0, 191, 255}
\definecolor{mycolor21}{RGB}{255, 215, 0}
\definecolor{mycolor22}{RGB}{50, 205, 50}
\definecolor{mycolor23}{RGB}{255, 69, 0}
\definecolor{mycolor24}{RGB}{127, 255, 212}
\definecolor{mycolor25}{RGB}{0, 255, 0}
\definecolor{mycolor26}{RGB}{255, 20, 147}
\definecolor{mycolor27}{RGB}{65, 105, 225}
\definecolor{mycolor28}{RGB}{255, 192, 203}
\definecolor{mycolor29}{RGB}{154, 205, 50}
\definecolor{mycolor30}{RGB}{255, 140, 105}
\definecolor{mycolor31}{RGB}{0, 250, 154}
\definecolor{mycolor32}{RGB}{255, 0, 255}
\definecolor{mycolor33}{RGB}{0, 128, 255}
\definecolor{mycolor34}{RGB}{255, 255, 102}
\definecolor{mycolor35}{RGB}{255, 127, 80}
\definecolor{mycolor36}{RGB}{72, 209, 204}
\definecolor{mycolor37}{RGB}{255, 182, 193}
\definecolor{mycolor38}{RGB}{124, 252, 0}
\definecolor{mycolor39}{RGB}{30, 144, 255}
\definecolor{mycolor40}{RGB}{255, 105, 0}
\definecolor{mycolor41}{RGB}{173, 255, 47}
\definecolor{mycolor42}{RGB}{255, 0, 102}
\definecolor{mycolor43}{RGB}{0, 255, 255}
\definecolor{mycolor44}{RGB}{255, 250, 0}
\definecolor{mycolor45}{RGB}{186, 85, 211}
\definecolor{mycolor46}{RGB}{255, 127, 0}
\definecolor{mycolor47}{RGB}{0, 206, 209}
\definecolor{mycolor48}{RGB}{255, 153, 204}
\definecolor{mycolor49}{RGB}{102, 255, 0}
\definecolor{mycolor50}{RGB}{0, 255, 170}
\definecolor{mycolor51}{RGB}{255, 51, 51}
\definecolor{mycolor52}{RGB}{51, 153, 255}
\definecolor{mycolor53}{RGB}{255, 204, 0}
\definecolor{mycolor54}{RGB}{153, 255, 51}
\definecolor{mycolor55}{RGB}{255, 102, 178}
\definecolor{mycolor56}{RGB}{102, 255, 255}
\definecolor{mycolor57}{RGB}{255, 153, 51}
\definecolor{mycolor58}{RGB}{153, 51, 255}
\definecolor{mycolor59}{RGB}{51, 255, 102}
\definecolor{mycolor60}{RGB}{255, 51, 153}
\definecolor{mycolor61}{RGB}{0, 230, 118}
\definecolor{mycolor62}{RGB}{255, 235, 59}
\definecolor{A}{RGB}{248, 248, 255}
\definecolor{B}{RGB}{245, 255, 250}
\definecolor{C}{RGB}{255, 250, 240}
\definecolor{D}{RGB}{240, 248, 255}
\definecolor{E}{RGB}{255, 245, 238}
\definecolor{F}{RGB}{240, 255, 240}
\definecolor{G}{RGB}{255, 250, 250}
\definecolor{H}{RGB}{248, 248, 247}
\definecolor{I}{RGB}{245, 245, 245}
\definecolor{J}{RGB}{250, 250, 250}
\definecolor{K}{RGB}{245, 248, 255}
\definecolor{L}{RGB}{255, 248, 240}
\definecolor{M}{RGB}{240, 252, 245}
\definecolor{N}{RGB}{250, 245, 255}
\definecolor{O}{RGB}{245, 250, 255}
\definecolor{P}{RGB}{255, 252, 245}
\definecolor{Q}{RGB}{248, 255, 248}
\definecolor{R}{RGB}{255, 248, 248}
\definecolor{S}{RGB}{245, 250, 245}
\definecolor{T}{RGB}{250, 248, 255}
\definecolor{U}{RGB}{252, 252, 250}
\definecolor{V}{RGB}{248, 250, 255}
\definecolor{W}{RGB}{255, 252, 250}
\definecolor{X}{RGB}{250, 252, 248}
\definecolor{Y}{RGB}{252, 250, 248}
\definecolor{Z}{RGB}{248, 252, 250}
\begin{table*}
\centering
\footnotesize
\setlength{\tabcolsep}{6.5pt}
\renewcommand{\arraystretch}{1.2}
\begin{tabular}{|*{10}{>{\columncolor{white}[\tabcolsep]}c|}}
\hline
\rowcolor{mycolor41!40} \textbf{Group} &
$[[\textbf{n, k,} (\le \textbf{d}_X, \le\textbf{d}_Z)]]$ &
$\textbf{H}_0$ &
$\textbf{H}_1$ &
$\overline{\textbf{w}_X}$ &
$\overline{\textbf{w}_Z}$ &
$\tfrac{\textbf{k}}{\textbf{n}}$ &
$\tfrac{\textbf{kd}_{\min}^2}{\textbf{n}}$ &
\shortstack{\textbf{Trials}\\(\texttt{QDistRnd}) \cite{pryadko2023qdistrnd}} &
\shortstack{\textbf{Trials}\\(\texttt{sQetch}) \cite{bhardwaj2026high}}
\\
\hline
\rowcolor{mycolor1!40} $S_3$ & $[[288, 8, (\le15,\le15)]]$ & [8,4,4] & [6,3,3] & 12 & 12 & 0.028 & 6.25 & 50K & 50M \\
\hline
\rowcolor{mycolor2!40}$S_3$ & $[[336, 4, (\le24,\le15)]]$ & [8,4,4] & [7,3,4] & 16 & 16 & 0.012 & 2.68 & 50K & 100M\\
\hline
\rowcolor{mycolor3!40} $S_3$ & $[[336, 12, (\le 14,\le14)]]$ & [8,4,4] & [7,3,4] & 16 & 16 & 0.036 & 7.00 & 50K & 50M \\
\hline
\rowcolor{mycolor4!40} $D_{10}$ & $[[360, 6, (\le18,\le18)]]$ & [6,3,3] & [6,3,3] & 9 & 9 & 0.017 & 5.40 & 50K & 50M\\
\hline
\rowcolor{mycolor5!40} $S_3$ & $[[384, 24, (\le 14,\le 14)]]$ & [8,4,4] & [8,4,4] & 16 & 16 & 0.063 & 12.25 & 50K & 50M\\
\hline
\rowcolor{mycolor6!40} $S_3$ & $[[384, 8, (\leq16,\leq16)]]$ & [8,4,4] & [8,4,4] & 16 & 16 & 0.021 & 5.33 & \shortstack{50K\\$(22,22)$} & \shortstack{100M\\$(16,16)$} \\
\hline
\rowcolor{mycolor7!40}$C_3 \times C_3$ & $[[432, 16, (\leq15,\leq15)]]$ & [8,4,4] & [6,3,3] & 12 & 12 & 0.037 & 8.33 & 50K & 100M\\
\hline
\rowcolor{mycolor8!40}$C_3 \times C_3$ & $[[432, 8, (\leq18,\leq18)]]$ & [8,4,4] & [6,3,3] & 12 & 12 & 0.019 & 6.00 & 50K & 100M \\
\hline
\rowcolor{mycolor9!40}$D_{10}$ & $[[480, 8, (\leq21,\leq18)]]$ & [8,4,4] & [6,3,3] & 12 & 12 & 0.017 & 5.40 & \shortstack{50K\\$(22,18)$} & \shortstack{100M\\$(21,18)$} \\
\hline
\rowcolor{mycolor10!40}$C_3 \ltimes C_4$ & $[[504, 4, (\leq34,\leq12)]]$ & [7,3,4] & [6,3,3] & 12 & 12 & 0.008 & 1.14 & \shortstack{50K\\$(36,12)$} & \shortstack{100M\\$(34,12)$} \\
\hline
\rowcolor{mycolor15!40} $D_{14}$ & $[[588, 6, (\leq 28, \leq 23)]]$ & [7,4,3] & [6,3,3] & 12 & 12 & 0.010 & 5.40 & \shortstack{50K\\$(28,25)$} & \shortstack{200M\\$(28,23)$} \\
\hline
\rowcolor{mycolor16!40}$D_{14}$ & $[[588, 14, (\le20,\le18)]]$ & [7,4,3] & [6,3,3] & 12 & 12 & 0.024 & 7.71 & \shortstack{50K\\$(20,18)$} & \shortstack{200M\\$(20,19)$} \\
\hline
\rowcolor{mycolor17!40}$D_{10}$ & $[[640, 24, (\leq 14,\leq 14)]]$ & [8,4,4] & [8,4,4] & 16 & 16 & 0.038 & 7.35 & 50K & 50M \\
\hline
\rowcolor{mycolor20!40} $C_3 \times S_3$ & $[[648, 26, (\leq12, \leq15)]]$ & [6,3,3] & [6,3,3] & 9 & 9 & 0.040 & 5.78 & 50K & 50M\\
\hline
\rowcolor{mycolor50!40}$C_3 \times S_3$ & $[[648, 14, (\leq16, \leq15)]]$ & [6,3,3] & [6,3,3] & 9 & 9 & 0.022 & 4.86 & 50K & 200M\\
\hline
\rowcolor{mycolor21!40}$C_3 \times S_3$ & $[[648, 2, (\leq18, \leq 32)]]$ & [6,3,3] & [6,3,3] & 9 & 9 & 0.003 & 1.00 & \shortstack{50K\\$(18,32)$} & \shortstack{100M\\$(18,36)$}\\
\hline
\rowcolor{mycolor29!40}$C_3 \times S_3$ & $[[648, 4, (\leq 26, \leq21)]]$ & [6,3,3] & [6,3,3] & 9 & 9 & 0.006 & 2.72 & \shortstack{50K\\$(26,21)$} & \shortstack{200M\\$(29,21)$} \\
\hline
\rowcolor{mycolor22!40} $C_3 \times S_3$ & $[[648, 10, (\le18, \le20)]]$ & [6,3,3] & [6,3,3] & 9 & 9 & 0.015 & 5.00 & \shortstack{50K\\$(18,20)$} & \shortstack{100M\\$(18,21)$}\\
\hline
\rowcolor{mycolor25!40}$C_3 \ltimes C_4$ & $[[756, 17, (\leq9,\leq21)]]$ & [7,3,4] & [9,5,3] & 20 & 20 & 0.022 & 1.82 & 50K & 50M \\
\hline
\rowcolor{mycolor26!40} $C_3 \ltimes C_4$ & $[[756, 30, (\leq21,\leq12)]]$ & [7,3,4] & [9,5,3] & 20 & 20 & 0.040 & 5.71 & 50K & 50M\\
\hline
\rowcolor{mycolor27!40}$C_3 \ltimes C_4$ & $[[756, 43, (\leq12,\leq12)]]$ & [7,3,4] & [9,5,3] & 20 & 20 & 0.057 & 8.19 & 50K & 50M \\
\hline
\rowcolor{mycolor29!40}$C_3 \ltimes C_4$ & $[[756, 10, (\leq9,\leq42)]]$ & [7,3,4] & [9,5,3] & 20 & 20 & 0.013 & 1.07 & \shortstack{50K\\$(9,46)$} & \shortstack{200M\\$(9,42)$} \\
\hline
\rowcolor{mycolor31!40}$C_3 \ltimes C_4$ & $[[768,24, (\leq16,\leq16)]]$ & [8,4,4] & [8,4,4] & 16 & 16 & 0.031 & 8.00 & \shortstack{50K\\$(18,18)$} & \shortstack{50M\\$(16,16)$} \\
\hline
\rowcolor{mycolor33!40} $A_4$ & $[[768,32, (\leq16,\leq16)]]$ & [8,4,4] & [6,3,3] & 16 & 16 & 0.042 & 10.67 & 50K & 50M \\
\hline
\rowcolor{mycolor35!40} $C_3 \times S_3$ & $[[864,24, (\leq21,\leq18)]]$ & [8,4,4] & [6,3,3] & 12 & 12 & 0.028 & 9.00 & 50K & 100M \\
\hline
\rowcolor{mycolor36!40}$C_3 \times S_3$ & $[[864,16, (\leq27,\leq18)]]$ & [8,4,4] & [6,3,3] & 12 & 12 & 0.019 & 6.00 & \shortstack{50K\\$(33,18)$} & \shortstack{100M\\$(27,18)$} \\
\hline
\rowcolor{mycolor47!40} $C_3 \times S_3$ & $[[864,8, (\leq 27,\leq32)]]$ & [8,4,4] & [6,3,3] & 12 & 12 & 0.009 & 6.75 & \shortstack{50K\\$(41,60)$} & \shortstack{200M\\$(27,32)$} \\
\hline
\rowcolor{mycolor37!40}$C_3 \times S_3$ & $[[864,8, (\leq 27,\leq 47)]]$ & [8,4,4] & [6,3,3] & 12 & 12 & 0.009 & 6.75 & \shortstack{50K\\$(134,139)$} & \shortstack{200M\\$(27,47)$} \\
\hline
\rowcolor{mycolor38!40}$C_4 \ltimes C_4$ & $[[896,40, (\leq16,\leq12)]]$ & [8,4,4] & [7,3,4] & 16 & 16 & 0.045 & 6.43 & \shortstack{50K\\$(24,12)$} & \shortstack{50M\\$(16,12)$}\\
\hline
\rowcolor{mycolor39!40}$C_4 \ltimes C_4$ & $[[1024, 16, (\leq 16, \leq 16)]]$ & [8,4,4] & [8,4,4] & 12 & 12 & 0.016 & 4.00 & \shortstack{50K\\$(186,174)$} & \shortstack{50M\\$(16,16)$}\\
\hline
\rowcolor{mycolor40!40} $C_5 \ltimes C_4$ & $[[1280, 24, (\leq 16, \leq 16)]]$ & [8,4,4] & [8,4,4] & 16 & 16 & 0.019 & 4.80 & \shortstack{50K\\$(30,24)$} & \shortstack{50M\\$(16,16)$} \\
\hline
\end{tabular}
\caption{\textbf{Selected instances of high-rate and high-distance QT codes.} For each instance we report the underlying group $\mathcal{G}$, the code parameters $[[n, k, (\leq d_X, \leq d_Z)]]$, the local codes $H_0$ and $H_1$, the maximum stabilizer row weights $\overline{w_X}$ and $\overline{w_Z}$, and the rate $k/n$. Distances are upper bounds from two independent randomized estimators: \texttt{DistRandCSS}~\cite{pryadko2023qdistrnd} with $50{,}000$ trials, and \texttt{sQetch}~\cite{bhardwaj2026high} with between $50$ and $200$ million trials on a single NVIDIA RTX~2080~Ti. We report the tighter of the two bounds for each code. Complete code data needed to reproduce these instances are provided in Appendix Tables~\ref{tab:multisets-of-lifted-quantum-codes} and~\ref{tab:permutations-of-lifted-quantum-codes}. Appendix~\ref{sec:reproducibility} provides an explicit example demonstrating the reproduction of the reported code parameters.}
\label{tab:qt-codes-main}
\end{table*}
\section{Numerical Results}
\label{sec:numerical-results-main-text}
We present explicit QT codes found via the lifting description Eq.~\eqref{eq:lifted}, using the search heuristics described below. Table~\ref{tab:qt-codes-main} collects the full set of codes produced by this search. From this catalogue we select four representative codes for decoding benchmarks (Table~\ref{pseudo_thresholds}) and highlight those whose minimum distance exceeds $\sqrt{n}$ in Table~\ref{tab:qt-codes-breaking-the-barrier}. We benchmark the four representative codes under both phenomenological and circuit-level noise models, decoded using the Tesseract decoder~\cite{beni2025tesseract} which we found to outperform BP+OSD~\cite{roffe2020decoding} on codes with high-weight stabilizer checks. Additional codes obtained via the original square-complex construction are provided in
Appendix~\ref{sec:codes-using-LRCC-method}.

\subsection{Search Heuristics}\label{sec:search-heuristics}
We adopt the lifting framework rather than the original square-complex picture for two reasons that directly inform the search.
First, restricting the lifted qubits to a single vertical $A$-slice yields a classical Tanner code on $\mathrm{Cay}_2(\mathcal{G},A)$ with local codes $C_i, C_i^\perp$, and restricting to a $B$-slice yields an analogous code on $\mathrm{Cay}_2(\mathcal{G},B)$; retaining only multisets that maximise the minimum distance of these classical codes eliminates many poor choices before the more expensive quantum distance estimation is run~\cite{leverrier2025small}.
Second, codewords of the $\mathcal{A}$-type Tanner codes give rise to logical $Z$-operators, and codewords of the $\mathcal{B}$-type codes give rise to logical $X$-operators, though they need not span the full logical operator space~\cite{leverrier2025small}; presenting codes in this form keeps that open problem in view.
\begin{figure*}[ht]
    \centering
    \includegraphics[width=1.0\linewidth]{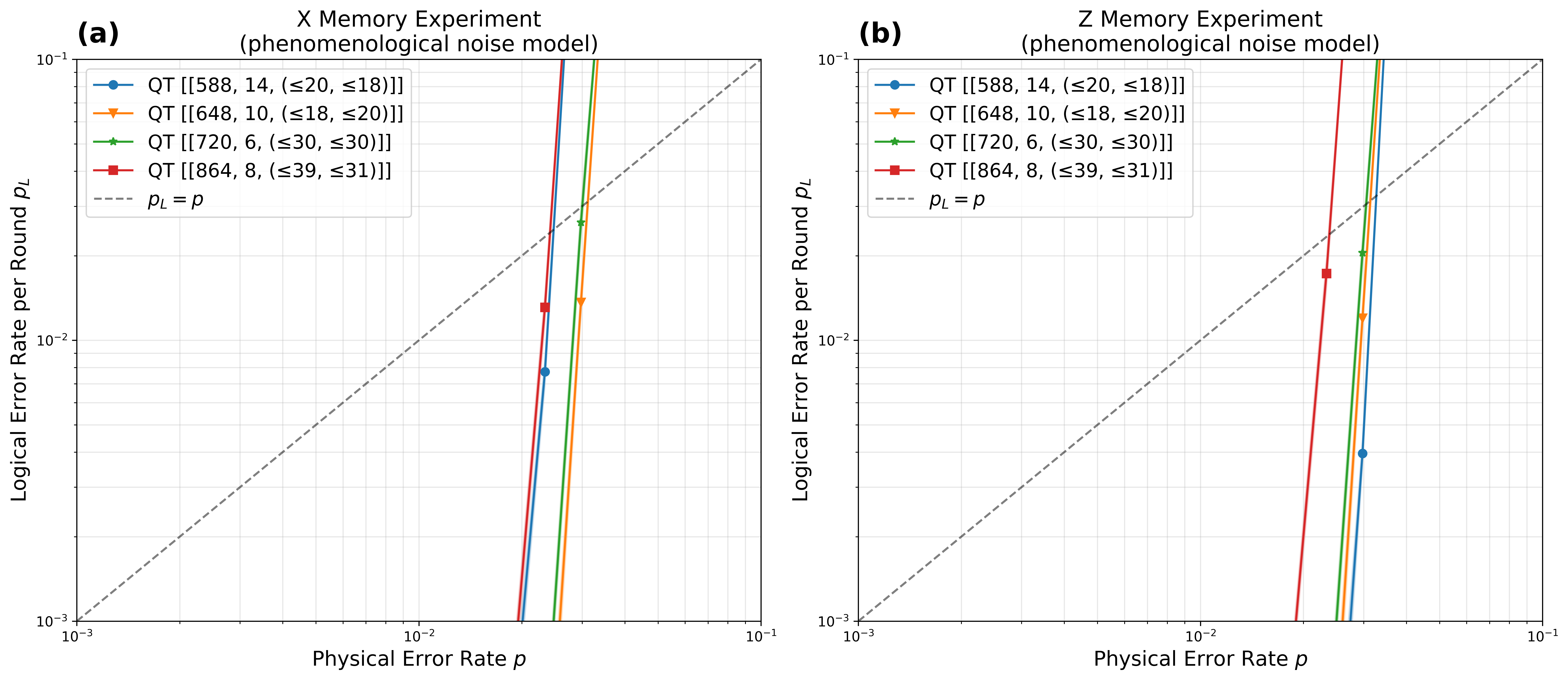}
    \caption{Logical error rate per round $p_L$ versus physical error rate $p$ under phenomenological noise, for the (a) $X$ and (b) $Z$ memory experiments, for the four quantum Tanner codes from our search. Data qubits and syndrome measurements are each flipped independently with probability $p$, and decoding is done with the Tesseract decoder.}
    \label{fig:pheno-xz}
\end{figure*}
Given a group $\mathcal{G}$ and two pairs of local codes $\mathcal{C}_i, \mathcal{C}_i'$ for $i \in \{0,1\}$, where $\mathcal{C}_i = \ker H_i = \mathrm{Im}(G_i^{\mathsf{T}})$ and $\mathcal{C}_i' = \ker H_i' = \mathrm{Im}(G_i'^{\mathsf{T}})$, a QT code of the form in Eq.~\eqref{eq:lifted} is specified by the multisets $\mathcal{A}, \mathcal{B} \subseteq \mathcal{G}$ and the column permutations $\pi_A, \pi_B$ that define
\begin{align}
H_1 = H_0[\cdot,\pi_A],\quad G_1 = G_0[\cdot,\pi_A], \\
H_1' = H_0'[\cdot,\pi_B],\quad G_1' = G_0'[\cdot,\pi_B].
\end{align}
We reduce this joint search space as follows.
Two permutations $\pi, \pi'$ yield the same code whenever $H_0[\cdot,\pi]$ and $H_0[\cdot,\pi']$ span the same row space over $\mathbb{F}_2$; we therefore compute the reduced row echelon form of $H_0[\cdot,\pi]$ for each $\pi \in S_{n_A}$ and retain one permutation per distinct result, applying the same permutation to $G_0$ to obtain $G_1$. For the length-$2$ repetition code, no search is needed since the $1\times 2$ matrix $(1\ 1)$ admits only one distinct row space. It suffices to consider multisets yielding inequivalent Cayley graphs
$\mathrm{Cay}_2(\mathcal{G},\mathcal{A})$ and $\mathrm{Cay}_2(\mathcal{G},\mathcal{B})$, since isomorphic graphs produce
isomorphic codes~\cite{leverrier2025small}. To reduce the search space, we use the fact that, by Lemma~3.7.3 of~\cite{godsil2013algebraic}, if $\theta$ is an automorphism of $\mathcal{G}$ then $X(\mathcal{G},\mathcal{A})\cong X(\mathcal{G},\theta(\mathcal{A}))$; we therefore enumerate all multisets of the appropriate size, group them into $\operatorname{Aut}(\mathcal{G})$-orbits, and retain one representative per orbit. For each surviving multiset $\mathcal{A}$, we form the classical Tanner code $\mathcal{C}(\mathcal{T}_A):=\ker_{\mathbb{F}_2}\mathcal{T}_A$ on $\mathrm{Cay}_2(\mathcal{G},\mathcal{A})$ with both local codes set to $\mathcal{C}_0$ (i.e.\ $\pi_A=\mathrm{id}$, the identity column permutation), whose parity-check matrix is~\cite{leverrier2025small}
\begin{align}
\mathcal{T}_A \;:=\;
\begin{pmatrix}
H_0 \otimes I_{|\mathcal{G}|}\\[2pt]
\bigl(H_0 \otimes I_{|\mathcal{G}|}\bigr)\,L_A
\end{pmatrix},
\end{align}
Since $\mathcal{T}_A$ depends only on $\mathcal{A}$ and $H_0$, and not on the permutation $\pi_A$, this decouples multiset selection from the
permutation search. Following~\cite{leverrier2025small}, we first rank the candidate multisets using an inexpensive proxy for the distance of the associated classical Tanner codes. For each multiset $\mathcal{A}$, we construct the classical Tanner code $\mathcal{C}(\mathcal{T}_A):=\ker_{\mathbb{F}_2}\mathcal{T}_A$ and use the minimum Hamming weight among the vectors of a computed kernel basis as a screening score. We retain all multisets maximizing this score independently on the two sides. The surviving combinations of multisets and column permutations are then assembled into quantum codes via~\eqref{eq:lifted}, and instances with $k=0$ are discarded. For the remaining codes, upper bounds on $d_X$ and $d_Z$ are obtained using \texttt{DistRandCSS}~\cite{pryadko2023qdistrnd} and \texttt{sQetch}~\cite{bhardwaj2026high}, with the latter run for between $50$ million and $350$ million trials.

\begin{figure*}[ht]
    \centering
    \includegraphics[width=1.0\linewidth]{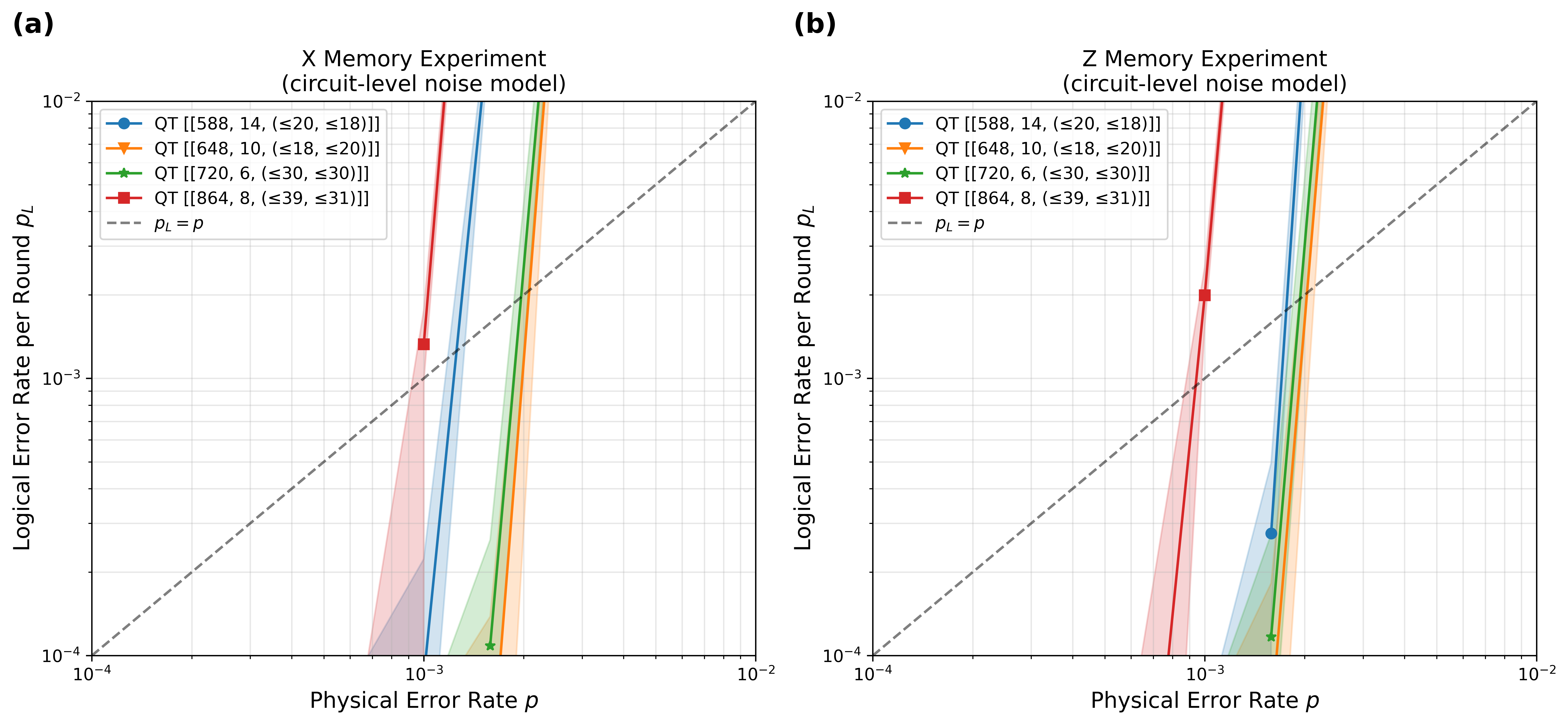}
\caption{Logical error rate per round $p_L$ versus physical error rate $p$ under circuit-level noise, for the (a) $X$ and (b) $Z$ memory experiments, for the four quantum Tanner codes from our search. Every gate, reset, and measurement is subject to depolarizing or flip noise at rate $p$, and idling qubits are subject to noise at rate $p/10$; decoding is done with the Tesseract decoder. The dashed line marks $p_L=p$ and serves only as a visual reference. The pseudo-thresholds reported in the text and in Table~\ref{pseudo_thresholds} are instead defined by the crossing
$p_L(p)=Tkp/10$, where $k$ is the number of logical qubits and $T$ is the number of Stim time steps (\texttt{TICK}s) in a single round of syndrome extraction.}
\label{fig:circuit-xz}
\end{figure*}
\subsection{Error Models}
\label{sec:error-models}

\setlength{\tabcolsep}{3pt}
\renewcommand{\arraystretch}{1.6}
\begin{table}[t]
\centering
\setlength{\tabcolsep}{4pt}
\renewcommand{\arraystretch}{1.4}
\small
\begin{tabular}{lcccc}
\toprule
& \multicolumn{2}{c}{\textbf{Phenom.}} & \multicolumn{2}{c}{\textbf{Circuit}} \\
\cmidrule(lr){2-3} \cmidrule(lr){4-5}
$[[n,k,(\le\! d_X,\le\! d_Z)]]$ & $X$ & $Z$ & $X$ & $Z$ \\
\midrule
$[[588,14,(20,18)]]$  & $4.17\%$ & $4.62\%$ & $0.19\%$ & $0.26\%$ \\
$[[648,10,(18,20)]]$  & $4.28\%$ & $4.28\%$ & $0.27\%$ & $0.27\%$ \\
$[[720,\:6,(30,30)]]$ & $3.99\%$ & $4.48\%$ & $0.24\%$ & $0.24\%$ \\
$[[864,\:8,(39,31)]]$ & $3.58\%$ & $3.57\%$ & $0.14\%$ & $0.13\%$ \\
\bottomrule
\end{tabular}
\caption{Pseudo-thresholds under phenomenological and circuit-level noise
for four quantum Tanner codes, decoded with the Tesseract
decoder~\cite{beni2025tesseract}. Distances are upper bounds.}
\label{pseudo_thresholds}
\end{table}
We benchmark four quantum Tanner codes from our search: $[[588, 14, (\leq 20, \leq 18)]]$, $[[648, 10, (\leq 18, \leq 20)]]$, $[[720, 6, (\leq 30, \leq 30)]]$, and $[[864, 8, (\leq 39, \leq 31)]]$. For each code we perform $X$- and $Z$-basis memory experiments under two noise models, decoded with Tesseract~\cite{beni2025tesseract}.

In the phenomenological noise model, each data qubit undergoes single-qubit depolarizing noise with probability $p$ at the start of every syndrome-extraction round, and each syndrome measurement is independently flipped with the same probability $p$; gates, resets, and idle locations are noiseless. In the circuit-level noise model, every Clifford gate is followed by depolarizing noise at rate $p$, resets and measurements each fail independently with probability $p$, data qubits are additionally subject to per-round depolarizing noise at rate $p$, and qubits idling during a gate layer or while ancilla measurements are inactive incur depolarizing noise at rate $p/10$.

We report the per-round logical error rate $p_L(p) = 1 - (1 - P_L)^{1/r}$, where $P_L$ is the total logical failure probability over $r$ syndrome-extraction rounds ($r = d_z$ for the $X$-basis experiment, $r = d_x$ for the $Z$-basis experiment). Pseudo-thresholds are defined by the crossing $p_L(p) = kp$ under phenomenological noise and $p_L(p) = Tkp/10$ under circuit-level noise, where $k$ is the number of logical qubits and $T$ is the depth of a single syndrome-extraction round; the dashed line in Figs.~\ref{fig:pheno-xz}--\ref{fig:circuit-xz} marks th break-even reference line $p_L = p$. The resulting pseudo-thresholds are collected in Table~\ref{pseudo_thresholds}: $3.6$--$4.6\%$ under phenomenological noise and $0.14$--$0.27\%$ under circuit-level noise. The $X$ and $Z$ bases behave similarly throughout, as expected from the near-symmetric distances of these codes.

\begin{table*}[ht!]
\centering
\footnotesize
\setlength{\tabcolsep}{10pt}
\renewcommand{\arraystretch}{2.2}
\begin{tabular}{|*{10}{>{\columncolor{white}[\tabcolsep]}c|}}
\hline
\rowcolor{mycolor41!50} \textbf{Group} &
$[[\textbf{n, k,} (\le \textbf{d}_X, \le\textbf{d}_Z)]]$ &
$\sqrt{\textbf{n}}$ &
$\textbf{H}_0$ &
$\textbf{H}_1$ &
$\overline{\textbf{w}_X}$ &
$\overline{\textbf{w}_Z}$ &
$\tfrac{\textbf{k}}{\textbf{n}}$ &
$\tfrac{\textbf{k}\,\textbf{d}_{\min}^2}{\textbf{n}}$ &
\shortstack{\textbf{Trials}\\(\texttt{sQetch}) \cite{bhardwaj2026high}} \\
\hline
\rowcolor{mycolor1!30} $D_{10}$          & $[[480,  8, (\le 21, \le 21)]]$ & $21.9$ & $[8,4,4]$ & $[6,3,3]$ & $12$ & $12$ & $0.017$ & $7.35$  & 250M \\ \hline
\rowcolor{mycolor2!30} $C_3 \times C_3$  & $[[504,  4, (\le 36, \le 27)]]$ & $22.4$ & $[8,4,4]$ & $[7,3,4]$ & $16$ & $16$ & $0.008$ & $5.79$  & 250M \\ \hline
\rowcolor{mycolor3!30} $C_3 \ltimes C_4$ & $[[672,  4, (\le 48, \le 28)]]$ & $25.9$ & $[8,4,4]$ & $[7,3,4]$ & $16$ & $16$ & $0.006$ & $4.67$  & 250M \\ \hline
\rowcolor{mycolor4!30} $C_5 \ltimes C_4$ & $[[720,  6, (\le 30, \le 30)]]$ & $26.8$ & $[6,3,3]$ & $[6,3,3]$ & $9$  & $9$  & $0.008$ & $7.50$  & 250M \\ \hline
\rowcolor{mycolor5!30} $C_3 \times S_3$  & $[[864,  8, (\le 39, \le 31)]]$ & $29.4$ & $[8,4,4]$ & $[6,3,3]$ & $12$ & $12$ & $0.009$ & $8.90$  & 250M \\ \hline
\rowcolor{mycolor6!30} $C_3 \times S_3$  & $[[864, 16, (\le 36, \le 32)]]$ & $29.4$ & $[8,4,4]$ & $[6,3,3]$ & $12$ & $12$ & $0.019$ & $18.96$ & 250M \\ \hline
\rowcolor{mycolor7!30} $C_3 \times S_3$  & $[[864,  8, (\le 42, \le 40)]]$ & $29.4$ & $[8,4,4]$ & $[6,3,3]$ & $12$ & $12$ & $0.009$ & $14.81$ & 350M  \\ \hline
\rowcolor{mycolor8!30} $C_5 \ltimes C_4$ & $[[1120, 4, (\le 80, \le 35)]]$ & $33.5$ & $[8,4,4]$ & $[7,3,4]$ & $16$ & $16$ & $0.004$ & $4.38$  & 350M  \\ \hline
\end{tabular}
\caption{\textbf{Quantum Tanner codes exceeding the $\sqrt{n}$ distance barrier at moderate blocklength.} For each instance we report the underlying group $\mathcal{G}$, the code parameters $[[n, k, (\le d_X, \le d_Z)]]$, the local codes $H_0$ and $H_1$, the maximum $X$- and $Z$-stabilizer row weights $\overline{w_X}$ and $\overline{w_Z}$, the rate $k/n$, the BPT figure of merit $k d_{\min}^2/n$ with $d_{\min} = \min(d_X, d_Z)$, and the number of trials used by \texttt{sQetch}~\cite{bhardwaj2026high} to obtain the distance upper bounds. Seven of the eight instances satisfy $d_{\min} > \sqrt{n}$; the $[[480,\,8,\,(\le 21,\le 21)]]$ code sits on the boundary, with $d_{\min} = 21$ against $\sqrt{n} \approx 21.9$. Complete code data needed to reproduce these instances are provided in Appendix Tables~\ref{tab:multisets-of-lifted-quantum-codes-breaking-the-barrier} and~\ref{tab:permutations-of-lifted-quantum-codes-breaking-the-barrier}. Appendix~\ref{sec:reproducibility} provides an explicit example demonstrating the reproduction of the reported code parameters.}
\label{tab:qt-codes-breaking-the-barrier}
\end{table*}
\subsection{Codes Beyond the \texorpdfstring{$\sqrt{n}$}{sqrt(n)} Barrier}
\label{subsec:sqrt-n}

The $\sqrt{n}$ distance ceiling that constrains 2D geometrically local codes via the BPT bound~\cite{PhysRevLett.104.050503}, and that persisted as a barrier for qLDPC codes more broadly~\cite{hastings2021fiber, panteleev2022asymptotically, leverrier2022quantum}, is an asymptotic statement; whether it can be crossed at practical blocklengths is a separate empirical question.

Table~\ref{tab:qt-codes-breaking-the-barrier} collects code instances from our search whose \texttt{sQetch}-estimated distance upper bounds lie near or above the $\sqrt{n}$ line at moderate blocklength. For the $[[480,\,8,\,(\le 21,\le 21)]]$ code, the reported value $d_{\min}=21$ lies just below $\sqrt{n}\approx21.9$; for the remaining seven instances, the reported ratios $d_{\min}/\sqrt{n}$ range from approximately $1.05$ for the $[[1120,\,4,\,(\le 80,\le 35)]]$ code to approximately $1.43$ for the $[[864,\,8,\,(\le 42,\le 42)]]$ code. The corresponding values of the BPT figure of merit $kd_{\min}^2/n$, computed from these distance upper bounds, exceed the surface-code value of $1$ throughout the table. Distances are upper bounds returned by \texttt{sQetch}~\cite{bhardwaj2026high} after between $50$ million and $350$ million trials; consequently, the true minimum distances, and hence the true values of $kd_{\min}^2/n$, may be smaller.

\section{Discussion and Conclusions}
\label{sec:discussion}
In this work, we have addressed the open question raised in~\cite{leverrier2025small} regarding whether QT codes with competitive parameters exist in the moderate blocklength regime $n \in [500, 1000]$. Through an extensive search over non-abelian groups in GAP's \texttt{SmallGrp} \cite{SmallGrp} library, we have identified several new code instances achieving distance $d > 20$ in this regime, providing an affirmative answer to this question.

Among the local codes explored in our search, the classical $[8,4,4]$ code consistently yields quantum codes with good parameters, emerging as a particularly effective building block in the moderate blocklength range. The new code instances discovered in this work include $[[480,  8, (\leq21, \leq21)]]$, $[[504,  4, (\leq36,\leq27)]]$, $[[672,  4, (\le 48, \le28)]]$, $[[720,  6, (\leq30, \leq30)]]$ and $[[864,  8, (\leq39, \leq31)]]$, demonstrating that the moderate blocklength regime admits codes with substantially larger distances than previously reported. Accompanying this work is \texttt{QuantumExpanders.jl}, an open-source \texttt{Julia} library implementing both the LRCC and lifted constructions alongside the search heuristics of Sec.~\ref{sec:search-heuristics}.

Our work complements the concurrent code search of Wang~\textit{et al.}~\cite{wang2026check}, which also identifies explicit QT code instances, primarily to populate the parameter space defined by their linear-programming (LP) upper bounds on check-weight-constrained codes. Our contribution differs in focus and scope: rather than benchmarking codes against LP bounds,
we conduct a comprehensive search specifically targeting the moderate blocklength regime $n\in[500,1000]$, sweeping over a substantially broader set of groups with particular emphasis on non-abelian groups, and approaching the construction from both the LRCC and lifting perspectives. Furthermore, the concurrent \emph{mitten codes} of Bhardwaj~\textit{et al.}~\cite{bhardwaj2026high} similarly exploit non-abelian groups, constructing Panteleev--Kalachev lifted product codes~\cite{panteleev2022asymptotically, Panteleev_2022} rather than QT codes, obtaining a rate-$1/5$ family of eight instances from $[[150,30,10]]$ to $[[975,195,24]]$. No distance guarantee exceeding $\Theta(\sqrt{n})$ is known in this regime, and the reported instances empirically scale as $d \sim \sqrt{n}$; QT codes, by contrast, admit an asymptotic linear-distance guarantee~\cite{leverrier2022quantum} that our search pursues at moderate blocklength.

Lifting has recently emerged as a productive way to build larger quantum codes from smaller ones. Gu\'emard~\cite{guemard2025lifts} shows that any CSS code admits a geometric representation (a Tanner cone-complex) whose coverings define a lift, recovering the Panteleev--Kalachev lifted product \cite{panteleev2022asymptoticallygoodquantumlocally} as a special case; Hirasaki and Lee~\cite{hirasaki2026lifting} instead lift Panteleev--Kalachev codes via group extensions, exactly preserving check weight and qubit degree. Both provide analytic control over how a lift's parameters relate to its seed, beyond what we attempt here. Closer to our setting, Gu\'emard and Z\'emor~\cite{guemard2025moderate} specialize this idea to quantum Tanner codes, lifting arbitrary 2D square complexes through odd-index coverings ($\tilde n = tn$, $\tilde k \geq k$, $\tilde d \leq td$, with $\tilde d \geq d$ when $\tilde k = k$), and search by sweeping indices $t=1$ to $30$ over a few seed complexes up to $n<800$, estimating distance with GAP and QDistRnd as we do. Our left-right group-action lift, following~\cite{leverrier2025small}, requires no such compatibility between local codes and complex (Appendix~\ref{app:multisets}); rather than lifting a fixed seed along a sequence of indices, we search a broad space of non-abelian groups directly in the regime $n \in [500,1000]$.

Looking ahead, an important direction for future work is a better understanding of the logical operator structure of QT codes. Finding a canonical basis of logical operators for QT codes would allow them to be incorporated into proposed designs for quantum LDPC processors \cite{bhardwaj2026high, zheng2026logical, hong2026quantum, lu2026quantum}. The code instances and open-source tools presented here provide a concrete starting point for such investigations.

\section{acknowledgments}
VLA started QuantumExpanders.jl a few years ago as a class project under the supervision of SK. FAM conceived the majority of the newer features and the search techniques described here, adopted and significantly expanded QuantumExpanders.jl, performed the numerical experiments, and wrote the manuscript with the input of SK and VLA. FAM performed the LRCC searches reported in the appendix, building on early attempts by AM and AC. We acknowledge support from NSF Grants No. 1941583, 2346089, 2402861, 2522101, and 2615830.

\bibliography{bibliography}

\appendix
\section{Reproducibility with QuantumExpanders.jl}
\label{sec:reproducibility}

The code instances reported in this work were constructed and analyzed with \texttt{QuantumExpanders.jl}, an open-source \texttt{Julia} library. The library provides deterministic implementations of both the LRCC-based QT code construction of Section~\ref{subsec:lrccconstruction} following \cite{radebold2025explicit} and the lifted QT code construction of Section~\ref{subsec:liftedtannercodes} following\cite{leverrier2025small}, together with the search heuristics of Section~\ref{sec:search-heuristics}. Groups can be specified directly or imported from GAP's \texttt{SmallGrp} library~\cite{SmallGrp} via \texttt{Oscar.jl}~\cite{OSCAR, OSCAR-book}, and distances of the resulting CSS codes are estimated via \texttt{sQetch}~\cite{bhardwaj2026high} and \texttt{DistRandCSS}~\cite{pryadko2023qdistrnd}. Both constructions, along with the search infrastructure, were added for the present work, extending an earlier version of the library that supported only the random QT construction of~\cite{gu2022efficient} and Morgenstern's $(q+1)$-regular Ramanujan graphs for even prime powers. The code data is available at \cite{feroz_ahmed_mian_2026_21904804}.

\subsection{Reproducing a code from Table~\ref{tab:qt-codes-main}}

As a self-contained example, the snippet below constructs the $[[756, 10, (\le 9, \le 42)]]$ code of Table~\ref{tab:qt-codes-main} through the lifted construction of Eq.~\eqref{eq:lifted}. The group $C_3 \ltimes C_4$ is imported from GAP's \texttt{SmallGrp}
library~\cite{SmallGrp} via \texttt{Oscar.jl}; the multisets $\mathcal{A}, \mathcal{B}$ and column permutations $\pi_A, \pi_B$ are those recorded in the corresponding row of the table. The same pattern applies to every code instance we report.

\begin{lstlisting}[language=JuliaREPL]
julia> using QuantumExpanders, QuantumClifford, Oscar;
julia> G734 = [1 0 1 1 1 0 0;
               1 1 1 0 0 1 0;
               0 1 1 1 0 0 1]; # [7,3,4]
julia> H734 = dual_code(G734);
julia> G953 = [1 0 0 0 0 1 1 1 1;
               0 1 0 0 0 1 1 1 0;
               0 0 1 0 0 1 1 0 1;
               0 0 0 1 0 1 0 1 1;
               0 0 0 0 1 0 1 1 1]; # [9,5,3]
julia> H953 = dual_code(G953);
julia> G = codomain(isomorphism(PermGroup, small_group(12, 1)));
julia> A = [one(G), one(G), cperm(G,[5,6,7]), cperm(G,[5,6,7]), cperm(G,[1,4,3,2],[6,7]), cperm(G,[1,4,3,2],[6,7]), cperm(G,[1,4,3,2],[5,7])];
julia> B = [one(G), one(G), cperm(G,[5,6,7]), cperm(G,[5,6,7]), cperm(G,[1,4,3,2],[6,7]), cperm(G,[1,4,3,2],[5,7]), cperm(G,[1,4,3,2],[5,6]), cperm(G,[1,2,3,4],[6,7]), cperm(G,[1,2,3,4],[5,7])];
julia> c = QuantumTannerViaLeftRightActions(G, A, B, H734, G734, H953, G953; p1 = [1,2,3,4,5,6,7], p2 = [1,2,3,4,5,7,8,9,6]);
\end{lstlisting}

The inputs \texttt{p1} and \texttt{p2} are the column permutations $\pi_A$ and $\pi_B$ of Section~\ref{sec:search-heuristics}, which reorder the columns of the $A$-side and $B$-side local codes to produce $H_1$ from $H_0$ and $H_1'$ from $H_0'$; the identity permutation on the $A$-side ($\pi_A = \mathrm{id}$) means $H_1 = H_0 = H_{7,3,4}$, while $\pi_B$ acts non-trivially on the $B$-side. Distances $(d_X, d_Z)$ can then be estimated by passing
\texttt{c.hx} and \texttt{c.hz} to \texttt{sQetch}~\cite{bhardwaj2026high} or \texttt{DistRandCSS}~\cite{pryadko2023qdistrnd}, matching the values reported in Table~\ref{tab:qt-codes-main}.

\subsection{Reproducing a code from Table~\ref{appendix:lrcc1}}

To make the results of Table~\ref{appendix:lrcc1} easy to verify, we walk through a concrete example: the $[[324,\,8,\,(17,\,14)]]$ code built from the group $C_3\times S_3$. The construction follows Section~\ref{subsec:lrccconstruction} and requires only the group, its generators, and the two local codes.

The group is loaded directly from GAP's \texttt{SmallGrp} library~\cite{SmallGrp} via \texttt{Oscar.jl} as \texttt{small\_group(18,\,3)}, whose three generators \texttt{r}, \texttt{s}, \texttt{t} correspond to the elements listed in the table. The generating multisets $\mathcal{A}$ and $\mathcal{B}$ are written as words in these generators, and both local codes are the $[6,3,3]$ Hamming code.

\begin{lstlisting}[language=JuliaREPL]
julia> using QuantumExpanders, QuantumClifford, Oscar;
julia> H633 = [1 0 0 0 1 1;
               0 1 0 1 0 1;
               0 0 1 1 1 0];   
julia> G633 = dual_code(H633);
julia> G = small_group(18, 3); # C_3 x S_3
julia> r, s, t = Oscar.gens(G);
julia> A = [s, s^2, t, t^2, r*t^2, r];
julia> B = [r*s, r*s^2, s*t, s^2*t^2, r*s*t^2, r*s^2*t^2];
julia> c = QuantumTannerCode(G, A, B, ((H633, G633), (H633, G633)));
julia> hx, hz = parity_matrix_xz(c);
julia> code_n(c), code_k(c)
(324, 8)
\end{lstlisting}

\noindent
This produces a $[[324, 8, (\leq17, \leq14)]]$ code. Passing \texttt{hx} and \texttt{hz} to \texttt{sQetch}~\cite{bhardwaj2026high} with $50$ million random-ISD trials recovers the distances reported in the table. The same steps applies to every entry in Table~\ref{appendix:lrcc1}.
\section{Codes using LRCC method}
\label{sec:codes-using-LRCC-method}
In addition to the lifted construction described in Section~\ref{subsec:liftedtannercodes}, we performed a complementary search using the original LRCC framework of Section~\ref{subsec:lrccconstruction}. While the lifted construction enables a more systematic search, the LRCC construction provides a direct geometric realization of the code and is useful as an independent check on the codes discovered.

For fixed $G$, $C_A$, and $C_B$, the LRCC construction is parameterized by the generating sets $\mathcal{A}$ and $\mathcal{B}$. These must satisfy four algebraic conditions: (i) $\mathcal{A} = \mathcal{A}^{-1}$ and $\mathcal{B} = \mathcal{B}^{-1}$ (symmetry); (ii) $\mathcal{A} \cap \mathcal{B} = \emptyset$; (iii) $g^{-1} a g \neq b$ for all $g \in G$, $a \in \mathcal{A}$, $b \in \mathcal{B}$ (non-conjugacy~\cite[Definition~3.6]{dinur2022locally}) when using bipartite LRCC construction; and (iv) $\langle \mathcal{A} \cup \mathcal{B} \rangle = \mathcal{G}$.

We sample candidate pairs $(A, B)$ by randomly permuting the non-identity elements of $G$ and assigning them to $A$ or $B$: involutions are added individually, while elements of order greater than $2$ are added in inverse pairs $\{g, g^{-1}\}$ to preserve symmetry by construction. Candidates failing conditions (ii)--(iv) are discarded. For each accepted pair, the parity-check matrices $H_X, H_Z$ are constructed as above, the number of logical qubits is computed as $k = n - \operatorname{rank}_{\mathbb{F}_2}(H_X) - \operatorname{rank}_{\mathbb{F}_2}(H_Z)$, instances with $k = 0$ are discarded, and the distances $d_X, d_Z$ are estimated via \texttt{sQetch}~\cite{bhardwaj2026high} and \texttt{DistRandCSS}~\cite{pryadko2023qdistrnd}. The search is repeated over independent random seeds, and sweeps over non-abelian groups from the GAP \texttt{SmallGrp} \cite{SmallGrp} library that have not been considered in prior LRCC searches.

\section{Quantum Tanner Codes using Morgenstern Ramanujan Graphs}
\label{sec:morgenstern}
Quantum Tanner codes require expander graphs with strong spectral properties to guarantee good distance. Morgenstern's construction~\cite{morgenstern1994existence} provides explicit families of Ramanujan graphs for even prime powers $q = 2^l$, making it possible to instantiate QT codes from provably optimal expanders. To our knowledge, this is the first time such an explicit family has been used to construct QT codes.
\begin{table*}
\begingroup
\begin{tabular}{|*{7}{>{\columncolor{white}[\tabcolsep]}c|}}
\hline
 \rowcolor{zebragray} $[[n,k,(\leq d_X, \leq d_Z)]]$ & $r$ & $\rho$ & $\frac{kd_{min}^2}{n}$ & $C_A$ & $C_B$ \\
\hline
 $[[360,8,(\leq 3, \leq 4)]]$ & 0.02 & 0.35 & 
0.2 & $\begin{pmatrix} 1 & 1 & 0 & 1\\ 1 & 1 & 1 &1\end{pmatrix}$
& $\begin{pmatrix}1 & 1 & 1\end{pmatrix}$\\
\hline
 $[[360,61,(\leq6,\leq3)]]$ & 0.17 & 0.75 &
1.5 & $\begin{pmatrix}
    1 & 1 & 1 & 0
\end{pmatrix}$
& $\begin{pmatrix}
    0 & 1 & 1 \\ 1 & 1 & 0
\end{pmatrix}$\\
\hline
\end{tabular}
\editcolor{\caption{Quantum Tanner codes from $\SL_2(\F_4)$ and Morgenstern generators for even prime power $q$. The generators of the this are as follows: A = $\Bigg\langle
\begin{matrix}
\begin{pmatrix}\omega+1 & \omega+1 \\ \omega & 0 \end{pmatrix}, 
\begin{pmatrix}0 & \omega+1 \\ \omega & \omega+1 \end{pmatrix}, 
\begin{pmatrix}0 & 1 \\ 1 & \omega+1 \end{pmatrix}, 
\begin{pmatrix}\omega+1 & 1 \\ 1 & 0 \end{pmatrix}
\end{matrix}
\Bigg\rangle$  and B = $\Bigg\langle
\begin{matrix}
\begin{pmatrix}\omega+1 & \omega+1 \\ 1 & \omega+1 \end{pmatrix}, 
\begin{pmatrix}\omega+1 & \omega \\ \omega & \omega+1 \end{pmatrix}, 
\begin{pmatrix}\omega+1 & 1 \\ \omega+1 & \omega+1 \end{pmatrix}
\end{matrix}
\Bigg\rangle$. The explicit construction of Morgenstern Ramanujan (q+1) graphs for even prime power q are implemented in our QuantumExpanders.jl library. $p$ is the rate of the classical local code that is randomly generated and we use the bipartite construction of QT codes. The rate and the efficieny ratio of the denoted by $r$, and $\frac{kd_{min}^2}{n}$ respectively. The classical local codes are defined by $C_A$ and $C_B$ which are classical parity check matrices are used to define the local constraints at vertices of LRCC. The minimum distance $d = min(d_X, d_Z)$ was computed using \texttt{DistRandCSS}~\cite{pryadko2023qdistrnd} using 50,000 trials.}}

\endgroup
\label{tab:morgenstern}
\end{table*}
\subsection{Morgenstern's Graph Construction}
We use the explicit $(q+1)$-regular Ramanujan graphs of
Morgenstern~\cite{morgenstern1994existence} for even prime powers
$q=2^l$. Let $\varepsilon\in\mathbb{F}_q$ be chosen such that
$t^2+t+\varepsilon$ is irreducible over $\mathbb{F}_q$, and let
$g(x)\in\mathbb{F}_q[x]$ be irreducible of even degree $d$. Writing
$\mathbb{F}_{q^d}\cong\mathbb{F}_q[x]/(g(x))$, let $\bar{x}$ denote
the residue class of $x$ and let $\iota\in\mathbb{F}_{q^d}$ be a
root of
\begin{align}
\iota^2+\iota+\varepsilon=0.
\end{align}
The $q+1$ Morgenstern generators are indexed by the solutions
$(\gamma,\delta)\in\mathbb{F}_q^2$ of
\begin{align}
\gamma^2+\gamma\delta+\varepsilon\delta^2=1,
\end{align}
of which there are exactly $q+1$, and are represented projectively
by
\begin{align}
b_{\gamma,\delta}
=
\left[
\begin{pmatrix}
1 & \gamma+\delta\iota \\
(\gamma+\delta\iota+\delta)\bar{x} & 1
\end{pmatrix}
\right].
\end{align}
For even $q$,
$\mathrm{PGL}_2(\mathbb{F}_{q^d})
=\mathrm{PSL}_2(\mathbb{F}_{q^d})$, and the resulting Cayley graph
\begin{align}
\Gamma_g
=
\operatorname{Cay}
\left(
\mathrm{PSL}_2(\mathbb{F}_{q^d}),B
\right)
\end{align}
is $(q+1)$-regular and non-bipartite. By
Theorem~5.13 of~\cite{morgenstern1994existence},
\begin{align}
|\Gamma_g| &= q^{3d}-q^d, &
\operatorname{girth}(\Gamma_g)
&\geq \frac{2}{3}\log_q|\Gamma_g|,
\end{align}
and every nontrivial adjacency eigenvalue satisfies
\begin{align}
|\mu|\leq2\sqrt{q}.
\end{align}
Hence $\Gamma_g$ is Ramanujan.
\subsection{Alternating Generating Sets}
The QT code construction requires two symmetric generating sets $A$ and $B$ satisfying the total non-conjugacy (TNC) condition. Starting from the Morgenstern generating set $B=\{b_0,b_1,\ldots,b_q\}$, Dinur~\cite{dinur2022locally} provides the alternative symmetric generating set
\begin{align}
A = \{\,b_0b_j,\;b_jb_0
\mid j=1,\ldots,q\,\},
\end{align}
of size $2q$. The pair $(A,B)$ satisfies the TNC condition. Moreover, $\mathrm{Cay}(G,A)$ is a quasi-Ramanujan spectral expander; writing $\lambda$ for the largest nontrivial eigenvalue in absolute value of the normalized adjacency operator,
\begin{align}
\lambda\bigl(\mathrm{Cay}(G,A)\bigr) < \frac{3\sqrt{2q-1}}{2q}.
\end{align}

These two generating sets are then passed directly to the LRCC construction of Section~\ref{subsec:lrccconstruction}. We implement two strategies for constructing $A$ from $B$. The \texttt{FirstOnly} strategy uses $\{b_0b_j,b_jb_0\mid j=1,\ldots,q\}$ and gives $|A|=2q$, whereas the \texttt{AllPairs} strategy uses all pairwise products $\{b_tb_s\mid t\neq s\}$ and gives $|A|=q^2+q$.

The following example uses \texttt{FirstOnly} to construct $\mathrm{SL}_2(\mathbb{F}_4)$ with the Morgenstern generating set $B$ of size $q+1=3$ and the alternative generating set $A$ of size $2q=4$:

\begin{lstlisting}[language=JuliaREPL]
julia> using Oscar; using QuantumExpanders;
julia> SL2, B = morgenstern_generators(l, i)
[ Info: |SL2($\mathbb{F}(4)$)| = 60
(SL(2,4), MatrixGroupElem{FqFieldElem, FqMatrix}[[o+1 o+1; 1 o+1], [o+1 1; o+1 o+1], [o+1 o; o o+1]])
julia> B
3-element Vector{MatrixGroupElem{FqFieldElem, FqMatrix}}:
 [o+1 o+1; 1 o+1]
 [o+1 1; o+1 o+1]
 [o+1 o; o o+1]
julia> A = alternative_morgenstern_generators(B, FirstOnly())
4-element Vector{MatrixGroupElem{FqFieldElem, FqMatrix}}:
 [0 1; 1 o+1]
 [o+1 1; 1 0]
 [o+1 o+1; o 0]
 [0 o+1; o o+1]
julia> H_A = [1 1 1 0];
julia> G_A = Matrix{Int}(lift.(dual_code(matrix(ZZ, H_A))));
julia> H_B = [0 1 1; 1 1 0];
julia> G_B = Matrix{Int}(lift.(dual_code(matrix(ZZ, H_B))));
julia> code_pair = ((H_A, G_A), (H_B, G_B));
julia> c = QuantumTannerCode(SL2, A, B, code_pair);
julia> code_n(c), code_k(c)
(360, 61)
Hx, Hz = parity_matrix_xz(c);
julia> gap_gf2(M) = GAP.Globals.Matrix(GAP.Globals.GF(2), GAP.Obj(mod.(M, 2)));
julia> trials = 50_000;
julia> dz = Int(GAP.Globals.DistRandCSS( gap_gf2(Hx), gap_gf2(Hz), GAP.Obj(trials), GAP.Obj(0), GAP.Obj(0)));
julia> dx = Int(GAP.Globals.DistRandCSS(gap_gf2(Hz), gap_gf2(Hx), GAP.Obj(trials), GAP.Obj(0), GAP.Obj(0)));
julia> (dx, hz)
(10, 3)
\end{lstlisting}

where \texttt{o} denotes the primitive element $\omega \in \mathbb{F}_4$.
The resulting QT code parameters are reported in
Table~\ref{tab:morgenstern}
\raggedbottom
\section{Multiplicity in the multisets}\label{app:multisets}
The multisets $\mathcal{A}$ and $\mathcal{B}$ in the lifted QT code construction Eq.~\eqref{eq:lifted} are not required to consist of distinct elements. Indeed, Eq.~\eqref{eq:lifted} depends on $\mathcal{A}$ and $\mathcal{B}$ only through the block-diagonal permutation matrices \cite{leverrier2025small}
\begin{align*}
L_A&=\bigoplus_{i=1}^{n_A}\bigoplus_{j=1}^{n_B}\lambda_{a_i},
&
R_B&=\bigoplus_{i=1}^{n_A}\bigoplus_{j=1}^{n_B}\rho_{b_j},
\end{align*}
where $\lambda_a,\rho_b\in\mathbb{F}_2^{|\mathcal{G}|\times|\mathcal{G}|}$ are the matrices of the left- and right-regular representations of $\mathcal{G}$, with entries $(\lambda_a)_{g,h}=\delta_{h,ag}$ and $(\rho_b)_{g,h}=\delta_{h,gb^{-1}}$ \cite{leverrier2025small}. The CSS condition $H_XH_Z^{\mathsf T}=0$ relies only on two properties of these matrices: each $\lambda_{a_i}$ and $\rho_{b_j}$ is a permutation matrix, and the left and right regular actions commute, $\lambda_a\rho_b=\rho_b\lambda_a$ for all $a,b\in\mathcal{G}$. Both properties hold for arbitrary group elements. This is in contrast with the original construction of~\cite{leverrier2022quantum}, where $\mathcal{A}$ and $\mathcal{B}$ are required to be symmetric generating sets of $\mathcal{G}$ of cardinality $\Delta$.

To illustrate this explicitly, consider $\mathcal{G}=C_2$. We write the elements of  $\mathcal{G}$ in cycle notation, identifying $\mathcal{G}$ with a permutation group via Cayley's theorem. In this case $\mathcal{G}=\{(),(1\,2)\}$, where $()$ denotes the identity permutation. Let all four local codes equal the $[2,1,2]$ repetition code,
\begin{align*}
H_0=H_1=G_0=G_1=H_0'=H_1'=G_0'=G_1'=(1\quad 1),
\end{align*}
so that the lifted code has $n=8$ physical qubits. Ordering the elements of $\mathcal{G}$ as $\bigl\{(),(1\,2)\bigr\}$, the matrices of the regular representations are
\begin{align*}
\lambda_{()}=\rho_{()}=I_2,
\quad
\lambda_{(1\,2)}=\rho_{(1\,2)}=
\begin{pmatrix}
0 & 1\\
1 & 0
\end{pmatrix},
\end{align*}
where $\lambda_{(1\,2)}=\rho_{(1\,2)}$ since $(1\,2)$ is its own inverse. The multisets of size two are $\bigl\{(),()\bigr\}$, $\bigl\{(),(1\,2)\bigr\}$, and $\bigl((1\,2),(1\,2)\bigr)$. Consider first $A=B=\{(),(1\,2)\}$. Using the definitions of $L_A$ and $R_B$ with $n_A=n_B=2$, and ordering the pairs $(i,j)$ as $(1,1),(1,2),(2,1),(2,2)$, we have
\begin{align*}
L_A &=\lambda_{()} \oplus\lambda_{()}
\oplus\lambda_{(1\,2)}
\oplus\lambda_{(1\,2)},
\\
R_B
&=\rho_{()}
\oplus\rho_{(1\,2)}
\oplus\rho_{()}
\oplus\rho_{(1\,2)} .
\end{align*}
The repeated blocks occur because the block $(i,j)$ of $L_A$ depends only on $i$, while the block $(i,j)$ of $R_B$ depends only on $j$. Writing these matrices explicitly in terms of their $2\times2$ blocks gives
\begingroup
\renewcommand{\arraystretch}{0.85}
\begin{align}
L_A&=
\left[
\begin{array}{cc|cc|cc|cc}
1 & 0 & 0 & 0 & 0 & 0 & 0 & 0\\
0 & 1 & 0 & 0 & 0 & 0 & 0 & 0\\ \hline
0 & 0 & 1 & 0 & 0 & 0 & 0 & 0\\
0 & 0 & 0 & 1 & 0 & 0 & 0 & 0\\ \hline
0 & 0 & 0 & 0 & 0 & 1 & 0 & 0\\
0 & 0 & 0 & 0 & 1 & 0 & 0 & 0\\ \hline
0 & 0 & 0 & 0 & 0 & 0 & 0 & 1\\
0 & 0 & 0 & 0 & 0 & 0 & 1 & 0
\end{array}
\right].
\end{align}
\endgroup

\begingroup
\renewcommand{\arraystretch}{0.85}
\begin{align}
R_B&=
\left[
\begin{array}{cc|cc|cc|cc}
1 & 0 & 0 & 0 & 0 & 0 & 0 & 0\\
0 & 1 & 0 & 0 & 0 & 0 & 0 & 0\\ \hline
0 & 0 & 0 & 1 & 0 & 0 & 0 & 0\\
0 & 0 & 1 & 0 & 0 & 0 & 0 & 0\\ \hline
0 & 0 & 0 & 0 & 1 & 0 & 0 & 0\\
0 & 0 & 0 & 0 & 0 & 1 & 0 & 0\\ \hline
0 & 0 & 0 & 0 & 0 & 0 & 0 & 1\\
0 & 0 & 0 & 0 & 0 & 0 & 1 & 0
\end{array}
\right].
\end{align}
\endgroup
Substituting these matrices into Eq.~\eqref{eq:lifted} and removing repeated rows gives
\begingroup
\begin{align}
H_X&=
\begin{pmatrix}
1 & 0 & 1 & 0 & 1 & 0 & 1 & 0\\
0 & 1 & 0 & 1 & 0 & 1 & 0 & 1\\
1 & 0 & 0 & 1 & 0 & 1 & 1 & 0\\
0 & 1 & 1 & 0 & 1 & 0 & 0 & 1
\end{pmatrix}.
\end{align}

\begingroup
\begin{align}
H_Z&=
\begin{pmatrix}
1 & 0 & 0 & 1 & 1 & 0 & 0 & 1\\
0 & 1 & 1 & 0 & 0 & 1 & 1 & 0\\
1 & 0 & 1 & 0 & 0 & 1 & 0 & 1\\
0 & 1 & 0 & 1 & 1 & 0 & 1 & 0
\end{pmatrix}.
\end{align}

The ranks satisfy
\begin{align}
\operatorname{rank}(H_X)
=
\operatorname{rank}(H_Z)
=
3.
\end{align}
and therefore $k=8-3-3=2$. An exhaustive search gives $d=2$, resulting in a $[[8,2,2]]$ quantum code.
We next consider the case where all elements in the multisets are repeated:
\begin{figure*}[ht!]
    \centering
    \includegraphics[width=1.0\linewidth]{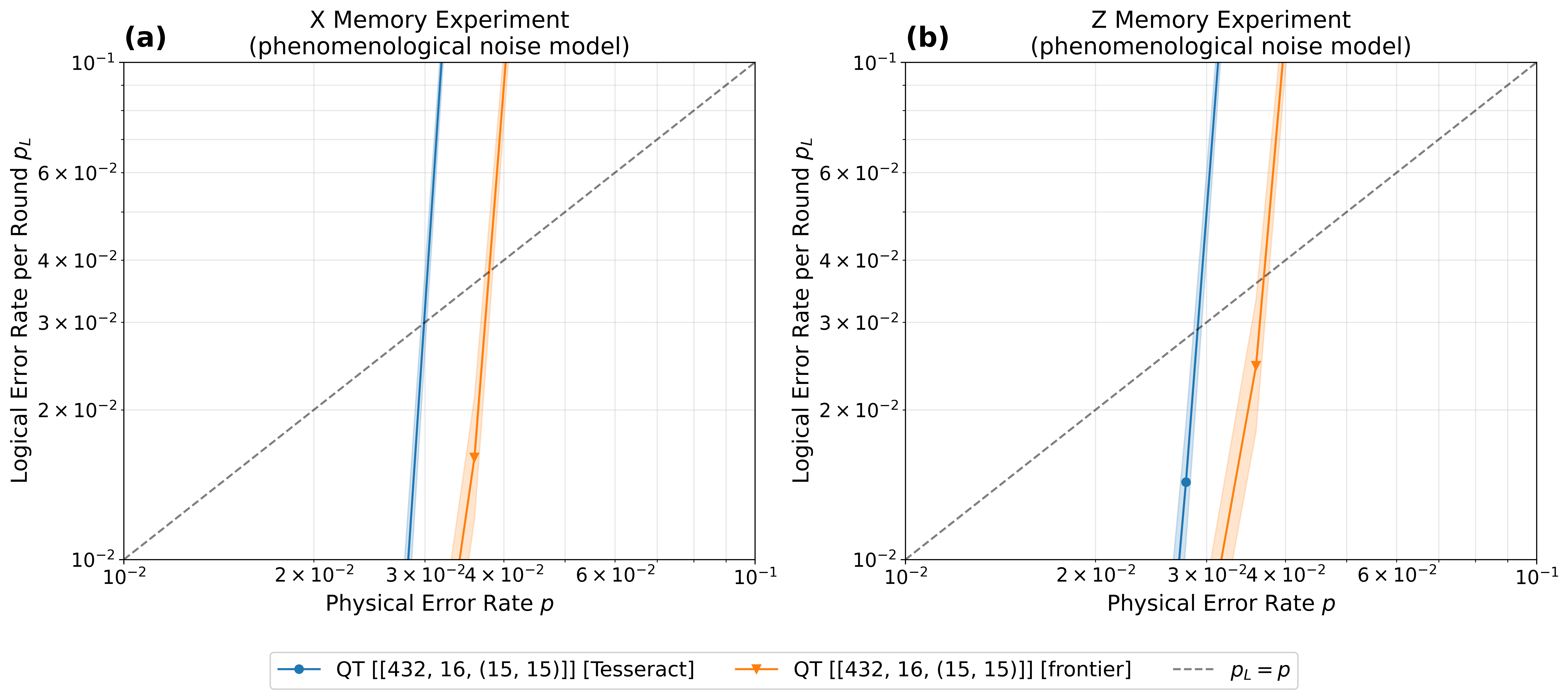}
    \caption{\textbf{Comparison of Tesseract and \texttt{Frontier} decoding for the
    $[[432,16,(\leq15,\leq15)]]$ QT code.}
    Per-round logical error rate $p_L^{(1)}(p)$ for (a) $X$-basis and (b) $Z$-basis memory experiments under phenomenological noise. Tesseract is shown in blue and \texttt{Frontier} in orange. For \texttt{Frontier}, we use $K=2^{12}=4096$, $\Delta=20$, and
    $\alpha=0.8$. The dashed line $p_L^{(1)}=p$ is included as a reference.}
    \label{fig:frontier-vs-tesseract}
\end{figure*}
\begin{align}
\mathcal{A}=\mathcal{B}=\bigl\{(1\,2),(1\,2)\bigr\}.
\end{align}
In this case, every block of $L_A$ and $R_B$ is the same nontrivial permutation matrix,
\begin{align}
\lambda_{(1\,2)}=\rho_{(1\,2)}
=
\begin{pmatrix}
0 & 1\\
1 & 0
\end{pmatrix},
\end{align}
and therefore
\begin{align}
L_A=R_B
&=
\lambda_{(1\,2)}
\oplus
\lambda_{(1\,2)}
\oplus
\lambda_{(1\,2)}
\oplus
\lambda_{(1\,2)} .
\end{align}
Explicitly,
\renewcommand{\arraystretch}{0.95}
\begin{align}
L_A=R_B
&=
\left[
\begin{array}{cc|cc|cc|cc}
0 & 1 & 0 & 0 & 0 & 0 & 0 & 0\\
1 & 0 & 0 & 0 & 0 & 0 & 0 & 0\\ \hline
0 & 0 & 0 & 1 & 0 & 0 & 0 & 0\\
0 & 0 & 1 & 0 & 0 & 0 & 0 & 0\\ \hline
0 & 0 & 0 & 0 & 0 & 1 & 0 & 0\\
0 & 0 & 0 & 0 & 1 & 0 & 0 & 0\\ \hline
0 & 0 & 0 & 0 & 0 & 0 & 0 & 1\\
0 & 0 & 0 & 0 & 0 & 0 & 1 & 0
\end{array}
\right].
\end{align}

Substituting these matrices into Eq.~\eqref{eq:lifted} gives
\begin{align}
H_X
&=
\begin{pmatrix}
1 & 0 & 1 & 0 & 1 & 0 & 1 & 0\\
0 & 1 & 0 & 1 & 0 & 1 & 0 & 1\\
1 & 0 & 1 & 0 & 1 & 0 & 1 & 0\\
0 & 1 & 0 & 1 & 0 & 1 & 0 & 1\\
1 & 0 & 1 & 0 & 1 & 0 & 1 & 0\\
0 & 1 & 0 & 1 & 0 & 1 & 0 & 1\\
1 & 0 & 1 & 0 & 1 & 0 & 1 & 0\\
0 & 1 & 0 & 1 & 0 & 1 & 0 & 1
\end{pmatrix}.
\end{align}

\begin{align}
H_Z
&=
\begin{pmatrix}
0 & 1 & 0 & 1 & 0 & 1 & 0 & 1\\
1 & 0 & 1 & 0 & 1 & 0 & 1 & 0\\
0 & 1 & 0 & 1 & 0 & 1 & 0 & 1\\
1 & 0 & 1 & 0 & 1 & 0 & 1 & 0\\
0 & 1 & 0 & 1 & 0 & 1 & 0 & 1\\
1 & 0 & 1 & 0 & 1 & 0 & 1 & 0\\
0 & 1 & 0 & 1 & 0 & 1 & 0 & 1\\
1 & 0 & 1 & 0 & 1 & 0 & 1 & 0
\end{pmatrix}.
\end{align}

After removing redundant rows, the resulting CSS code has parameters
\begin{align}
[[8,4,2]],
\end{align}
with distances
\begin{align}
d_X=2,\qquad d_Z=2.
\end{align}
Even when the multisets contain only repeated elements, the construction still produces a valid CSS code.

\section{Decoding with \texttt{Frontier}}
\label{app:decoding-with-frontier}
As a comparison with Tesseract, we also evaluated the \texttt{Frontier} decoder of Leverrier and Urbanke~\cite{leverrier2026approximating} on the $[[432,16,(\leq15,\leq15)]]$ quantum Tanner (QT) code. We use \texttt{Frontier} with a retained-frontier cap $K=4096=2^{12}$, score-gap parameter $\Delta=20$, and score weight $\alpha=0.8$. We refer the reader to Ref.~\cite{leverrier2026approximating} for details of the decoding algorithm.

Using the memory-experiment protocol and pseudo-threshold convention described in Sec.~\ref{sec:error-models}, we compared Tesseract with
\texttt{Frontier} on the $[[432,16,(\leq15,\leq15)]]$ QT code in both the $X$ and $Z$ bases. Since $d_X=d_Z=15$, both experiments use $r=15$ rounds. We swept $p$ over $10$ logarithmically spaced values between $10^{-2}$ and $10^{-1}$, using up to $10{,}000$ shots and a stopping criterion of $200$ logical failures per point. For \texttt{Frontier}, we used $K=4096=2^{12}$, $\Delta=20$, and $\alpha=0.8$.

Using the pseudo-threshold convention $p_L^{(1)}(p^\ast)=k p^\ast$ described in the main text, with $k=16$, we obtain pseudo-threshold estimates of $5.96\%$ and $4.47\%$ for \texttt{Frontier} and Tesseract, respectively, in the $X$-basis experiment, and $5.96\%$ and $4.46\%$, respectively, in the $Z$-basis experiment. The dashed line $p_L^{(1)}=p$ in Fig.~\ref{fig:frontier-vs-tesseract} is included only as a reference.

The \texttt{Frontier} parameters were not optimized for this QT code; the results therefore correspond to the fixed decoder configuration used for this comparison. in Fig.~\ref{fig:frontier-vs-tesseract}.

\definecolor{A}{RGB}{248, 248, 255}
\definecolor{B}{RGB}{245, 255, 250}
\definecolor{C}{RGB}{255, 250, 240}
\definecolor{D}{RGB}{240, 248, 255}
\definecolor{E}{RGB}{255, 245, 238}
\definecolor{F}{RGB}{240, 255, 240}
\definecolor{G}{RGB}{255, 250, 250}
\definecolor{H}{RGB}{248, 248, 247}
\definecolor{I}{RGB}{245, 245, 245}
\definecolor{J}{RGB}{250, 250, 250}
\definecolor{K}{RGB}{245, 248, 255}
\definecolor{L}{RGB}{255, 248, 240}
\definecolor{M}{RGB}{240, 252, 245}
\definecolor{N}{RGB}{250, 245, 255}
\definecolor{O}{RGB}{245, 250, 255}
\definecolor{P}{RGB}{255, 252, 245}
\definecolor{Q}{RGB}{248, 255, 248}
\definecolor{R}{RGB}{255, 248, 248}
\definecolor{S}{RGB}{245, 250, 245}
\definecolor{T}{RGB}{250, 248, 255}
\definecolor{U}{RGB}{252, 252, 250}
\definecolor{V}{RGB}{248, 250, 255}
\definecolor{W}{RGB}{255, 252, 250}
\definecolor{X}{RGB}{250, 252, 248}
\definecolor{Y}{RGB}{252, 250, 248}
\definecolor{Z}{RGB}{248, 252, 250}
\setlength{\tabcolsep}{0.005pt}
\renewcommand{\arraystretch}{2.3}
\begin{table*}
\begin{tabular}{|*{6}{>{\columncolor{white}[\tabcolsep]}c|}}
\hline
 \rowcolor{zebragray}Group & $[[n, k, (\leq d_X, \leq d_Z)]]$ & $H_0 = H_1$ & $\mathcal{A} \subseteq \mathcal{G}$ & $\mathcal{B}  \subseteq \mathcal{G}$ & $\tilde{w} =\overline{w}$  \\
\hline
 $C_3 \times S_3$ & $[[324, \,8,\,(\leq 17,\,\leq 14)]]$ & [6,3,3] & $\{s,s^2,t,t^2,rt^2,r
\}$ & $\{
rs,rs^2,st,s^2t^2,rst^2,rs^2t^2\}$ &9 \\
\hline
 \rowcolor{zebragray} $C_3 \times S_3$ & $[[576,\,22,\,(\leq 16,\,\leq 16)]]$ & [8,4,4] & $\{rst^2,rs^2t^2,rs^2t,rst,t^2,t,rs,rs^2
\}$ & $\{rt^2,s^2,s,st,s^2t^2,s^2t,st^2,r\}$ &16 \\
\hline
 $C_2 \times A_4$ & $[[432,\,16,\,(\leq 12,\,\leq12)]]$ & [6,3,3] & $\{s^2tu,st,t,s^2u,stu,rt
\}$ & $\{rs^2tu,rst,rs^2u,rstu,rs^2,rs\}$ &9 \\
\hline
 \rowcolor{zebragray} $S_4$ & $[[768,\,32,\,(\leq 16,\,\leq 16)]]$ & [8,4,4] & $\{s^2t,su,s^2,s,stu,s^2u,st,s^2tu
\}$ & $\{rs^2u,rs^2tu,rt,ru,rs^2,rst,rstu,r\}$ &16  \\
\hline
 $C_2 \times A_4$ & $[[768,\,42,\,(\leq16,\,\leq16)]]$ & [8,4,4] & $\{r,rs^2tu,rst,rs^2t,rsu,rs^2u,rstu,t
\}$ & $\{s,s^2,st,s^2tu,stu,s^2u,ru,rtu\}$ &16   \\
\hline
 \rowcolor{zebragray} $C_2 \times A_4$ & $[[768,\,20,\,(\leq 16,\,\leq16)]]$ & [8,4,4] & $\{s,s^2,s^2tu,st,rt,stu,s^2u,r
\}$ & $\{t,rstu,rs^2u,rs,rs^2,rst,rs^2tu,tu\}$ &16 \\
\hline
 $C_{12} \times C_2$ & $[[768,\,40,(\leq12,\,\leq12)]]$ & [8,4,4] & $\{su,rt^2,rtu,rst^2,rstu,t^2u,tu,u
\}$ & $\{st^2u,stu,rsu,rs,t^2,t,ru,r\}$ &16 \\
\hline
 \rowcolor{zebragray} $(C_6 \times C_2) \ltimes C_2$ & $[[768,\,38,\,(\leq 16,\,\leq16)]]$ & [8,4,4] & $\{ru,u^2,u,st,ru^2,tu,tu^2,rt
\}$ & $\{rstu^2,rsu^2,rsu,rstu,rst,rs,stu^2,stu\}$ &16  \\
\hline
 $C_2 \times (C_3 \ltimes C_4)$ & $[[768,\,34,\,(\leq 12,\,\leq 12)]]$ & [8,4,4] & $\{ru^2,rtu^2,tu,tu^2,stu^2,stu,su,su^2
\}$ & $\{u^2,u,rsu^2,rstu^2,t,s,rst,rs\}$ &16 \\
\hline
 \rowcolor{zebragray} $C_4 \times S_3$ &$[[768,\,48,\,(\leq 16,\,\leq16)]]$ & [8,4,4] & $\{rstu^2,rsu^2,rst,rs,s,st,rstu,rsu
\}$ & $\{rt,tu^2,tu,su^2,stu,ru^2,t,rtu^2\}$ &16  \\
\hline 
 $C_3 \ltimes Q_8$ &$[[768,\,28,\,(\leq16,\,\leq16)]]$ & [8,4,4] & $\{rstu^2,rsu^2,rst,rs,s,st,rstu,rsu
\}$ & $\{rt,r,tu^2,tu,su^2,stu,ru^2,rtu^2\}$ &16  \\
\hline 
 \rowcolor{zebragray} $SL(2,3)$ &$[[768,\,32,\,(\leq 16,\,\leq 16)]]$ & [8,4,4] & $\{r^2stu,rs,r^2,r,r^2su,rt,s,su
\}$ & $\{rsu,r^2st,rstu,r^2t,ru,r^2u,rtu,r^2s\}$ &16 \\
\hline 
 $C_3 \ltimes C_8$ &$[[768,\,32,\,(\leq 16,\,\leq16)]]$ &[8,4,4] & $\{ru,rstu,ru^2,rstu^2,s,st,tu^2,tu
\}$ & $\{rs,rt,su^2,stu,rsu^2,rtu^2,u,u^2\}$ &16  \\
\hline 
 \rowcolor{zebragray}$C_3 \ltimes C_8$ &$[[768,\,24,\,(\leq16,\,\leq16)]]$ & [8,4,4] & $\{rst,r,stu,su^2,rstu,ru,stu^2,su\}$ & $\{rs,rt,tu^2,tu,s,st,rtu,rsu\}$ &16  \\
\hline
 $C_{5} \times C_5$ & $[[800,\,24,\,(\leq20,\,\leq20)]]$ & [8,4,4] & $\{rs^2,r^4s^3,r^3s,r^2s^4,r^4s^2,rs^3,r,r^4
\}$ & $\{r^3,r^2,r^4s,rs^4,s,s^4,r^2s,r^3s^4\}$ &16 \\
\hline
\rowcolor{zebragray} $C_9 \times C_3$ &$[[864,\,32,\,(\leq12,\,\leq12)]]$ & [8,4,4] & $\{r^2t^2,r,t,t^2,r^2,rt^2,rs^2,r^2st^2\}$ & $\{r^2s^2,rst^2,r^2st,rs^2t,s^2t,st^2,r^2s,rs^2t^2\}$ &16 \\
\hline 
 $(C_3 \times C_3) \ltimes C_3$ &$[[864,\,20,\,(\leq24,\,\leq24)]]$ & [8,4,4]&  $\{r^2t,rt^2,t^2,t,rt,r^2t^2,r,r^2\}$ & $\{r^2st,rs^2t,rst^2,r^2s^2t^2,r^2s^2t,rs,st^2,s^2t\}$ &16 \\
\hline 
 \rowcolor{zebragray} $(C_3 \times C_3) \ltimes C_3$ &$[[864,\,16,\,(\leq24,\,\leq24)]]$ & [8,4,4]& $\{r^2s,rs^2t^2,r^2st,rs^2t,s^2,s,st,s^2t^2\}$ & $\{rs,r^2s^2t,r^2s^2t^2,rst^2,rt^2,r^2t,r^2s^2,rst\}$ &16 \\
\hline 
 $C_9 \ltimes C_3$ &$[[864,\,28,\,(\leq18,\,\leq18)]]$ & [8,4,4]& $\{r,r^2t^2,r^2,rt^2,r^2t,rt,r^2s^2,rs\}$ & $\{s^2,s,st^2,s^2t,s^2t^2,st,r^2s,rs^2t\}$ &16 \\
\hline 
 \rowcolor{zebragray}$C_3 \times C_3 \times C_3$ &$[[864,\,24,\,(\leq 24,\,\leq 24)]]$ & [8,4,4]& $\{s^2t,st^2,rt,r^2t^2,s,s^2,r,r^2\}$ & $\{st,s^2t^2,t^2,t,rs,r^2s^2,r^2t,rt^2\}$ &16 \\
\hline
$C_7 \ltimes C_4$ &$[[896,\,40,\,(\leq16,\,\leq16)]]$ & [8,4,4]&  $\{rst^3,rt^3,r,rs,rst^5,rt^5,rt^4,rst^4\}$ & $\{st,st^6,t,t^6,st^2,st^5,t^2,t^5\}$ &16 \\
\hline
 \rowcolor{zebragray}$C_{14} \times C_{2}$ &$[[896,\,16,\,(\leq,16\,\leq16)]]$ & [8,4,4]& $\{st^2,st^5,t^6,t,rst^4,rst^3,rt,rt^6\}$ & $\{r,rs,t^5,t^2,rst^2,rst^5,rt^2,rt^5\}$ &16  \\
\hline
$C_3 \times D_{10}$ &$[[960,\,24,\,(\leq 16,\,\leq 16)]]$ & [8,4,4] & $\{rs^2t,rst,s^2t^3,st^2,rst^4,rs^2t^4,rs^2,rs\}$ & $\{t^3,t^2,st,s^2t^4,t^4,t,s,s^2\}$ &16\\
\hline
 \rowcolor{zebragray} $C_3 \times D_{10}$ &$[[960,\,32,\,(\leq16,\leq16)]]$ & [8,4,4] & $\{r,s^2t,st^4,s,s^2,s^2t^3,st^2,rt^4\}$ & $\{t^2,t^3,rst^4,rs^2t^4,t^4,t,rs^2t^3,rst^3\}$ &16  \\
\hline
\end{tabular}
\caption{\textbf{Quantum Tanner Codes via the LRCC construction.} For each instance we report the underlying group $\mathcal{G}$, the code parameters $[[n, k, (\leq d_X, \leq d_Z)]]$, the local codes $H_0$ and $H_1$, the generating multisets $\mathcal{A}$ and $\mathcal{B}$ and the maximum stabilizer row weight $\tilde{w}=\bar{w}$. Distances are upper bounds estimated using 50 million random-ISD trials of \texttt{sQetch}~\cite{bhardwaj2026high} on a single NVIDIA RTX~2080~Ti, with $1{,}000$ trials of
\texttt{DistRandCSS}~\cite{pryadko2023qdistrnd} used as a cross-check.}
\end{table*}

\definecolor{A}{RGB}{248, 248, 255}
\definecolor{B}{RGB}{245, 255, 250}
\definecolor{C}{RGB}{255, 250, 240}
\definecolor{D}{RGB}{240, 248, 255}
\definecolor{E}{RGB}{255, 245, 238}
\definecolor{F}{RGB}{240, 255, 240}
\definecolor{G}{RGB}{255, 250, 250}
\definecolor{H}{RGB}{248, 248, 247}
\definecolor{I}{RGB}{245, 245, 245}
\definecolor{J}{RGB}{250, 250, 250}
\definecolor{K}{RGB}{245, 248, 255}
\definecolor{L}{RGB}{255, 248, 240}
\definecolor{M}{RGB}{240, 252, 245}
\definecolor{N}{RGB}{250, 245, 255}
\definecolor{O}{RGB}{245, 250, 255}
\definecolor{P}{RGB}{255, 252, 245}
\definecolor{Q}{RGB}{248, 255, 248}
\definecolor{R}{RGB}{255, 248, 248}
\definecolor{S}{RGB}{245, 250, 245}
\definecolor{T}{RGB}{250, 248, 255}
\definecolor{U}{RGB}{252, 252, 250}
\definecolor{V}{RGB}{248, 250, 255}
\definecolor{W}{RGB}{255, 252, 250}
\definecolor{X}{RGB}{250, 252, 248}
\definecolor{Y}{RGB}{252, 250, 248}
\definecolor{Z}{RGB}{248, 252, 250}
\setlength{\tabcolsep}{2.5pt}
\renewcommand{\arraystretch}{2.4}
\begin{table*}
\begin{tabular}{|*{11}{>{\columncolor{white}[\tabcolsep]}c|}}
\hline
 \rowcolor{zebragray}Group & $[[n, k, (\leq d_X, \leq d_Z)]]$ & $H_0$ & $H_1$ & $\mathcal{A} \subseteq \mathcal{G}$ & $\mathcal{B}  \subseteq \mathcal{G}$ & $\tilde{w} =\overline{w}$ & $\frac{k}{n}$ \\
\hline
 $(C_4 \times C_2) \ltimes C_2$ & $[[384, 24, (\leq12,\leq12)]]$ & [6,3,3] & [8,4,4] & \{$r,u,stu,st,rs,rsu$\}& \{$s,rtu,rt,rst,rstu,t,tu,su$\} & 12 & 0.06 \\
\hline
 \rowcolor{zebragray} $(C_4 \times C_2) \ltimes C_2$ & $[[384, 8, (\leq12,\leq12)]]$ & [6,3,3] & [8,4,4] & \{$st,stu,rt,rtu,u,su$\} & \{$rs,rsu,r,ru,rstu,tu,t,rst$\}& 12 & 0.02 \\
\hline
 $C_2 \times D_8$ & $[[384, 44, (\leq10,\leq10)]]$ & [6,3,3] & [8,4,4] & \{$rs,rsu,s,rstu,rst,su$\} & \{$ru,t,rtu,r,tu,stu,rt,u$\} & 12 & 0.11 \\
\hline
 \rowcolor{zebragray} $C_2 \times D_8$ & $[[384, 38, (\leq12,\leq12)]]$ & [6,3,3] & [8,4,4] & $
\{s,rs,rsu,ru,su,u$\} & \{$rt,tu,t,stu,rstu,rst,st,rtu$\}& 12 & 0.10 \\
\hline
 $D_{16}$ & $[[384, 32, (\leq12,\leq12)]]$ & [6,3,3] & [8,4,4] & \{$s,stu,t,tu,st,su$\}& \{$rstu,rsu,rt,ru,rs,rst,r,rtu$\}& 12 & 0.08 \\
\hline
 \rowcolor{zebragray} $C_{4} \times C_2 \times C_2$ & $[[384, 16, (\leq12,\leq12)]]$ & [6,3,3] & [8,4,4] & \{$s,rtu,rt,rst,rstu,u$\}& \{$rs,rsu,st,ru,r,stu,t,tu$\}& 12 & 0.04 \\
\hline
 $C_8 \ltimes C_2$ & $[[384, 32, (\leq12,\leq12)]]$ & [6,3,3] & [8,4,4] & \{$stu,st,u,tu,t,su$\} & \{$rsu,rstu,rst,rs,r,rtu,ru,rt$\}& 12 & 0.08 \\
\hline
 \rowcolor{zebragray} $C_8 \ltimes C_2$ & $[[384, 8, (\leq12,\leq12)]]$ & [6,3,3] & [8,4,4] & \{$rst,rs,rstu,rsu,su,s$\} & \{$rt,ru,t,tu,st,stu,r,rtu$\}& 12 & 0.02 \\
\hline
$C_8 \times C_2$ & $[[384, 18, (\leq12,\leq12)]]$ & [6,3,3] & [8,4,4] & \{$su,rtu,r,tu,t,s
$\} & \{$rstu,rs,rt,ru,stu,st,rsu,rst$\}& 12 & 0.05 \\
\hline
 \rowcolor{zebragray} $C_4 \ltimes C_4$ & $[[384, 16, (\leq12,\leq12)]]$ & [6,3,3] & [8,4,4] & \{$rsu,rs,rst,rstu,tu,u$ \}& \{$su,stu,st,s,r,ru,rt,rtu$\}& 12 & 0.04 \\
\hline
 $C_4 \ltimes C_4$ & $[[384, 48, (\leq12,\leq12)]]$ & [6,3,3] & [8,4,4] & \{$su,stu,t,s,st,u
$\} & \{$rstu,rst,rtu,rt,rsu,rs,ru,r$\} & 12 & 0.13 \\
\hline
 \rowcolor{zebragray} $(C_4 \times C_2) \ltimes C_2$ & $[[384, 40, (\leq10,\leq10)]]$ & [6,3,3] & [8,4,4] & \{$st,rst,rsu,s,rstu,rs
$ \} & \{$rt,rtu,su,t,r,ru,u,tu$\} & 12 & 0.10\\
\hline
$C_4 \times C_4$ & $[[392, 12, (\leq13,\leq12)]]$ & [7,4,3] & [7,3,4] & $\{stu,st,rt,r,tu,ru,rtu\}
$ & $\{s,su,t,rs,rstu,rsu,rst\}$ & 16 & 0.03 \\
\hline
 \rowcolor{zebragray} $C_4 \ltimes C_4$ & $[[392, 15, (\leq14,\leq10)]]$ & [7,4,3] & [7,3,4] & \{$t,s,st,ru,r,rt,rtu$\}& \{$su,stu,rstu,rst,tu,rs,rsu$\} & 16 & 0.04 \\
 $C_2 \times D_8$ & $[[392, 36, (\leq12,\leq12)]]$ & [7,4,3] & [7,3,4] & \{$r,s,ru,rsu,rs,u,su$\}& \{$rstu,rst,t,st,rtu,stu,rt$\}& 16 & 0.09 \\
\hline
 \rowcolor{zebragray} $C_2 \times Q_8$ & $[[392, 8, (\leq19,\leq20)]]$ & [7,4,3] & [7,3,4] & \{$rstu,rst,s,su,st,stu,u$\} & \{$rsu,rs,r,ru,tu,rt,rtu$\} & 16 & 0.02 \\
\hline
 $C_2 \times Q_8$ & $[[392, 15, (\leq20,\leq10)]]$ & [7,4,3] & [7,3,4] & \{$stu,st,rt,rtu,t,su,s$\} & \{$rstu,rst,rs,rsu,u,r,ru$\}& 16 & 0.04 \\
\hline
 \rowcolor{zebragray} $(C_4 \times C_2) \ltimes C_2$ & $[[392, 33, (\leq12,\leq12)]]$ & [7,4,3] & [7,3,4] & \{$u,rtu,rt,rs,rsu,stu,st$\}& \{$t,tu,s,rst,r,rstu,ru$\} & 16 & 0.08 \\
\hline
 $(C_4 \times C_2) \ltimes C_2$ & $[[392, 36, (\leq12,\leq10)]]$ & [7,4,3] & [7,3,4] & \{$stu,st,u,tu,t,rsu,rs$\} & \{$rstu,rst,r,ru,rt,rtu,su$\} & 16 & 0.09 \\
\hline
 \rowcolor{zebragray} $(C_3 \times C_3) \ltimes C_2$ & $[[432, 4, (\leq24,\leq24)]]$ & [6,3,3] & [8,4,4] & \{$t,t^2,s^2,s,st^2,s^2t$\}& \{$rs,rs^2,r,rt,rst^2,rst,rt^2,rs^2t^2$\}& 12 & 0.01 \\
\hline
 $C_6 \times C_3$ & $[[432, 12, (\leq20,\leq20)]]$ & [6,3,3] & [8,4,4] & \{$s^2t,st^2,rs^2t,rst^2,t^2,t$\}& \{$rs^2,rs,rs^2t^2,rst,rt^2,rt,s^2t^2,st$\} & 12 & 0.03 \\
\hline
\end{tabular}
\label{appendix:lrcc1}
\caption{\textbf{Quantum Tanner Codes via the LRCC construction.} For each instance we report the underlying group $\mathcal{G}$, the code parameters $[[n, k, (\leq d_X, \leq d_Z)]]$, the local codes $H_0$ and $H_1$, the generating multisets $\mathcal{A}$ and $\mathcal{B}$, the maximum stabilizer row weight $\tilde{w}=\bar{w}$, and the rate $k/n$. Distances are upper bounds estimated using 50 million random-ISD trials of \texttt{sQetch}~\cite{bhardwaj2026high} on a single NVIDIA RTX~2080~Ti, with $1{,}000$ trials of
\texttt{DistRandCSS}~\cite{pryadko2023qdistrnd} used as a cross-check.}
\end{table*}

\definecolor{A}{RGB}{248, 248, 255}
\definecolor{B}{RGB}{245, 255, 250}
\definecolor{C}{RGB}{255, 250, 240}
\definecolor{D}{RGB}{240, 248, 255}
\definecolor{E}{RGB}{255, 245, 238}
\definecolor{F}{RGB}{240, 255, 240}
\definecolor{G}{RGB}{255, 250, 250}
\definecolor{H}{RGB}{248, 248, 247}
\definecolor{I}{RGB}{245, 245, 245}
\definecolor{J}{RGB}{250, 250, 250}
\definecolor{K}{RGB}{245, 248, 255}
\definecolor{L}{RGB}{255, 248, 240}
\definecolor{M}{RGB}{240, 252, 245}
\definecolor{N}{RGB}{250, 245, 255}
\definecolor{O}{RGB}{245, 250, 255}
\definecolor{P}{RGB}{255, 252, 245}
\definecolor{Q}{RGB}{248, 255, 248}
\definecolor{R}{RGB}{255, 248, 248}
\definecolor{S}{RGB}{245, 250, 245}
\definecolor{T}{RGB}{250, 248, 255}
\definecolor{U}{RGB}{252, 252, 250}
\definecolor{V}{RGB}{248, 250, 255}
\definecolor{W}{RGB}{255, 252, 250}
\definecolor{X}{RGB}{250, 252, 248}
\definecolor{Y}{RGB}{252, 250, 248}
\definecolor{Z}{RGB}{248, 252, 250}
\setlength{\tabcolsep}{0.5pt}
\renewcommand{\arraystretch}{2.6}
\editcolor{\begin{table*}
\begin{tabular}{|c|c|c|c|c|c|c|c|c|c|}
\hline
 \rowcolor{zebragray}Group & $[[n, k, (\leq d_X, \leq d_Z)]]$ & $H_0$ & $H_1$ & $\mathcal{A} \subseteq \mathcal{G}$ & $\mathcal{B}  \subseteq \mathcal{G}$ & $\tilde{w} =\overline{w}$ & $\frac{k}{n}$ \\
\hline
 $C_3 \ltimes C_8$ &  $[[576,16,(\leq12, \leq12)]]$ & [6,3,3] & [8,4,4] & \{$su^2,stu,stu^2,su,tu^2,tu$\} & \{$rst,r,rtu^2,rsu^2,u,u^2,rstu^2,ru^2$\}& 12 & 0.03\\
\hline
 \rowcolor{zebragray} $SL(2,3)$ &  $[[576,12,(\leq12, \leq12)]]$ & [6,3,3] & [8,4,4] & \{$rs,r^2stu,rst,r^2tu,rt,r^2su$\}& \{$r^2st,rsu,r^2s,rtu,t,tu,r^2t,rstu$\}& 12 & 0.02\\
\hline
 $C_3 \ltimes Q_8$ &  $[[576,8,(\leq12, \leq12)]]$ & [6,3,3] & [8,4,4] & \{$stu,su^2,rstu^2,rsu^2,rsu,rstu$\}& \{$ru,rtu,rtu^2,ru^2,tu,tu^2,rt,r$\}& 12 & 0.01 \\
\hline
 \rowcolor{zebragray} $C_4 \times S_3$ &  $[[576,28,(\leq12, \leq12)]]$ & [6,3,3] & [8,4,4] & \{$rtu^2,ru^2,rt,tu^2,tu,ru
$ \}& \{$rsu^2,rstu^2,su,stu^2,rstu,rsu,rst,rs$ \}& 12 & 0.05\\
\hline
 $D_{24}$ &  $[[576,26,(\leq12, \leq12)]]$ & [6,3,3] & [8,4,4] & \{$rtu^2,rst,tu^2,tu,rstu^2,t
$\} & \{$u^2,u,su^2,stu,s,st,su,stu^2$\}& 12 & 0.05\\
\hline
 \rowcolor{zebragray} $C_2 \times (C_3 \ltimes C_4)$ &  $[[576,16,(\leq 12,\leq12)]]$ & [6,3,3] & [8,4,4] & \{$tu^2,tu,stu^2,stu,t,s
$\} & \{$rt,r,u^2,u,rsu^2,rstu^2,rsu,rstu$\}& 12 & 0.03\\
\hline
$C_2 \times (C_3 \ltimes C_4)$ &  $[[576,40,(\leq12, \leq9)]]$ & [6,3,3] & [8,4,4] & \{$stu,stu^2,t,u^2,u,st$ & \{$rstu^2,rsu^2,su,su^2,rtu^2,ru^2,ru,rtu$\}& 12 & 0.07\\
\hline
 \rowcolor{zebragray} $(C_6 \times C_2) \ltimes C_2$ &  $[[576,8,(\leq12, \leq12)]]$ & [6,3,3] & [8,4,4] & \{$tu^2,tu,u^2,u,stu^2,stu$\} & \{$rt,ru^2,rtu,ru,rsu,rstu,rstu^2,rsu^2$\}& 12 & 0.01\\
\hline
 $(C_6 \times C_2) \ltimes C_2$ &  $[[576,28,(\leq12, \leq12)]]$ & [6,3,3] & [8,4,4] & \{$ru,u^2,u,st,ru^2,rt$\}& \{$tu,tu^2,rstu^2,rsu^2,rsu,rstu,rst,rs$\} & 12 & 0.05\\
\hline
 \rowcolor{zebragray} $C_{12} \times C_2$ &  $[[576,8,(\leq12, \leq12)]]$ & [6,3,3] & [8,4,4] & \{$su,st^2,st,tu,t^2u,s$\} & \{$
stu,st^2u,t^2,t,r,ru,rsu,rs$\} & 12 & 0.01\\
\hline
 $C_{12} \times C_2$ &  $[[576,30,(\leq12, \leq12)]]$ & [6,3,3] & [8,4,4] & \{$rst,rst^2u,u,rs,rsu,su$\} & $rtu,rt^2,stu,st^2u,rst^2,rstu,ru,r$ \}& 12 & 0.05\\
\hline
 \rowcolor{zebragray} $C_3 \times D_8$ &  $[[576,16,(\leq12, \leq12)]]$ & [6,3,3] & [8,4,4] & \{$rtu,rt^2u,su,rs,rsu,s$\} & \{$rst,rst^2u,t,t^2,st,st^2,rst^2,rstu$\} & 12 & 0.03 \\
\hline
 $C_3 \times D_8$ &  $[[576,32,(\leq12, \leq12)]]$ & [6,3,3] & [8,4,4] & \{$r,rsu,rs,su,ru,s$\}& \{$rt^2u,rtu,rst^2,rstu,st^2,st,st^2u,stu$\}& 12 & 0.06\\
\hline
 \rowcolor{zebragray} $S_4$ &  $[[576,23,(\leq12, \leq12)]]$ & [6,3,3] & [8,4,4] & \{$rs^2tu,rs^2u,r,rs,rst,rstu$\}& \{ $s,s^2,su,s^2t,u,tu,st,s^2tu$\}& 12 & 0.04\\
\hline
 $C_2 \times A_4$ &  $[[576, 6, (\leq12, \leq12)]]$ & [6,3,3] & [8,4,4] & \{$s^2t,su,s^2u,stu,rtu,ru$\} & \{$rsu,rs^2t,rs^2u,rstu,r,rs^2tu,rst,t$\}& 12 & 0.01\\
\hline
 \rowcolor{zebragray} $C_2 \times A_4$ &  $[[576, 28, (\leq12, \leq12)]]$ & [6,3,3] & [8,4,4] & \{$tu,rs,rs^2,rsu,rs^2t,t$\}& \{$su,s^2t,s^2,s,stu,s^2u,s^2tu,st$\}& 12 & 0.05\\
\hline
 $C_2 \times C_2 \times S_3$ &  $[[576, 20, (\leq12, \leq12)]]$ & [6,3,3] & [8,4,4] & \{$stu^2,stu,u^2,u,st,ru^2
$\} & \{$rstu^2,rst,su,su^2,rsu^2,rt,rstu,rs$\}& 12 & 0.03\\
\hline
 \rowcolor{zebragray} $C_2 \times C_2 \times S_3$ &  $[[576, 16, (\leq12, \leq12)]]$ & [6,3,3] & [8,4,4] & \{$u^2,u,rtu^2,t,su^2,su$\}& \{$r,rstu,rst,rs,tu,tu^2,st,rsu^2$ \}& 12 & 0.03\\
\hline
 $C_6 \times C_2 \times C_2$ &  $[[576, 26, (\leq12, \leq12)]]$ & [6,3,3] & [8,4,4] & \{$rsu^2,rsu,u,u^2,rs,rt$\} & \{$st,stu^2,stu,rstu,rstu^2,su,su^2,rst$\}& 12 & 0.05\\
\hline
 \rowcolor{zebragray} $(C_3 \times C_3) \ltimes C_3$ &  $[[648, 8, (\leq20, \leq20]]$ & [6,3,3] & [8,4,4] & \{$rs^2t^2,r^2s,rt^2,r^2t,t,t^2$\}& \{$rst^2,r^2s^2t^2,s^2t,st^2,s,s^2,s^2t^2,st$\}& 12 & 0.01\\
\hline
\end{tabular}
\caption{\textbf{Quantum Tanner Codes via the LRCC construction.} For each instance we report the underlying group $\mathcal{G}$, the code parameters $[[n, k, (\leq d_X, \leq d_Z)]]$, the local codes $H_0$ and $H_1$, the generating multisets $\mathcal{A}$ and $\mathcal{B}$, the maximum stabilizer row weight $\tilde{w}=\bar{w}$, and the rate $k/n$. Distances are upper bounds estimated using 50 million random-ISD trials of \texttt{sQetch}~\cite{bhardwaj2026high} on a single NVIDIA RTX~2080~Ti, with $1{,}000$ trials of
\texttt{DistRandCSS}~\cite{pryadko2023qdistrnd} used as a cross-check.}
\end{table*}}

\definecolor{A}{RGB}{248, 248, 255}
\definecolor{B}{RGB}{245, 255, 250}
\definecolor{C}{RGB}{255, 250, 240}
\definecolor{D}{RGB}{240, 248, 255}
\definecolor{E}{RGB}{255, 245, 238}
\definecolor{F}{RGB}{240, 255, 240}
\definecolor{G}{RGB}{255, 250, 250}
\definecolor{H}{RGB}{248, 248, 247}
\definecolor{I}{RGB}{245, 245, 245}
\definecolor{J}{RGB}{250, 250, 250}
\definecolor{K}{RGB}{245, 248, 255}
\definecolor{L}{RGB}{255, 248, 240}
\definecolor{M}{RGB}{240, 252, 245}
\definecolor{N}{RGB}{250, 245, 255}
\definecolor{O}{RGB}{245, 250, 255}
\definecolor{P}{RGB}{255, 252, 245}
\definecolor{Q}{RGB}{248, 255, 248}
\definecolor{R}{RGB}{255, 248, 248}
\definecolor{S}{RGB}{245, 250, 245}
\definecolor{T}{RGB}{250, 248, 255}
\definecolor{U}{RGB}{252, 252, 250}
\definecolor{V}{RGB}{248, 250, 255}
\definecolor{W}{RGB}{255, 252, 250}
\definecolor{X}{RGB}{250, 252, 248}
\definecolor{Y}{RGB}{252, 250, 248}
\definecolor{Z}{RGB}{248, 252, 250}
\setlength{\tabcolsep}{0.5pt}
\renewcommand{\arraystretch}{2.6}
\editcolor{\begin{table*}
\begin{tabular}{|c|c|c|c|c|c|c|c|c|c|}
\hline
 \rowcolor{zebragray}Group & $[[n, k, (\leq d_X, \leq d_Z)]]$ & $H_0$ & $H_1$ & $\mathcal{A} \subseteq \mathcal{G}$ & $\mathcal{B}  \subseteq \mathcal{G}$ & $\tilde{w} =\overline{w}$ & $\frac{k}{n}$ \\
\hline
 $C_7 \ltimes C_4$ &  $[[672,18,(\leq24, \leq24)]]$ & [6,3,3] & [8,4,4] & \{$st^2,st^5,st^4,st^3,t,t^6\}
$ & \{$rt^5,rst^5,rst,rt,st,st^6,rt^6,rst^6$\}& 12 & 0.03\\
\hline
 \rowcolor{zebragray} $D_{28}$ &  $[[672,12,(\leq22, \leq22)]]$ & [6,3,3] & [8,4,4] &\{$rst,st^4,st^3,t^3,t^4,rst^5$\} & \{$t^2,t^5,st^2,st^5,t^6,t,st,st^6$ \}& 12 & 0.02 \\
\hline
$C_{14} \times C_2$ &  $[[672,20,(\leq21, \leq21)]]$ & [6,3,3] & [8,4,4] & \{$s,st^3,st^4,t^3,t^4,rs$\}& \{$st^2,st^5,rst^6,rst,t^2,t^5,t^6,t$\}& 12 & 0.03 \\
\hline
 \rowcolor{zebragray} $C_5 \times S_3$ &  $[[720,32,(\leq15, \leq15)]]$ & [6,3,3] & [8,4,4] & \{$s^2t,s^3t^2,rt,rs^2t,rs^3t,r$\} & \{$st^2,s^4t,rs^4,rs,t,t^2,rst,rs^4t$\}& 12 & 0.04 \\
\hline
 $D_{30}$ &  $[[720,16,(\leq20, \leq20)]]$ & [6,3,3] & [8,4,4] & \{$st^2,s^2t^3,rst^2,s^2t,st^4,rs^2t^3$\} & \{$s^2,s,s^2t^4,st,s^2t^2,st^3,t^2,t^3$\}&  12 & 0.02 \\
\hline
 \rowcolor{zebragray} $C_2 \times A_4$ &  $[[588,33,(\leq19,\leq18)]]$ & [7,4,3] & [7,3,4] & \{$rs^2tu,rst,rt,rs^2,rs,rs^2u,rstu$\} & \{$s^2u,stu,s,s^2,st,s^2tu,t$\}&  16 & 0.06 \\
\hline
 $C_2 \times C_2 \times S_3$ &  $[[588,29,(\leq18,\leq12)]]$ & [7,4,3] & [7,3,4] & \{$su,su^2,stu,stu^2,t,st,r
$\} & \{$tu,tu^2,u^2,u,s,rsu,rtu$\}&  16 & 0.05  \\
\hline
 \rowcolor{zebragray}$C_2 \times C_2 \times S_3$ &  $[[588,39,(\leq12,\leq9)]]$ & [7,4,3] & [7,3,4] & \{$st,rstu,rtu,su^2,su,rstu^2,rst$\} & \{$tu,tu^2,u^2,u,rsu^2,r,s$\}&  16 & 0.07 \\
\hline
 $C_6 \times C_2 \times C_2$ &  $[[588,40,(\leq9,\leq18)]]$ & [7,4,3] & [7,3,4] & \{$rsu^2,rsu,u,u^2,rs,rt,st$\} & \{$stu^2,stu,rstu,rstu^2,su,su^2,rst$\}&  16 & 0.07 \\
\hline
 \rowcolor{zebragray} $C_6 \times C_2 \times C_2$ &  $[[588,38,(\leq14,\leq10)]]$ & [7,4,3] & [7,3,4] & \{$su,su^2,rtu,rtu^2,rstu,rstu^2,rt$\} & \{$tu^2,tu,rst,u,u^2,stu^2,stu$\}&  16 & 0.06 \\
\hline
$C_6 \times C_2 \times C_2$ &  $[[588,21,(\leq21,\leq12)]]$ & [7,4,3] & [7,3,4] & \{$rt,su,su^2,rs,rsu^2,rsu,s$\} & \{$u^2,u,rstu,rstu^2,stu^2,stu,st$\}&  16 & 0.04 \\
\hline
 \rowcolor{zebragray} $C_8 \times C_4$ &  $[[784,16,(\leq19,\leq25)]]$ & [7,4,3] & [7,3,4] & \{$stuv,st,tuv,tu,rs,rstuv,v$\} & \{$rt,rv,rsv,rstu,rtu,ruv,uv$\}&  16 & 0.02 \\
\hline
 $(C_8 \times C_2) \ltimes C_2$ &  $[[784,37,(\leq14,\leq12)]]$ & [7,4,3] & [7,3,4] & \{$rtu,rtv,t,tu,tuv,rstu,rsv$\} & \{$stv,v,rstuv,rs,rt,rtuv,sv$\}&  16 & 0.05 \\
\hline
 \rowcolor{zebragray} $(C_2 \times C_2 \times C_2) \ltimes C_4$ &  $[[784,40,(\leq14,\leq12)]]$ & [7,4,3] & [7,3,4] & \{$rst,rsuv,v,rs,rstu,rstv,rsu$\} & \{$rt,rtuv,ruv,rv,uv,t,stv$\}&  16 & 0.05 \\
\hline
 $(C_8 \ltimes C_2) \ltimes C_2$ &  $[[784,24,(\leq16,\leq13)]]$ & [7,4,3] & [7,3,4] & \{$rst,rsu,v,rs,rstuv,rstv,rsuv$\} & \{$rt,rtu,ruv,r,uv,u,t$\}&  16 &  0.03 \\
\hline
 \rowcolor{zebragray} $(C_8 \ltimes C_2) \ltimes C_2$ &  $[[784,58,(\leq12,\leq10)]]$ & [7,4,3] & [7,3,4] & \{$rtu,rt,rv,ru,s,ruv,r$\} & \{$suv,tv,rs,rstuv,stu,t,v$\}&  16 & 0.07 \\
\hline
\end{tabular}
\caption{\textbf{Quantum Tanner Codes via the LRCC construction.} For each instance we report the underlying group $\mathcal{G}$, the code parameters $[[n, k, (\leq d_X, \leq d_Z)]]$, the local codes $H_0$ and $H_1$, the generating multisets $\mathcal{A}$ and $\mathcal{B}$, the maximum stabilizer row weight $\tilde{w}=\bar{w}$, and the rate $k/n$. Distances are upper bounds estimated using 50 million random-ISD trials of \texttt{sQetch}~\cite{bhardwaj2026high} on a single NVIDIA RTX~2080~Ti, with $1{,}000$ trials of
\texttt{DistRandCSS}~\cite{pryadko2023qdistrnd} used as a cross-check.}
\end{table*}}

\clearpage
\onecolumngrid
\begingroup
\setlength{\tabcolsep}{12pt}
\renewcommand{\arraystretch}{1.4}
\fontsize{16}{12}\selectfont
\begin{longtable}{|*{3}{>{\columncolor{white}[\tabcolsep]}c|}}
\hline
\rowcolor{zebragray} $[[\mathbf{n},\mathbf{k},(\leq\mathbf{d}_X,\leq\mathbf{d}_Z)]]$ & $\mathcal{A}$ & $\mathcal{B}$ \\
\hline
\endfirsthead

\multicolumn{3}{c}{\tablename\ \thetable{} -- \textit{continued from previous page}} \\
\hline
\rowcolor{zebragray} $[[\mathbf{n},\mathbf{k},(\leq\mathbf{d}_X,\leq\mathbf{d}_Z)]]$ & $\mathcal{A}$ & $\mathcal{B}$ \\
\hline
\endhead

$[[288, 8, (\leq15,\leq15)]]$ &
\shortstack{$\{();();(2,3);(2,3);$\\$(1,2,3);(1,2);(1,2);(1,3,2)\}$} &
$\{();();(2,3);(1,2,3);(1,2);(1,3,2)\}$ \\
\hline
\rowcolor{zebragray} $[[336, 4, (\leq24,\leq15)]]$ &
\shortstack{$\{();();(2,3);(2,3);$\\$(1,2,3);(1,2);(1,2);(1,3,2)\}$} &
\shortstack{$\{();();(2,3);(2,3);$\\$(1,2,3);(1,2,3);(1,2)\}$} \\
\hline
$[[336, 12, (\leq14,\leq14)]]$ &
\shortstack{$\{();();(2,3);(2,3);$\\$(1,2,3);(1,2);(1,2);(1,3,2)\}$} &
\shortstack{$\{();();(2,3);(2,3);$\\$(1,2,3);(1,2,3);(1,2)\}$} \\
\hline
\rowcolor{zebragray} $[[360, 6, (\leq18,\leq18)]]$ &
\shortstack{$\{();(2,5)(3,4);(1,5,4,3,2);$\\$(1,5)(2,4);(1,4)(2,3);(1,2,3,4,5)\}$} &
\textbf{same as $\mathcal{A}$} \\
\hline
$[[384, 24, (\leq14,\leq14)]]$ &
\shortstack{$\{();();(2,3);(2,3);$\\$(1,2,3);(1,2);(1,2);(1,3,2)\}$} &
\textbf{same as $\mathcal{A}$} \\
\hline
\rowcolor{zebragray} $[[384, 8, (\leq16,\leq16)]]$ &
\shortstack{$\{();();(2,3);(2,3);$\\$(1,2,3);(1,2);(1,2);(1,3,2)\}$} &
\textbf{same as $\mathcal{A}$} \\
\hline
$[[432, 16, (\leq15,\leq15)]]$ &
\shortstack{$\{();();(4,6,5);(4,6,5);(1,3,2);$\\$(1,3,2)(4,6,5);(1,2,3);(1,2,3)(4,5,6)\}$} &
\shortstack{$\{();(4,6,5);(4,5,6);$\\$(1,3,2);(1,3,2)(4,6,5);(1,2,3)\}$} \\
\hline
\rowcolor{zebragray} $[[432, 8, (\leq18,\leq18)]]$ &
\shortstack{$\{();();(4,6,5);(4,6,5);(1,3,2);$\\$(1,3,2)(4,6,5);(1,2,3);(1,2,3)(4,5,6)\}$} &
\shortstack{$\{();(4,6,5);(4,5,6);$\\$(1,3,2);(1,3,2)(4,6,5);(1,2,3)\}$} \\
\hline
$[[480, 8, (\leq21,\leq18)]]$ &
\shortstack{$\{();();(2,5)(3,4);(2,5)(3,4);(1,5,4,3,2);$\\$(1,5)(2,4);(1,5)(2,4);(1,4,2,5,3)\}$} &
\shortstack{$\{();(2,5)(3,4);(1,5,4,3,2);$\\$(1,5)(2,4);(1,4)(2,3);(1,2,3,4,5)\}$} \\
\hline
\rowcolor{zebragray} $[[504, 4, (\leq34,\leq12)]]$ &
\shortstack{$\{();();(5,6,7);(5,6,7);(1,4,3,2)(6,7);$\\$(1,4,3,2)(6,7);(1,4,3,2)(5,7)\}$} &
\shortstack{$\{();(1,3)(2,4)(5,6,7);(1,4,3,2)(6,7);$\\$(1,4,3,2)(5,7);(1,4,3,2)(5,6);(1,2,3,4)(6,7)\}$} \\
\hline
$[[588, 6, (\leq28,\leq23)]]$ &
\shortstack{$\{();(2,7)(3,6)(4,5);(1,5,2,6,3,7,4);$\\$(1,5)(2,4)(6,7);(1,2)(3,7)(4,6);$\\$(1,6,4,2,7,5,3);(1,3)(4,7)(5,6)\}$} &
\shortstack{$\{();(2,7)(3,6)(4,5);(1,5,2,6,3,7,4);$\\$(1,5)(2,4)(6,7);(1,2)(3,7)(4,6);$\\$(1,6,4,2,7,5,3)\}$} \\
\hline
\rowcolor{zebragray} $[[588, 14, (\leq20,\leq18)]]$ &
\shortstack{$\{();(2,7)(3,6)(4,5);(1,5,2,6,3,7,4);$\\$(1,5)(2,4)(6,7);(1,2)(3,7)(4,6);$\\$(1,6,4,2,7,5,3);(1,3)(4,7)(5,6)\}$} &
\shortstack{$\{();(2,7)(3,6)(4,5);(1,5,2,6,3,7,4);$\\$(1,5)(2,4)(6,7);(1,2)(3,7)(4,6);$\\$(1,6,4,2,7,5,3)\}$} \\
\hline
$[[640, 24, (\leq14,\leq14)]]$ &
\shortstack{$\{();();(2,5)(3,4);(2,5)(3,4);(1,5,4,3,2);$\\$(1,5)(2,4);(1,5)(2,4);(1,4,2,5,3)\}$} &
\textbf{same as $\mathcal{A}$} \\
\hline
\rowcolor{zebragray} $[[648, 26, (\leq12,\leq15)]]$ &
\shortstack{$\{();(5,6);(4,5,6);$\\$(1,3,2)(4,5);(1,2,3)(4,5);(1,3,2)(4,6)\}$} &
\textbf{same as $\mathcal{A}$} \\
\hline
$[[648, 2, (\leq18,\leq32)]]$ &
\shortstack{$\{();(5,6);(4,5,6);$\\$(1,3,2)(4,5);(1,2,3)(4,5);(1,3,2)(4,6)\}$} &
\textbf{same as $\mathcal{A}$} \\
\hline
\rowcolor{zebragray} $[[648, 4, (\leq25,\leq21)]]$ &
\shortstack{$\{();(5,6);(4,5,6);$\\$(1,3,2)(4,5);(1,2,3)(4,5);(1,3,2)(4,6)\}$} &
\textbf{same as $\mathcal{A}$} \\
\hline
$[[648, 10, (\leq18,\leq20)]]$ &
\shortstack{$\{();(5,6);(4,5,6);$\\$(1,3,2)(4,5);(1,2,3)(4,5);(1,3,2)(4,6)\}$} &
\textbf{same as $\mathcal{A}$} \\
\hline
\rowcolor{zebragray} $[[756, 17, (\leq9,\leq21)]]$ &
\shortstack{$\{();();(5,6,7);(5,6,7);(1,4,3,2)(6,7);$\\$(1,4,3,2)(6,7);(1,4,3,2)(5,7)\}$} &
\shortstack{$\{();();(5,6,7);(5,6,7);(1,4,3,2)(6,7);$\\$(1,4,3,2)(5,7);(1,4,3,2)(5,6);$\\$(1,2,3,4)(6,7);(1,2,3,4)(5,7)\}$} \\
\hline
$[[756, 30, (\leq21,\leq12)]]$ &
\shortstack{$\{();();(1,3)(2,4)(5,6,7);(1,3)(2,4)(5,7,6);$\\$(1,4,3,2)(6,7);(1,4,3,2)(6,7);(1,4,3,2)(5,7)\}$} &
\shortstack{$\{();();(5,6,7);(5,6,7);(1,4,3,2)(6,7);$\\$(1,4,3,2)(5,7);(1,4,3,2)(5,6);$\\$(1,2,3,4)(6,7);(1,2,3,4)(5,7)\}$} \\
\hline
\rowcolor{zebragray} $[[756, 43, (\leq12,\leq12)]]$ &
\shortstack{$\{();();(1,3)(2,4)(5,6,7);(1,3)(2,4)(5,7,6);$\\$(1,4,3,2)(6,7);(1,4,3,2)(6,7);(1,4,3,2)(5,7)\}$} &
\shortstack{$\{();();(5,6,7);(5,6,7);(1,4,3,2)(6,7);$\\$(1,4,3,2)(5,7);(1,4,3,2)(5,6);$\\$(1,2,3,4)(6,7);(1,2,3,4)(5,7)\}$} \\
\hline
$[[756, 10, (\leq9,\leq42)]]$ &
\shortstack{$\{();();(5,6,7);(5,6,7);(1,4,3,2)(6,7);$\\$(1,4,3,2)(6,7);(1,4,3,2)(5,7)\}$} &
\shortstack{$\{();();(5,6,7);(5,6,7);(1,4,3,2)(6,7);$\\$(1,4,3,2)(5,7);(1,4,3,2)(5,6);$\\$(1,2,3,4)(6,7);(1,2,3,4)(5,7)\}$} \\
\hline
\rowcolor{zebragray} $[[768, 24, (\leq16,\leq16)]]$ &
\shortstack{$\{();();(5,6,7);(5,6,7);(1,4,3,2)(6,7);$\\$(1,4,3,2)(5,7);(1,4,3,2)(5,6);(1,2,3,4)(6,7)\}$} &
\textbf{same as $\mathcal{A}$} \\
\hline
$[[768, 32, (\leq16,\leq16)]]$ &
\shortstack{$\{();(2,4,3);(2,4,3);(1,4)(2,3);$\\$(1,4,2);(1,4,3);(1,2,3);(1,2,3)\}$} &
\shortstack{$\{();(2,4,3);(2,3,4);(1,4)(2,3);$\\$(1,4,2);(1,4,3);(1,2)(3,4);(1,3,2)\}$} \\
\hline
\rowcolor{zebragray} $[[864, 24, (\leq21,\leq18)]]$ &
\shortstack{$\{();();(5,6);(5,6);(1,3,2)(4,5,6);$\\$(1,3,2)(4,5);(1,3,2)(4,5);(1,3,2)(4,6,5)\}$} &
\shortstack{$\{();(5,6);(4,5,6);(1,3,2)(4,5);$\\$(1,2,3)(4,5);(1,3,2)(4,6)\}$} \\
\hline
$[[864, 16, (\leq27,\leq18)]]$ &
\shortstack{$\{();();(5,6);(5,6);(1,3,2)(4,5,6);$\\$(1,3,2)(4,5);(1,3,2)(4,5);(1,3,2)(4,6,5)\}$} &
\shortstack{$\{();(5,6);(4,5,6);(1,3,2)(4,5);$\\$(1,2,3)(4,5);(1,3,2)(4,6)\}$} \\
\hline
\rowcolor{zebragray} $[[864, 8, (\leq27,\leq32)]]$ &
\shortstack{$\{();();(5,6);(5,6);(1,3,2)(4,5,6);$\\$(1,3,2)(4,5);(1,3,2)(4,5);(1,3,2)(4,6,5)\}$} &
\shortstack{$\{();(5,6);(4,5,6);(1,3,2)(4,5);$\\$(1,2,3)(4,5);(1,3,2)(4,6)\}$} \\
\hline
$[[864, 8, (\leq27,\leq47)]]$ &
\shortstack{$\{();();(5,6);(5,6);(1,3,2)(4,5,6);$\\$(1,3,2)(4,5);(1,3,2)(4,5);(1,3,2)(4,6,5)\}$} &
\shortstack{$\{();(5,6);(4,5,6);(1,3,2)(4,5);$\\$(1,2,3)(4,5);(1,3,2)(4,6)\}$} \\
\hline
\rowcolor{zebragray} $[[896, 40, (\leq16,\leq12)]]$ &
\shortstack{$\{(5,8,7,6);(1,3)(2,4);(1,3)(2,4)(5,7)(6,8);$\\$(1,3)(2,4)(5,6,7,8);(1,4,3,2)(6,8);$\\$(1,4,3,2)(5,6)(7,8);(1,2,3,4)(6,8);$\\$(1,2,3,4)(5,6)(7,8)\}$} &
\shortstack{$\{();();(5,8,7,6);(5,8,7,6);$\\$(1,4,3,2)(6,8);(1,4,3,2)(5,6)(7,8);$\\$(1,2,3,4)(5,6)(7,8)\}$} \\
\hline
$[[1024, 16, (\leq16,\leq16)]]$ &
\shortstack{$\{(5,8,7,6);(1,3)(2,4);(1,3)(2,4)(5,7)(6,8);$\\$(1,3)(2,4)(5,6,7,8);(1,4,3,2)(6,8);$\\$(1,4,3,2)(5,6)(7,8);(1,2,3,4)(6,8);$\\$(1,2,3,4)(5,6)(7,8)\}$} &
\shortstack{$\{();();(5,8,7,6);(5,6,7,8);$\\$(1,4,3,2)(6,8);(1,4,3,2)(5,6)(7,8);$\\$(1,2,3,4)(6,8);(1,2,3,4)(5,6)(7,8)\}$} \\
\hline
\rowcolor{zebragray} $[[1280, 24, (\leq16,\leq16)]]$ &
\shortstack{$\{();();(5,9,8,7,6);(5,9,8,7,6);$\\$(1,4,3,2)(6,9)(7,8);(1,4,3,2)(5,6)(7,9);$\\$(1,4,3,2)(5,7)(8,9);(1,4,3,2)(5,8)(6,7)\}$} &
\shortstack{$\{();();(5,9,8,7,6);(5,9,8,7,6);$\\$(1,4,3,2)(6,9)(7,8);(1,4,3,2)(5,6)(7,9);$\\$(1,4,3,2)(5,7)(8,9);(1,4,3,2)(5,8)(6,7)\}$} \\
\hline
\caption{\textbf{Multisets $\mathcal{A}, \mathcal{B} \subseteq \mathcal{G}$ for the code instances of Table~\ref{tab:qt-codes-main}.} For each code $[[n, k, (\le d_X, \le d_Z)]]$ we list the multisets $\mathcal{A}$ and $\mathcal{B}$ used in the lifted construction of Eq.~\eqref{eq:lifted}, with elements written in cycle notation under the identification $\mathcal{G} \hookrightarrow S_m$ obtained via \texttt{isomorphism(PermGroup, small\_group($|\mathcal{G}|$, $\cdot$))} in \texttt{Oscar.jl}; $()$ denotes the identity. Rows are ordered by blocklength $n$. Together with the corresponding local codes $H_0, H_1$ recorded in Table~\ref{tab:qt-codes-main} and the column permutations $\pi_A, \pi_B$, these multisets fully specify the parity-check matrices $H_X, H_Z$ of each code.}
\label{tab:multisets-of-lifted-quantum-codes}
\end{longtable}
\clearpage
\endgroup
\onecolumngrid

\clearpage 
\onecolumngrid 
\begingroup 
\setlength{\tabcolsep}{12pt} 
\renewcommand{\arraystretch}{1.2} 
\fontsize{14}{18}\selectfont
\begin{longtable}{|*{3}{>{\columncolor{white}[\tabcolsep]}c|}} 
\hline 
\rowcolor{zebragray} 
$[[\mathbf{n},\mathbf{k},(\leq\mathbf{d}_X,\leq\mathbf{d}_Z)]]$ & $\mathcal{A}$ & $\mathcal{B}$ \\ 
\hline 
\endfirsthead 
\multicolumn{3}{c}{\tablename\ \thetable{} -- \textit{continued from previous page}} \\ 
\hline 
\rowcolor{zebragray} 
$[[\mathbf{n},\mathbf{k},(\leq\mathbf{d}_X,\leq\mathbf{d}_Z)]]$ & $\mathcal{A}$ & $\mathcal{B}$ \\ 
\hline 
\endhead 
$[[480, 8, (\leq 21, \leq 21)]] $ & 
\shortstack{$\{();();(2,5)(3,4);(2,5)(3,4);$\\$(1,5,4,3,2);(1,5)(2,4);(1,5)(2,4);(1,4,2,5,3)\}$} & 
\shortstack{$\{();(2,5)(3,4);(1,5,4,3,2);$\\$(1,5)(2,4);(1,4)(2,3);(1,2,3,4,5)\}$} \\ 
\hline
\rowcolor{zebragray} $[[504, 4, (\leq 36, \leq 27)]]$ & 
\shortstack{$\{();();(4,6,5);(4,6,5);(1,3,2);$\\$(1,3,2)(4,6,5);(1,2,3);(1,2,3)(4,5,6)\}$} & 
\shortstack{$\{();();(4,6,5);(4,6,5);(1,3,2);$\\$(1,3,2);(1,3,2)(4,6,5)\}$} \\ 
\hline
 $[[672, 4, (\leq 48, \leq 28)]]$ & 
\shortstack{$\{();();(5,6,7);(5,6,7);(1,4,3,2)(6,7);$\\$(1,4,3,2)(5,7);(1,4,3,2)(5,6);(1,2,3,4)(6,7)\}$} & 
\shortstack{$\{();();(5,6,7);(5,6,7);(1,4,3,2)(6,7);$\\$(1,4,3,2)(6,7);(1,4,3,2)(5,7)\}$} \\ 
\hline
\rowcolor{zebragray} $[720, 6, (\leq 30, \leq 30)]]$ & 
\shortstack{$\{();(2,4,5,3);(1,5,4,3,2);$\\$(1,5,2,3);(1,3,4,2);(1,2,3,4,5)\}$} & 
\shortstack{$\{();(2,4,5,3);(1,5,4,3,2);$\\$(1,5,2,3);(1,3,4,2);(1,2,3,4,5)\}$} \\ 
\hline
 $[[864, 8, (\leq 39, \leq 31)]]$ & 
\shortstack{$\{();();(5,6);(5,6);(1,3,2)(4,5,6);$\\$(1,3,2)(4,5);(1,3,2)(4,5);(1,3,2)(4,6,5)\}$} & 
\shortstack{$\{();(5,6);(4,5,6);(1,3,2)(4,5);$\\$(1,2,3)(4,5);(1,3,2)(4,6)\}$} \\ 
\hline
\rowcolor{zebragray} $[[864, 16, (\leq 36, \leq 32)]]$ & 
\shortstack{$\{();();(5,6);(5,6);(1,3,2)(4,5,6);$\\$(1,3,2)(4,5);(1,3,2)(4,5);(1,3,2)(4,6,5)\}$} & 
\shortstack{$\{();(5,6);(4,5,6);(1,3,2)(4,5);$\\$(1,2,3)(4,5);(1,3,2)(4,6)\}$} \\ 
\hline
 $[[864, 8, (\leq 42, \leq 40)]]$ & 
\shortstack{$\{();();(5,6);(5,6);(1,3,2)(4,5,6);$\\$(1,3,2)(4,5);(1,3,2)(4,5);(1,3,2)(4,6,5)\}$} & 
\shortstack{$\{();(5,6);(4,5,6);(1,3,2)(4,5);$\\$(1,2,3)(4,5);(1,3,2)(4,6)\}$} \\ 
\hline
\rowcolor{zebragray} $[1120, 4, (\leq 80, \leq 35)]]$ & 
\shortstack{$\{();();(5,9,8,7,6);(5,9,8,7,6);$\\$(1,4,3,2)(6,9)(7,8);(1,4,3,2)(5,6)(7,9);$\\$(1,4,3,2)(5,7)(8,9);(1,4,3,2)(5,8)(6,7)\}$} & 
\shortstack{$\{();();(5,9,8,7,6);(5,9,8,7,6);$\\$(1,4,3,2)(6,9)(7,8);(1,4,3,2)(6,9)(7,8);$\\$(1,4,3,2)(5,6)(7,9)\}$} \\ 
\hline
\caption{\textbf{Multisets $\mathcal{A}, \mathcal{B} \subseteq \mathcal{G}$ for the code instances of Table~\ref{tab:qt-codes-breaking-the-barrier}.} As Table~\ref{tab:multisets-of-lifted-quantum-codes}, but for the codes of Table~\ref{tab:qt-codes-breaking-the-barrier}; together with the corresponding local codes $H_0, H_1$ and column permutations $\pi_A, \pi_B$ recorded there, these multisets fully specify $H_X, H_Z$.}
\label{tab:multisets-of-lifted-quantum-codes-breaking-the-barrier}
\end{longtable} 

\begin{table*}[t]
\centering
\label{tab:perms}
\centering
\footnotesize
\setlength{\tabcolsep}{12pt}
\renewcommand{\arraystretch}{2}
\begin{tabular}{|*{3}{>{\columncolor{white}[\tabcolsep]}c|}}
\hline
\rowcolor{zebragray} $[[n, k, (\le d_X, \le d_Z)]]$ & $\pi_A$ & $\pi_B$ \\
\hline
$[[480, 8, (\leq 21, \leq 21)]] $ & [1,2,3,4,5,6,7,8] & [1,2,5,6,3,4]\\ 
\hline
\rowcolor{zebragray} $[[504, 4, (\leq 36, \leq 27)]]$ & [1,2,3,4,5,6,7,8] & [1,2,3,5,6,7,4]\\ 
\hline
 $[[672, 4, (\leq 48, \leq 28)]]$ & [1,2,3,4,5,6,7,8] & [1,2,3,5,6,7,4] \\  
\hline
\rowcolor{zebragray} $[720, 6, (\leq 30, \leq 30)]]$ & [1,2,3,4,5,6] & [1,2,5,4,6,3] \\ 
\hline
 $[[864, 8, (\leq 39, \leq 31)]]$ & [1,2,3,4,5,6,7,8] & [1,2,4,5,3,6]\\  
\hline
\rowcolor{zebragray} $[[864, 16, (\leq 36, \leq 32)]]$ & [1,2,3,4,5,6,7,8] & [1,2,6,4,5,3] \\ 
\hline
 $[[864, 8, (\leq 42, \leq 40)]]$ & [1,2,3,4,5,6;7,8] & 1,2,6,5,4,3] \\ 
\hline
\rowcolor{zebragray} $[1120, 4, (\leq 80, \leq 35)]]$ & [1,2,3,4,5,6,7,8] & [1,2,3,6,4,7,5] \\ 
\hline
\end{tabular}
\caption{Column permutations $\pi_A, \pi_B$ for the codes o Table~\ref{tab:qt-codes-breaking-the-barrier}. As Table~\ref{tab:permutations-of-lifted-quantum-codes}, but for the codes of Table~\ref{tab:qt-codes-breaking-the-barrier}; together with the group $\mathcal{G}$, the multisets $\mathcal{A}, \mathcal{B}$ of
Table~\ref{tab:multisets-of-lifted-quantum-codes-breaking-the-barrier}, and the local codes $H_0, H_0'$ recorded there, these permutations fully specify each code via Eq.~\eqref{eq:lifted}.}
\label{tab:permutations-of-lifted-quantum-codes-breaking-the-barrier}
\end{table*}
\clearpage 
\endgroup 
\onecolumngrid

\begin{table*}[t]
\centering
\label{tab:perms}
\centering
\footnotesize
\setlength{\tabcolsep}{12pt}
\renewcommand{\arraystretch}{2}
\begin{tabular}{|*{3}{>{\columncolor{white}[\tabcolsep]}c|}}
\hline
\rowcolor{zebragray} $[[n, k, (\le d_X, \le d_Z)]]$ & $\pi_A$ & $\pi_B$ \\
\hline
$[[288, 8, (\le 15, \le 15)]]$ & $[1,2,3,4,5,6,7,8]$ & $[1,2,4,3,6,5]$ \\
\hline
\rowcolor{zebragray} $[[336, 4, (\le 24, \le 15)]]$ & $[1,2,3,4,5,6,7,8]$ & $[1,2,3,5,6,7,4]$ \\
\hline
$[[336, 12, (\le 14, \le 14)]]$ & $[1,2,3,4,5,6,7,8]$ & $[1,2,3,5,4,7,6]$ \\
\hline
\rowcolor{zebragray}$[[360, 6, (\le 18, \le 18)]]$ & $[1,2,3,4,5,6]$ & $[1,2,4,5,3,6]$ \\
\hline
$[[384, 24, (\le 14, \le 14)]]$ & $[1,2,3,4,5,6,7,8]$ & $[1,2,3,4,5,6,8,7]$ \\
\hline
\rowcolor{zebragray} $[[384, 8, (\le 16, \le 16)]]$ & $[1,2,3,4,5,6,7,8]$ & $[1,2,3,4,6,7,8,5]$ \\
\hline
\rowcolor{zebragray}$[[432, 16, (\le 15, \le 15)]]$ & $[1,2,3,4,5,6,7,8]$ & $[1,2,4,3,5,6]$ \\
\hline
$[[432, 8, (\le 18, \le 18)]]$ & $[1,2,3,4,5,6,7,8]$ & $[1,2,4,6,5,3]$ \\
\hline
\rowcolor{zebragray}$[[480, 8, (\le 21, \le 18)]]$ & $[1,2,3,4,5,6,7,8]$ & $[1,2,3,6,4,5]$ \\
\hline
$[[504, 4, (\le 34, \le 12)]]$ & $[1,2,3,4,5,6,7]$ & $[1,2,6;3,4,5]$ \\
\hline
\rowcolor{zebragray}$[[588, 6, (\le 28, \le 23)]]$ & $[1,2,3,4,5,6,7]$ & $[1,2,4,5,6,3]$ \\
\hline
$[[588, 14, (\le 20, \le 18)]]$ & $[1,2,3,4,5,6,7]$ & $[1,2,3,6,4,5]$ \\
\hline
\rowcolor{zebragray}$[[640, 24, (\le 14, \le 14)]]$ & $[1,2,3,4,5,6,7,8]$ & $[1,2,3,4,5,6,8,7]$ \\
\hline
$[[648, 26, (\le 12, \le 15)]]$ & $[1,2,3,4,5,6]$ & $[1,2,3,4,6,5]$ \\
\hline
\rowcolor{zebragray}$[[648, 14, (\le 16, \le 15)]]$ & $[1,2,3,4,5,6]$ & $[1,2,4,3,5,6]$ \\
\hline
$[[648, 2, (\le 18, \le 32)]]$ & $[1,2,3,4,5,6]$ & $[1,2,4,6,3,5]$ \\
\hline
\rowcolor{zebragray}$[[648, 4, (\le 26, \le 21)]]$ & $[1,2,3,4,5,6]$ & $[1,2,6,3,4,5]$ \\
\hline
$[[648, 10, (\le 18, \le 20)]]$ & $[1,2,3,4,5,6]$ & $[1,2,5,6,4,3]$ \\
\hline
\rowcolor{zebragray}$[[756, 17, (\le 9, \le 21)]]$ & $[1,2,3,4,5,6,7]$ & $[1,2,3,4,5,6,8,9,7]$ \\
\hline
$[[756, 30, (\le 21, \le 12)]]$ & $[1,2,3,4,5,6,7]$ & $[1,2,3,4,5,7,6,9,8]$ \\
\hline
\rowcolor{zebragray}$[[756, 43, (\le 12, \le 12)]]$ & $[1,2,3,4,5,6,7]$ & $[1,2,3,4,5,6,7,8,9]$ \\
\hline
$[[756, 10, (\le 9, \le 42)]]$ & $[1,2,3,4,5,6,7]$ & $[1,2,3,4,5,7,8,9,6]$ \\
\hline
\rowcolor{zebragray}$[[768, 24, (\le 16, \le 16)]]$ & $[1,2,3,4,5,6,7,8]$ & $[1,2,3,4,5,7,8,6]$ \\
\hline
$[[768, 32, (\le 16, \le 16)]]$ & $[1,2,3,4,5,6,7,8]$ & $[1,2,3,4,5,6,7,8]$ \\
\hline
\rowcolor{zebragray}$[[864, 24, (\le 21, \le 18)]]$ & $[1,2,3,4,5,6,7,8]$ & $[1,2,3,4,5,6]$ \\
\hline
$[[864, 16, (\le 27, \le 18)]]$ & $[1,2,3,4,5,6,7,8]$ & $[1,2,3,5,4,6]$ \\
\hline
\rowcolor{zebragray}$[[864, 8, (\le 27, \le 32)]]$ & $[1,2,3,4,5,6,7,8]$ & $[1,2,3,5,6,4]$ \\
\hline
$[[864, 8, (\le 27, \le 47)]]$ & $[1,2,3,4,5,6,7,8]$ & $[1,2,3,6,4,5]$ \\
\hline
\rowcolor{zebragray}$[[896, 40, (\le 16, \le 12)]]$ & $[1,2,3,4,5,6,7,8]$ & $[1,2,3,4,5,6,7]$ \\
\hline
$[[1024, 16, (\le 16, \le 16)]]$ & $[1,2,3,4,5,6,7,8]$ & $[1,2,3,4,5,8,6,7]$ \\
\hline
\rowcolor{zebragray}$[[1280, 24, (\le 16, \le 16)]]$ & $[1,2,3,4,5,6,7,8]$ & $[1,2,3,4,5,6,8,7]$ \\
\hline
\end{tabular}
\caption{Column permutations $\pi_A, \pi_B$ for the codes of Table~\ref{tab:qt-codes-main}. Each permutation is given as an integer vector of the same length as the corresponding local codelength ($n_A$ or $n_B$), read as follows: the entry at position
$i$ specifies the index of the column of $H_0$ that becomes column $i$ of $H_1$, so that $H_1 = H_0[\,\cdot\,,\,\pi_A]$ and, in parallel, $H_1' = H_0'[\,\cdot\,,\,\pi_B]$ in the standard Julia column-indexing convention. For example, $\pi_A = [1,2,4,3,6,5]$
means the first two columns of $H_1$ are the first two columns of $H_0$, the third column of $H_1$ is the fourth column of $H_0$ (and vice versa), and the fifth and sixth are similarly swapped; $\pi_A = [1,2,\ldots,n_A]$ corresponds to $H_1 = H_0$. The same
permutation acts in parallel on the associated generator matrices, $G_1 = G_0[\,\cdot\,,\,\pi_A]$ and $G_1' = G_0'[\,\cdot\,,\,\pi_B]$. Together with the group $\mathcal{G}$, the multisets $\mathcal{A}, \mathcal{B}$ of Table~\ref{tab:multisets-of-lifted-quantum-codes}, and the local codes $H_0, H_0'$ of Table~\ref{tab:qt-codes-main}, these permutations fully specify each code via Eq.~\eqref{eq:lifted}.}
\label{tab:permutations-of-lifted-quantum-codes}
\end{table*}

\end{document}